\documentclass[pra,aps,twocolumn,nopacs]{revtex4}

\usepackage{amsmath}
\usepackage{amssymb,mathrsfs}
\usepackage{graphicx}
\usepackage[dvipsnames]{xcolor}

\usepackage[utf8]{inputenc}
\usepackage[english,main=russian]{babel}

\newcommand{\prt}{\partial}
\newcommand{\om}{\omega}

\newcommand{\eps}{\varepsilon}
\newcommand{\ox}{{\overline{x}}}

\newcommand {\sn} {\mathrm{s n}}

\newcommand {\ra} {\rightarrow}

\begin{document}
\title{Gurevich-Pitaevskii problem and its development\footnote{{\it Usp. Fiz. Nauk}
{\bf 191} 52-87 (2021); {\it Phys.-Uspekhi} {\bf 64} 48-82 (2021)}}

\author{A. M. Kamchatnov\\
{\it
Institute of Spectroscopy,
Russian Academy of Sciences\\
ul. Fizicheskaya 5, 108840 Troitsk, Moscow, Russia}}


\begin{abstract}
We present an introduction to the theory of dispersive
shock waves in the framework of the approach proposed by
Gurevich and Pitaevskii (Zh. Eksp. Teor. Fiz., {\bf 65,} 590
(1973) [Sov. Phys. JETP, {\bf 38,} 291 (1974)]) based on the
Whitham theory of modulation of nonlinear waves. We explain
how Whitham equations for a periodic solution can be
derived for the Korteweg-de Vries equation and outline some
elementary methods to solve them. We illustrate this approach
with solutions to the main problems discussed by Gurevich and
Pitaevskii. We consider a generalization of the theory to systems
with weak dissipation and discuss the theory of dispersive
shock waves for the Gross-Pitaevskii equation.
\end{abstract}

\maketitle

{\it Dedicated to the 90th birthday of A.~V.~Gurevich}

\bigskip

{\bf\small{ Content

\smallskip\noindent
1. Introduction

\smallskip\noindent
2. Korteweg-de Vries equation

\smallskip\noindent
3. Modulation of linear waves

\smallskip\noindent
4.  Whitham theory

\smallskip\noindent
5. Generalized hodograph method

\smallskip\noindent
6. Formulation of the Gurevich-Pitaevskii problem

\smallskip\noindent
7.  Evolution of the initial discontinuity in the Korteweg-de Vries theory

\smallskip\noindent
8. Breaking of the wave with a parabolic profile

\smallskip\noindent
9. Breaking of a cubic profile

\smallskip\noindent
10. Motion of edges of dispersive shock waves

\smallskip\noindent
11.  Theorem on the number of oscillations in dispersive shock waves

\smallskip\noindent
12. Theory of dispersive shock waves for the Korteweg-de Vries equation
with dissipation

\smallskip\noindent
13. Gross-Pitaevskii equation

\smallskip\noindent
14. Evolution of the initial discontinuity in the Gross-Pitaevskii theory

\smallskip\noindent
15. Piston problem

\smallskip\noindent
16. Uniformly accelerated piston problem

\smallskip\noindent
17. Motion of edges of `quasi-simple' dispersive shock waves

\smallskip\noindent
18. Breaking of a cubic profile in the Gross-Pitaevskii theory

\smallskip\noindent
19.  Conclusions

\smallskip\noindent
References }}

\section{Introduction}\label{intro}

Any physical theory grows out of particular observations and
attempts to interpret them, solving specific problems and
gradually constructing generalizations. But at the same time,
studies can be singled out in the development of each theory
that served to transform a collection of particular results and
vague ideas into a field of science, with its own physical ideas
and tools that allow posing and solving problems characteristic
of just that field. In the field of nonlinear physics, known
under its modern name as the theory of {\it  dispersive shock waves} (DSWs),
this role goes to Gurevich and Pitaevskii's 1973 paper \cite{gp-73}.
They formulated a general approach to constructing
a theoretical picture of the formation and evolution of such
waves based on the Whitham theory~\cite{whitham-65} of modulation
of nonlinear waves, and solved several typical
problems that yielded a quantitative description of typical
DSW structures. The Gurevich-Pitaevskii problem can therefore be
understood both as the general approach to the DSW
theory proposed by these authors and as the particular
problems that were posed and solved in \cite{gp-73}
and have since then found numerous applications in explaining various
physical observations underlying the subsequent development of the theory.

The aim of this paper is to give a sufficiently detailed
introduction to that domain of nonlinear studies concentrated on a
detailed presentation of Gurevich and Pitaevskii's work \cite{gp-73}
and related studies. But first we discuss the principal
stages in the formation of the DSW theory that eventually
resulted in the appearance of paper \cite{gp-73}.

Dispersive shock waves are not very common in the world around us.Their
first observations were apparently associated with the
formation of wave-like structures near the tidal wave front
when a wave was advancing sufficiently fast into river beds or
narrow straits. This effect was called {\it the undular bore} and for
an extended period of time was apparently studied by a
dedicated community of researchers and engineers dealing
with river hydrodynamics. Still, some fundamental facts
about such bores have been revealed. In particular, the
leading swell of water at the bore front was identified with a
solitary wave that had first been observed by Scott Russell~\cite{russel-1844}
and then explained by Boussinesq~\cite{bouss-1871a},
Lord Rayleigh~\cite{rayleigh-1876}, and Korteweg and de Vries~\cite{kdv}.
Benjamin and Lighthill~\cite{bl-54} attempted to clarify the
conditions under which the undular bore can be described as
a modulated periodic solution of the Korteweg-de Vries
(KdV) equation. It was then assumed that the modulation of
a periodic solution called the `cnoidal wave' by the authors of
\cite{kdv} was caused by dissipative processes in the wave-like flow of
the liquid. It nevertheless transpired from those early works that
explaining the formation of an undular bore requires taking
the interplay of dispersion and nonlinearity effects into
account for shallow-water waves, assuming an essential
role of dissipation effects in explaining the wave modulation
and the formation of turbulent bores at sufficiently
high amplitudes of the tidal wave. However, the problem of
a theoretical description of undular bores did not garner
much attention outside the community of experts. For
example, in classic books \cite{lamb,stoker}, where various phenomena
related to water waves are described in detail, that problem
is not even mentioned.

The situation changed due to the development of modern nonlinear physics.
Back the early 1960s, it became clear that
solitary waves, or `solitons' if using modern terminology, can
propagate in different physical systems, in plasmas in
particular \cite{gm-60,vvs-61}, and the KdV equation has a universal
character and finds applications in very diverse physical
situations with weak dispersion and small nonlinearity.
Soliton solutions of the equations of plasma dynamics, in
both their original form and in the KdV approximation
without dissipation, propagate with their shape being
unchanged. If there is dissipation in the system, then
propagation of {\it shock waves} becomes possible, such that the
transition layer width is proportional to the dissipation level.
Therefore, the width of such a layer can reach a magnitude of
the order of the characteristic width of the soliton. Competition
then occurs between dispersive and dissipative effects,
and the transition layer is also formed due to the occurrence
of a domain of soliton-type nonlinear oscillations.
As a result, we arrive at the notion of a shock wave in
which the transition from one state of the plasma to another
occurs via a stationary wave structure of strong nonlinear
oscillations. The wave length in this structure is determined by
the balance of dispersion and nonlinearity, and the general
width of the shock wave, i.e., the characteristic length at
which oscillations are modulated, is inversely proportional to
the magnitude of dissipation effects. Such a picture of shock
waves was proposed by Sagdeev~\cite{sagdeev}, and it was observed in
the evolution of ion-sound pulses in plasmas~\cite{ABS-68,TBI-70}.

Gurevich and Pitaevskii took a different path to approach
the problem. In the second half of the 1960s and early 1970s,
they published (in part jointly with Pariiskaya) a series of
papers \cite{gpp-65a,gpp-65b,gp-69,gp-71},  on the dynamics of
rarefied plasmas in the framework of kinetic theory. In this theory,
the plasma state is described by a distribution function of ions
over positions and
velocities, and hot electrons are in thermal equilibrium and
are distributed over space in accordance with the Boltzmann
distribution, with the potential determined by the Poisson
equation, with the charge density equal to the difference
between ion and electron charge distributions. Particle
collisions are disregarded in this theory, and hence dissipative
effects are absent, but it is nevertheless obvious that
nonlinear and dispersive effects are entirely present. A
characteristic feature of this problem setting compared with
that considered above is that the focus is shifted to the
non-stationary dynamics, different from the stationary propagation
of periodic waves, solitons, or stationary DSWs, in
which modulation of an oscillating structure was caused by
dissipation. In their consecutive treatment of problems starting with a
simple self-similar expansion of plasma into a vacuum \cite{gpp-65a,gpp-65b}
and further on to more complicated dynamics of simple waves \cite{gp-69},
where the formation of an infinitely steep front of the
distribution function had already been observed, Gurevich
and Pitaevskii concluded in \cite{gp-71} that, in the kinetics of
rarefied plasmas, the breaking of an analogue of a simple
hydrodynamic wave leads to the formation of an evolving
oscillation domain with the wavelength of the order of the
Debye radius; moreover, if the wave amplitude is small (but
not infinitesimally small), then the dynamics of that domain
are described by the KdV equation, which, ignoring the
dispersion, also leads to breaking solutions.
A natural conclusion was that when taking dispersion into
account the domain of multi-valuedness is to be superseded by
an oscillatory domain, with a series of solitons forming on its
front in accordance with the balance between nonlinear and
dispersive effects, whereas, farther away from the front, the
oscillation amplitude decreases, and the solution approaches
the dispersionless one. The list of references on the theory of
the KdV equation given in\cite{gp-71}, contains a reference to
Whitham's paper~\cite{whitham-65}.

Such were the preparations to create the DSW theory in~\cite{gp-73}:
on the one hand, the problem was reduced to the theory of
waves satisfying the KdV equation, which made that paper
part of the theory of nonlinear waves that was vigorously
being developed at the time, and on the other hand, a new
problem setup was focused on the question of non-stationary
evolution of the wave after its breaking without taking
dissipative processes into account. Just that problem was
solved in \cite{gp-73} for waves whose evolution is governed by the
KdV equation. Subsequently, this theory was extended to
numerous other equations and has found diverse applications,
ranging from the physics of water waves to nonlinear
optics and the dynamics of the Bose-Einstein condensate.
This is why paper~\cite{gp-73} has many
times been cited in both the physical and mathematical
literature. In this paper, we present the basic ideas of
Gurevich and Pitaevskii's approach to the DSW theory,
while staying within methods that are standard for theoretical physics.

\section{Korteweg-de Vries equation}

As noted in the Introduction, the KdV equation is a universal
equation for nonlinear waves, which often arises in the
leading approximation in small nonlinearity and weak
dispersion. Because Gurevich and Pitaevskii's work that
resulted in creating the DSW theory is written in the context
of plasma wave physics, we here give a simple derivation of
the KdV equation for ion-sound waves in a two-temperature
plasma, with the electron temperature $T_e$ being much higher
than the ion temperature. The thermal motion of ions can
then be disregarded and their dynamics can be described by
standard hydrodynamic equations, with the separation of ion
and electron charges taken into account.

We let $\rho$ denote the number of ions per unit volume and
$M$ denote their mass, and assume for simplicity that they have a
unit charge $e$ and the plasma moves along the $x$ axis with a
speed $u$. As is known (see, e.g., \cite{LL-10}), such a plasma has an
intrinsic parameter with the dimension of length, the Debye
radius
\begin{equation}\label{eq2.8}
 r_D=\sqrt{{T_e}/{4\pi e^2\rho_0}},
\end{equation}
whose ratio to the characteristic wavelength determines the
magnitude of dispersive effects ($\rho_0$ is the equilibrium density
in the absence of a wave). For convenience, we discuss the
nonlinear and dispersive effects separately.

Small deviations from equilibrium are described by linear
harmonic waves with $\rho-\rho_0, u\propto\exp[i(k x-\omega t)]$, and we
easily find their dispersion law as \cite{LL-10}
\begin{equation}\label{eq2.9}
 \omega=\pm\sqrt\frac{T_e}{M}\frac{k}{\sqrt{1+r_D^2 k^2}},
\end{equation}
where the choice of sign is determined by the wave propagation direction.
Hence, it follows that dispersive effects are
small when the wavelength $2\pi/k$ is much greater than the
Debye radius $r_D$. The first terms of the expansion in the small
parameter $kr_D$ give
\begin{equation}\label{eq2.10}
 \omega=\pm c_0k\left(1-\tfrac12{r_D^2}k^2\right), \quad kr_D\ll 1,
\end{equation}
where $c_0=\sqrt{{T_e}/{M}}$ is the speed of ion-sound waves in the
long-wavelength limit. Each harmonic with dispersion law (\ref{rev2.1})
satisfies the equation
\begin{equation}\label{rev2.1}
  u_t\pm\left(c_0u_x+\tfrac12c_0r_D^2u_{xxx}\right)=0,
\end{equation}
where we still understand $u$ as the speed of the plasma flow. In
the linear approximation, any pulse can be represented as a
sum of harmonics, and therefore the evolution of any wave
propagating in a certain direction is governed by Eqn.~(\ref{rev2.1})
the leading approximation in the dispersive effects. Plasma
density perturbations $\rho'$ are then related to the flow speed $u$ as
\begin{equation}\label{rev2.2}
  \frac{\rho'}{\rho_0}=\pm\frac{u}{c_0}
\end{equation}
with the same choice of sign as in (\ref{eq2.10}).

If the wavelength is much greater than the Debye radius,
then charge separation can be ignored, the electron and ion
densities coincide, and their deviation from the equilibrium
density $\rho_0$ is related to the electric potential by Boltzmann's
formula $\rho=\rho_0\exp({e\phi}/{T_e})$. Using it to eliminate the
potential $\phi$ from the dynamic equations leads to a system of
hydrodynamic equations \cite{LL-10},
\begin{equation}\label{eq2.13}
\begin{split}
& \rho_t+(\rho u)_x=0,\\
& u_t+u u_x+({T_e}/{M})\,({\rho_x}/{\rho})=0,
 \end{split}
\end{equation}
which describe the dynamics of an isothermal gas when the
pressure $p$ depends on the density $\rho$ as $p=({T_e}/{M})\rho$. The
local speed of sound, determined by the formula $c^2=dp/d\rho=T_e/M=c_0^2$,
coincides with the above speed of long
linear waves and is independent of the local density.

If we now consider some suitably arbitrary initial
localized pulse, then, as is known from basic gas dynamics,
it splits after some time into two pulses running in opposite
directions. In each such wave, the local change in density $\delta\rho$
on the background of $\rho$ is related to the local change in the
flow speed $\delta u$ as $\delta\rho\approx\pm(\rho/c_0)\delta u$,
which follows from (\ref{rev2.2}),
whence $\rho_x=\pm(\rho/c_0)u_x$; because the speed of sound is
constant, we do not have to take its dependence on density
into account in this case. Substituting this expression into (\ref{eq2.13})
gives a nonlinear equation for smooth pulses with the
dispersion disregarded:
\begin{equation}\label{eq2.15}
 u_t\pm(c_0+u)u_x=0.
\end{equation}

We have thus found two equations, (\ref{rev2.1}) and (\ref{eq2.15}), which
separately describe the evolution of ion-sound waves in the
case of either low dispersion or small nonlinearity. In both
cases, the dispersive or nonlinear correction amounts to the
addition of a small term, in the corresponding approximation,
to the simplest equation $u_t\pm c_0u_x=0$ for one-dimensional
wave propagation. In the leading approximation,
therefore, simultaneously taking both corrections into
account amounts to combining them into a single equation.
Assuming for definiteness that the wave propagates in the
positive direction of the $x$ axis, we obtain the KdV equation
for ion-sound waves in plasma:
\begin{equation}\label{eq2.16}
 u_t+(c_0+u)\,u_x+\tfrac12{c_0 r_D^2}\,u_{x x x}=0.
\end{equation}
To simplify the notation, it is convenient to transform this
equation by introducing the dimensionless variables
$x'=(x-c_0t)/r_D$, $t'=c_0t/(2r_D)$, and $u=3c_0u'$. Substituting
them into (\ref{eq2.16}) and omitting the primes on the new
variables, we obtain the currently most popular dimensionless
form of the KdV equation:
\begin{equation}\label{eq1.270}
  u_t+6uu_x+u_{xxx}=0.
\end{equation}
The coefficient 6 in front of the nonlinear term is chosen here
so as to simplify the formulas in what follows.

\begin{figure}[t]
\begin{center}
\includegraphics[width=6cm]{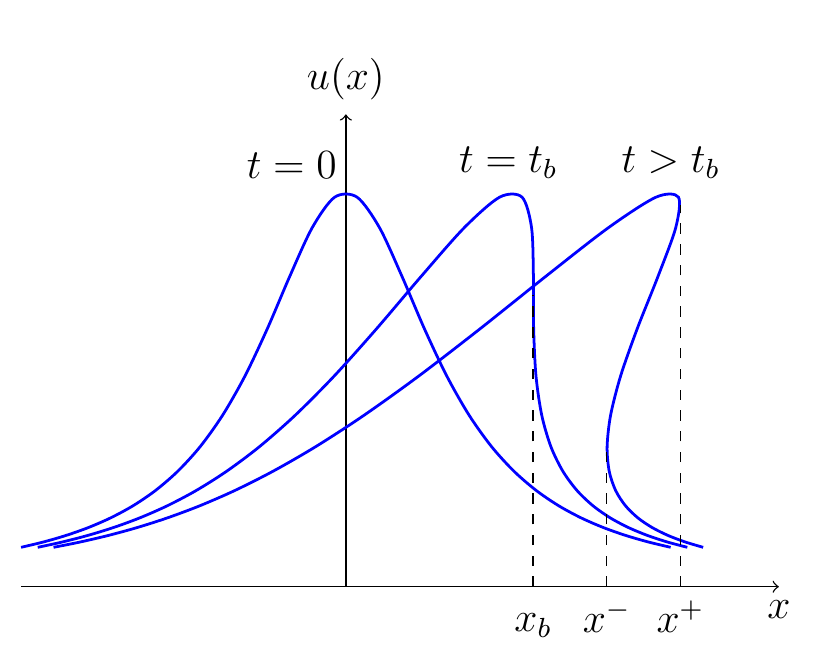}
\caption{ Evolution of a typical pulse in accordance with Hopf equation
(\ref{rev3.6}). After the instant of breaking, $t>t_b$, the distribution $u(x,t)$
formally becomes a three-valued function of the coordinate $x$ in the domain
$x^-<x<x^+$.
}
\label{fig3}
\end{center}
\end{figure}

With dispersion ignored, Eq.~(\ref{eq1.270}) becomes the Hopf equation
\begin{equation}\label{rev3.6}
u_t+6uu_x=0,
\end{equation}
which is a dimensionless form of Eq.~(\ref{eq2.15}). It readily follows
that $u$ is constant along the characteristics $x-6ut=\mathrm{const}$,
which are solutions of the equation $dx/dt=6u$. Therefore, if
the initial distribution $u$ is described by a function $u=u_0(x)$ at $t=0$
and $x=\overline{x}(u)$ is the inverse function, then the implicit solution
of the Hopf equation is given by
\begin{equation}\label{rev3.7}
  x-6ut=\overline{x}(u),
\end{equation}
which describes the distribution $u(x,t)$ at subsequent times.

The most significant feature of these solutions is that the
transfer speed of $u$ values increases as $u$ increases and, for
typical initial distributions $u_0(x)$, the solution becomes
multi-valued after a certain instant $t=t_b$, as is shown in
Fig.~\ref{fig3}. Evidently, we have gone outside the applicability
domain of the dispersionless approximation: at the instant
of breaking $t=t_b$, the derivative of the distribution with
respect to $x$ becomes infinitely large at the point $x_b$, and the
dispersion term with the third-order derivative in KdV
equation (\ref{eq1.270}) is by no means small in the vicinity of $x_b$.
As noted in the Introduction, taking dispersion into
account suppresses this nonphysical behavior, and in the
solution of the full KdV equation the multi-valuedness
domain is superseded with an oscillatory domain evolving
with time, i.e., a {\it dispersive shock wave}. Gurevich and
Pitaevskii assumed that this oscillatory domain can be
approximately represented as a modulated periodic solution
of the KdV equation, which means that the next step in
constructing the DSW theory must consist of deriving such
periodic solutions---which was done by Korteweg and
de Vries themselves in \cite{kdv}. Here, we give the necessary
background.
As usual, we seek a solution of Eq.~(\ref{eq1.270}) as a traveling
wave $u=u(\xi)$, $\xi=x-V t$, where $V$ is the wave propagation
speed; we then find that $u(\xi)$ satisfies the ordinary differential
equation $ u_{\xi\xi\xi}=V u_\xi-6u u_\xi$, which, after two elementary
integrations, takes the form of the equation
\begin{equation}\label{eq1.273}
\begin{split}
  \tfrac{1}{2}u_\xi^2&=-A+B u+\tfrac{1}{2}V u^2-u^3=\\
 & =-\mathcal{R}(u)=-(u-\nu_1)(u-\nu_2)(u-\nu_3),
 \end{split}
\end{equation}
where $A$ and $B$ are constants of integration. This equation has
real solutions if the polynomial $\mathcal{R}(u)$ has three real zeros:
$\nu_1$, $\nu_2$, and $\nu_3$ with $\nu_1\leq\nu_2\leq\nu_3$.
Evidently, the oscillating
solution corresponds to the motion of $u$ between two zeros
in the interval
\begin{equation}\label{eq1.274}
 \nu_2\leq u\leq \nu_3,
\end{equation}
where $\mathcal{R}(u)\leq 0$. The constants $A$, $B$, and $V$
can be expressed in terms of $\nu_1$, $\nu_2$, and $\nu_3$ as
\begin{equation}\label{eq1.275}
\begin{split}
 & A=-\nu_1\nu_2\nu_3, \quad B=-(\nu_1\nu_2+\nu_2\nu_3+\nu_3\nu_1), \\
 & V=2(\nu_1+\nu_2+\nu_3).
 \end{split}
\end{equation}
It now follows from Eq.~(\ref{eq1.273}) that the periodic solution of the
KdV equation can be expressed as
\begin{equation}\label{eq1.276}
 \sqrt2\,\xi=\int\limits_u^{\nu_3}\frac{d u'}{\sqrt{(u'-\nu_1)(u'-\nu_2)(\nu_3-u')}},
\end{equation}
where the integration constant that is additive with respect to
$\xi$ is chosen such that $u(\xi)$ takes the maximum value $\nu_3$ at
$\xi=0$. Integral (\ref{eq1.276}) can be standardly expressed in terms of
elliptic integrals, and their inversion gives the dependence
$u=u(\xi)$ in terms of elliptic functions. Omitting the calculations
that are routine for nonlinear physics, we get the result
\begin{equation}\label{eq1.279}
 u=\nu_3-(\nu_3-\nu_2)\,\sn^2\left(\sqrt{({\nu_3-\nu_1})/{2}}\,(x-V t),m\right),
\end{equation}
where $\sn$  is the elliptic sine, and the parameter $m$ is defined as
\begin{equation}\label{eq1.278}
 m=\frac{\nu_3-\nu_2}{\nu_3-\nu_1}
\end{equation}
in accordance with the notation in handbook \cite{AS-2}. Using the
identity $\mathrm{s n}^2z+\mathrm{c n}^2z=1$ allows expressing this solution in
terms of the elliptic cosine $\mathrm{cn}$, which is why Korteweg and de
Vries called their solution the `cnoidal wave', similarly to the
cosine wave in the linear theory. The properties of such a
cnoidal wave are determined by the three zeros, $\nu_1,\nu_2,$ and $\nu_3$,
of the polynomial $\mathcal{R}(u)$. In particular, the speed of the wave $V$
and the parameter $m$ are expressed by formulas (\ref{eq1.275}) and (\ref{eq1.278}).
The wavelength $L$ can be defined as the distance between two
neighboring maxima of $u(\xi)$, and it is then expressed through
the full elliptic integral of the first kind $K(m)$ as
\begin{equation}\label{eq1.287}
L=\oint\frac{du}{\sqrt{-2\mathcal{R}(u)}}=\frac{2\sqrt{2}K(m)}{\sqrt{\nu_3-\nu_1}}
\end{equation}
The cnoidal wave amplitude can be defined by the relation
\begin{equation}\label{eq1.281}
a=(u_{max}-u_{min})/2=(\nu_3-\nu_2)/2.
\end{equation}

Solution (\ref{eq1.279}) passes into a harmonic linear-approximation wave
\begin{equation}\label{t3-51.14}
  u\cong \nu_2+\frac12(\nu_3-\nu_2)\cos\left(\sqrt{2(\nu_2-\nu_1)}(x-Vt)\right).
\end{equation}
for a small wave amplitude $a\ll \nu_2-\nu_1$, when $m\ll1$. The
wave number $k=2\pi/L=\sqrt{2(\nu_2-\nu_1)}$
and the phase velocity $V=2\nu_1+4\nu_2=6\nu_2-k^2$
of the wave are then related as $V=\om/k$,
which follows from the dispersion law $\om=6\nu_2k-k^3$
that corresponds to the linearized KdV equation
$u'_t+6\nu_2u'_x+u'_{xxx}=0$ for a wave propagating along the
uniform state with $u=\nu_2$.

In the opposite limit $\nu_2\to\nu_1$ and $m\to1$, the wavelength
tends to infinity and $\sn(z,1)=\th(z)$, and hence solution
(\ref{eq1.279}) becomes
\begin{equation}\label{t3-52.20}
  u=\nu_1+\frac{\nu_3-\nu_1}{\ch^2\left(\sqrt{(\nu_3-\nu_1)/2}\,(x-Vt)\right)}.
\end{equation}
In this case, the profile $u=u(x-Vt)$ has the shape of a
solitary wave propagating along the uniform state $u=\nu_1$.
Thus, in the limit $m\to1$, the periodic wave transforms into
solitary pulses, or solitons (\ref{t3-52.20}), separated by an infinitely
long distance.

The fundamental assumption of Gurevich and Pitaevskii's approach
to the DSW theory was that at sufficiently
large times after the instant of breaking, when the length of
the emerging oscillatory domain becomes much greater than
the local wavelengths $L$, the DSW evolution can be represented as
a slow variation of the parameters $\nu_1,\nu_2$, and $\nu_3$ in a
modulated cnoidal wave (\ref{eq1.279}). The `slowness' condition here
means that the relative change in the modulation parameters
$\nu_1,\nu_2$, and $\nu_3$ or the equivalent variables is small either at
distances of the order of the wavelength $L$ or over a time of the
order of one oscillation period.

Thus, the problem of constructing the theory of DSWs
reduces to deriving equations for the evolution of modulation
parameters and to obtaining their solutions in specific
physical situations. Fortunately, by that time, equations for
the modulation of a cnoidal KdV wave had already been
derived by Whitham~\cite{whitham-65}. Unfortunately, in
both ~\cite{whitham-65} and his later book \cite{whitham-74},
Whitham only gave the final result of the
calculations, having omitted all the details. Because these
calculations are highly nontrivial, we briefly describe them in
Section~4 for completeness, but first, with methodological
purposes in mind, we discuss a linear-approximation analogue of
Whitham's modulation theory.

\section{Modulation of linear waves}

A well-known result in the theory of modulation of linear
waves is that the envelope of a modulated wave packet
propagates with the group velocity of the carrier wave.
Methods for deriving asymptotic solutions of linear equations
have also been developed in much detail to describe such
behavior of waves. But we look at problems of this sort from
another standpoint, which is very transparent physically and
allows an extension to the dynamics of nonlinear waves.

\begin{figure}[t]
\begin{center}
\includegraphics[width=6cm]{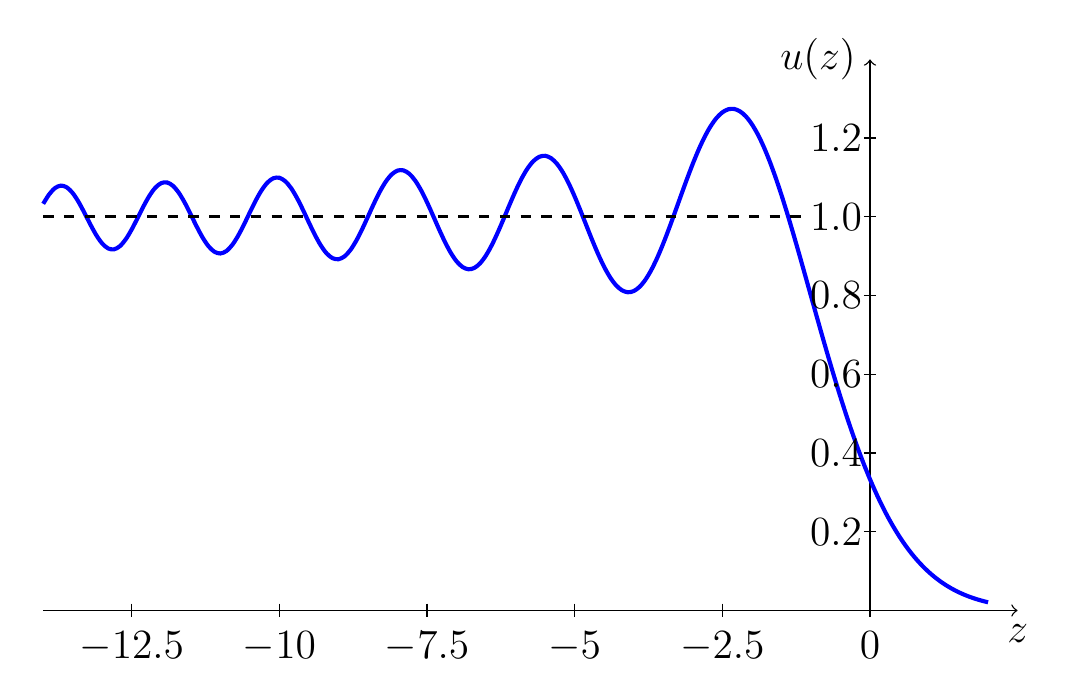}
\caption{ Profile of wave (\ref{eq1.48k}) plotted as a function of
the self-similar variable $z=x/(3t)^{1/3}$. }
\label{fig4}
\end{center}
\end{figure}

As an example, we consider the evolution of a wave
described by the linearized KdV equation $u'_t+6\nu_2u'_x+u'_{xxx}=0$
and having the initial shape of a `step'. Because the
$6\nu_2u'_x$ can easily be eliminated by passing to the reference
frame $x'=x-6\nu_2t,t'=t$, we write the linear KdV equation as
\begin{equation}\label{eq1.41k}
  u_t + u_{xxx} = 0,
\end{equation}
and take the initial condition in the form
\begin{equation} \label{eq1.47k}
 u_0(x)=
 \begin{cases} 1, & \text{  $x\le 0$}, \\
  0, & \text{   $x> 0$.}
 \end{cases}
\end{equation}
This problem can easily be solved exactly by the Fourier
method, and the result can be brought to the form
\begin{equation}\label{eq1.48k}
  u(x,t)= \int\limits_{x/(3t)^{1/3}}^\infty \mathrm{A i}(z)\,d z,
\end{equation}
where $\mathrm{A i}(z)$ is the standard notation for the Airy function
\cite{AS-2}. As we can see, the wave profile depends only on the
self-similar variable $ z={x}/{(3t)^{1/3}}$ (Fig.~\ref{fig4}).
At large $x$, when $z \gg 1$,
the wave amplitude decreases exponentially into the `shadow'
domain, and in the opposite limit of large negative $x$, we can
use the known asymptotic form of the Airy function to obtain
($-z \gg 1$)
\begin{equation}\label{eq1.51k}
 u(z) \cong 1-\frac{1}{\sqrt\pi}(-z)^{-3/4} \cos\left(\tfrac{2}
 {3}(-z)^{3/2} +\tfrac{\pi}{4}\right).
\end{equation}
The obtained results confirm the general idea that dispersive
effects manifest themselves in oscillatory wave structures
originating from pulses with sufficiently sharp fronts. But
the shape of the resultant wave structure suggests another
approach to its description.

Both Fig.~\ref{fig4} and formula (\ref{eq1.51k}) suggest that,
as $x\to-\infty$,
this wave can be interpreted as a modulated harmonic wave
with a variable wave number and variable frequency and
amplitude of oscillations. We represent such a wave as
\begin{equation}\label{125.1}
    u(x,t)=1+a(x,t)\cos[\theta(x,t)+\theta_0],
\end{equation}
where we introduce the wave phase
\begin{equation}\label{125.3}
    \theta(x,t)=\frac23\left(\frac{-x}{(3t)^{1/3}}\right)^{3/2},
\end{equation}
having for simplicity dropped the constant term $\theta_0=\pi/4$
from its definition. For such a modulated wave, it is natural to
define the wave number $k(x,t)$ and the frequency $\om(x,t)$ as
\begin{equation}\label{125.4a}
\begin{split}
    & k(x,t)=\theta_x(x,t)=-\left(\frac{-x}{3t}\right)^{1/2},\\
    & \om(x,t)=-\theta_t(x,t)=\left(\frac{-x}{3t}\right)^{3/2},
    \end{split}
\end{equation}
which are locally related by the dispersion law $\om=-k^3$ that
follows from linear KdV equation (\ref{eq1.41k}). In other words, wave
(\ref{125.1}) is locally a harmonic wave that is an exact solution of this
equation if modulation is ignored. If we consider a piece of the
structure with a fixed wave number $k(x,t)$, it immediately
follows from the first formula in (\ref{125.4a}) that this piece moves
along the $x$ axis with the group velocity
\begin{equation}\label{125.8a}
    v_g=-3k^2=\frac{d\om}{dk}
\end{equation}
in accordance with the known property of the group velocity.
It is clear that this way of introducing the group velocity into
the theory of modulation of linear waves has a general
character.

We assume that the modulated linear wave is represented as
\begin{equation}\label{125.2}
    u(x,t)=a(x,t)\cos[\theta(x,t)],
\end{equation}
and that this wave is locally harmonic with good accuracy,
with local values of the wave number and frequency defined as
\begin{equation}\label{125.4}
    k(x,t)=\theta_x(x,t),\qquad \om(x,t)=-\theta_t(x,t),
\end{equation}
and related by the dispersion law for harmonic waves
\begin{equation}\label{125.6}
    \om=\om(k).
\end{equation}
In view of (\ref{125.4}), the consistency condition for cross derivatives
of the phase $(\theta_x)_t=(\theta_t)_x$ leads to the equation
\begin{equation}\label{125.12}
    k_t+\om_x=0\quad\text{или}\quad k_t+(kV)_x=0,
\end{equation}
where $V=V(k)$  is the phase velocity of the wave. Because a
unit-length interval along the $x$ axis contains $1/L=k/(2\pi)$
waves, Eq.~(\ref{125.12}) can be interpreted as the conservation law for
the number of waves, with $k$ playing the role of the density of
waves and $\om=kV$ the flux. Substituting dispersion law (\ref{125.12})
into (\ref{125.6}), we arrive at the equation
\begin{equation}\label{125.13}
    k_t+v_g(k)k_x=0,
\end{equation}
which again states that the wave number $k$ propagates at the
speed $v_g(k)=\om'(k)$ and preserves its value along the
characteristic $x-v_g(k)t=\mathrm{const}$. Therefore, if changes in
the shape of the wave packet are disregarded, a wave packet
made of harmonics with the wave numbers close to $k=k_0$
propagates with the group velocity $v_g(k_0)=\om'(k_0)$.

We can now return to the problem of the decay of a step-like
profile with initial distribution (\ref{eq1.47k}) and use Eq.~(\ref{125.13})
instead of the exact solution expressed in terms of the Airy
function. The key role here is played by the observation that
the initial distribution does not contain parameters with the
dimension of length, but the original problem has some
characteristic value of speed $c_0$. Therefore, a solution of
Eq.~(\ref{125.13}) can depend only on the self-similar variable
$\xi=x/t$ (in dimensional units, on  $\xi=x/(c_0t)$). Substituting
$k=k(\xi)$ into (\ref{125.13}), we find $(dk/d\xi)(v_g(k)-\xi)=0$. Because
$dk/d\xi\neq0$ along the modulated wave, the dependence
$k=k(\xi)$ is defined implicitly by the equation
\begin{equation}\label{126.3}
    v_g(k)=\xi=\frac{x}t.
\end{equation}
Having used this to find $k=k(x/t)$, we can express the
phase $\theta(x,t)$ from the equation $\theta_x=k$ if we recall that the
frequency $\om=-\theta_t$, which is a function of $k$, can also
depend only on the self-similar variable. For the linear
KdV equation, the obtained results immediately reproduce
the known relations $-3k^2  ={x}/t$, $k=\theta_x=-(-x/(3t))^{1/2}$,
$\theta  =\frac23({-x}/{(3t)^{1/3}})^{3/2}$. Thus, modulation equation
(\ref{125.13}) has allowed us to easily find some characteristics of the
emergent wave structure.

To derive the modulation equation for the amplitude
$a(x,t)$ of wave (\ref{125.2}), it is natural to use the energy conservation
law, because expansion of the wave structure with time
leads to a redistribution of energy over a progressively larger
volume, and in linear systems the energy density is proportional to the
amplitude squared. After averaging over the
wavelength, the local energy density $a^2(x,t)$ is transported
with the group velocity $v_g$ corresponding to the local value of
the wave number $k$, and we can therefore write the energy
conservation law as
\begin{equation}\label{l-19.58}
  \frac{\prt(a^2)}{\prt t}+\frac{\prt(v_ga^2)}{\prt x}=0.
\end{equation}
In the case of a linear KdV equation and asymptotic regime
(\ref{126.3}) of the wave packet evolution, Eq.~(\ref{l-19.58}) becomes
$ta_t+xa_x=-\frac12a$. This can readily be solved using the
standard method of characteristics, with the result
$ a(x,t)=(1/{\sqrt{t}})f({x}/t)$, where $f$ is an arbitrary function. Assuming
that in the problem of the evolution of a step-like shape the
amplitude also depends only on the same self-similar variable
$z={-x}/{(3t)^{1/3}}$ as the wave number $k$ does, it is easy to find
that $f(x/t)=\mathrm{const}\cdot(-x/t)^{-3/4}$, which defines the modulated
wave shape up to a constant factor:
\begin{equation}\nonumber 
  u(x,t)\cong \frac{\mathrm{const}}{\sqrt{t}}\left(\frac{-x}{t}\right)^{-3/4}
  \cos\left[\frac23\frac{(-x)^{3/2}}{(3t)^{1/2}}\right].
\end{equation}
Thus, we have reproduced the main features of solution (\ref{eq1.51k})
without relying on any information on the properties of the
Airy function, but rather by just solving modulation equations (\ref{125.13})
and (\ref{l-19.58}) of the linear theory. Evidently, the idea of
this approach involving the wave number conservation law
and other conservation laws with averaged densities and
fluxes allows a generalization to nonlinear waves. Exactly
that was done by Whitham for modulated cnoidal waves of
the KdV equation, and we discuss his theory in Section~4.

\section{Whitham theory}

We restrict ourselves to describing the general idea of
Whitham \cite{whitham-65} on averaging conservation laws in the simple
case where the evolution of a wave is described by a nonlinear
equation for a single variable $u$,
\begin{equation}\label{eq3.1}
  \Phi(u,u_t,u_x,u_{tt}, u_{tx}, u_{xx},\ldots)=0.
\end{equation}
We assume that Eq.~(\ref{eq3.1}) has traveling-wave solutions when
$u(x,t)$ depends on $x$ and $t$ only through the combination
$u=u(\xi)$, $\xi=x-Vt$, and for such solutions, Eq.~(\ref{eq3.1}) can be
reduced to the form
\begin{equation}\label{eq3.3}
  u_{\xi}^2=F(u,V,A_i),
\end{equation}
where $A_i$ is a collection of parameters occurring in deriving
(\ref{eq3.3}) from (\ref{eq3.1}). In a periodic traveling wave, the variable $u$
oscillates between two zeros of $F(u)$. We let $u_1(V,A_i)$ and
$u_2(V,A_i)$, with $u_1<u_2$, denote these zeros, assuming that $F$ is
positive in the interval $u_1<u<u_2$. Obviously, the wavelength is
\begin{equation}\label{eq3.4}
  L=L(V,A_i)= 2\int_{u_1}^{u_2}\frac{du}{\sqrt{F(u;V,A_i)}},
\end{equation}
and the wave number $k$ and the frequency $\omega$ are
\begin{equation}\label{eq3.5}
\begin{split}
  & k=k(V,A_i)=1/L(V,A_i),\\
  & \omega=\omega(V,A_i)=Vk(V,A_i),
  \end{split}
\end{equation}
where we dropped the factor $2\pi$ in the definition of the wave
number because it is only needed in the nonlinear theory for
maintaining correspondence with the low-amplitude limit,
and this factor can easily be restored whenever necessary. As a
result, the wave number $k$ becomes exactly equal to the
density of the number of waves. In a modulated wave
$u(\xi;V,A_i)$, the parameters $V$ and $A_i$ are slowly varying
functions of $x$ and $t$, changing little over distances of the
order of the wavelength $L$ and over a time of the order of $1/\omega$.
This implies that there is an interval $\Delta$, much longer than the
wavelength $L$ but much shorter than a certain size $l$
characterizing the wave structure overall: $L\ll\Delta\ll l$. It is
clear that, up to small quantities of the order of $\eps\sim\Delta/l$,
averaging over the interval $\Delta$ is equivalent to averaging over
the wavelength $L$. Therefore, we average physical quantities
over fast oscillations in the wave in accordance with the rule
\begin{equation}\label{eq3.6}
  \langle\mathcal{F}\rangle \approx   \frac{1}{L}\int_{0}^{L}\mathcal{F}(x',t)dx'.
\end{equation}
If a conservation law $\mathcal{P}_t+\mathcal{Q}_x=0$ is known, then, after
the averaging, it takes the form
\begin{equation}\label{eq3.11}
  \frac{\prt}{\prt t}\langle\mathcal{P}\rangle+
  \frac{\prt}{\prt x}\langle\mathcal{Q}\rangle=0,
\end{equation}
where the dependence on $x$ and $t$ is only present in slowly
varying modulation parameters $V$ and $A_i$ that enter the
averaged quantities. We can regard Eqs~(\ref{eq3.11}) as differential
equations for these parameters, similarly to how we viewed
modulation equations in the linear theory.

We can now turn to the derivation of the modulation
equations for the cnoidal KdV wave. In a weakly modulated
wave, the parameters $A,\,B,\,V$ or $\nu_1,\,\nu_2,\,\nu_3$ become slowly
varying functions of $x$ and $t$, and we wish to find the equations
governing the evolution of these parameters. Calculations can
be simplified by recalling that one of the modulation
equations is already known. Replacing the elliptic function
argument in periodic solution (\ref{eq1.279}) with the phase $\theta$
that can be defined up to an appropriate numerical factor, we introduce
local values of the wave number and frequency via formulas
(\ref{125.4}), just as in the linear case; they must then satisfy the
conservation law for the number of waves in Eq.~(\ref{125.12}). In a
weakly modulated wave, the values of $k$ and $\om$ are given by
Eqs.~(\ref{eq3.5}) with variable parameters $V$ and $A_i$, and hence
variations of these parameters under the evolution of the
wave must satisfy the equation
\begin{equation}\label{rev4-1}
    k_t+(kV)_x=0, \qquad k=1/L.
\end{equation}
As two missing modulation equations, we use the averaged
conservation laws:
\begin{equation}\label{eq3.127}
  \begin{split}
  & u_t  +(3u^2+u_{xx})_x=0,\\
  & (\tfrac12u^2)_t  +(2u^3+uu_{xx}-\tfrac12u_x^2)_x=0,\\
  \end{split}
\end{equation}
which can be straightforwardly verified by substituting $u_t$
from the KdV equation.

We first derive the modulation equations for the para-
meters $A,\,B$, and $V$. Following Whitham, we express the
averaged quantities in terms of the function
\begin{equation}\label{eq3.128}
\begin{split}
   \mathcal{W}&=-\sqrt{2}\oint\sqrt{-A+Bu+\tfrac12Vu^2-u^3}\,\,du= \\
  & =-\sqrt{2}\oint\sqrt{-\mathcal{R}(u)}\,du,
  \end{split}
\end{equation}
where the integral is taken over a closed contour encompassing the
interval $\nu_2\leq u\leq\nu_3$. The wavelength $L=1/k$ is then
expressed through $\mathcal{W}$ as
\begin{equation}\label{eq3.129}
  L=\frac1{\sqrt{2}}\oint \frac{du}{\sqrt{-\mathcal{R}(u)}}
  =\frac{\prt \mathcal{W}}{\prt A}\equiv \mathcal{W}_A.
\end{equation}
We readily calculate the averaged quantities:
\begin{equation}\label{eq3.130}
  \begin{split}
&  \langle u\rangle=k\int_0^Lu\, d\xi=\frac{k}{\sqrt{2}}
\oint \frac{u\,du}{\sqrt{-\mathcal{R}(u)}}=-k\mathcal{W}_B,\\
&  \langle\tfrac12u^2\rangle=k\int_0^L\frac{\tfrac12u^2du}{u_{\xi}}=-k\mathcal{W}_V,\\
&  \langle u_{\xi}^2\rangle=k\oint \frac{u_{\xi}^2du}{u_{\xi}}=-k\mathcal{W}.
  \end{split}
\end{equation}
The second derivatives $u_{\xi\xi}$ can be eliminated from the
conservation laws with the help of the formula $u_{\xi\xi}=B+Vu-3u^2$.
After simple calculations using the relation
$k\mathcal{W}_A=1$ and the averaged values found above, we obtain the
averaged conservation laws:
\begin{equation}\label{eq3.131}
  \begin{split}
  (k\mathcal{W}_B)_t&+(kV\mathcal{W}_B-B)_x=0,\\
  (k\mathcal{W}_V)_t&+(kV\mathcal{W}_V-A)_x=0,\\
  \end{split}
\end{equation}
Having substituted $k=1/\mathcal{W}_A$ and introduced the `long'
derivative $D/Dt=\prt/\prt t+V\prt/\prt x$, we obtain the modulation
equations
\begin{equation}\label{eq3.133}
\begin{split}
  & \frac{D\mathcal{W}_A}{Dt}=\mathcal{W}_A\frac{\prt V}{\prt x},\quad
  \frac{D\mathcal{W}_B}{Dt}=\mathcal{W}_A\frac{\prt B}{\prt x},\\
  & \frac{D\mathcal{W}_V}{Dt}=\mathcal{W}_A\frac{\prt A}{\prt x},
  \end{split}
\end{equation}
the first of which is the conservation law (\ref{rev4-1}) with the wave
number expressed as $k=1/\mathcal{W}_A$.

Despite the apparent simplicity of the obtained equations,
they are not extremely useful in practice. We therefore
reexpress them in terms of $\nu_1,\,\nu_2$, and $\nu_3$. From (\ref{eq1.275}),
we find the relations between differentials:
\begin{equation*}
  \begin{split}
&  dV=2(d\nu_1+d\nu_2+d\nu_3),\\
&  dB=-[(\nu_2+\nu_3)d\nu_1+(\nu_1+\nu_3)d\nu_2+(\nu_1+\nu_2)d\nu_3],\\
&  dA=-(\nu_2\nu_3\cdot d\nu_1+\nu_1\nu_3\cdot d\nu_2+\nu_1\nu_2\cdot d\nu_3).
  \end{split}
\end{equation*}
Hence, Eqs.~(\ref{eq3.133}) expressed in the variables $\nu_1,\,\nu_2$, and $\nu_3$
take the form
\begin{equation}\label{eq3.134}
    \begin{split}
    &\mathcal{W}_{A,\nu_1}\frac{D\nu_1}{Dt}+\mathcal{W}_{A,\nu_2}\frac{D\nu_2}{Dt}+\mathcal{W}_{A,\nu_3}\frac{D\nu_3}{Dt}=\\
    &=2\mathcal{W}_A(\nu_{1,x}+\nu_{2,x}+\nu_{3,x}),\\
    &\mathcal{W}_{B,\nu_1}\frac{D\nu_1}{Dt}+\mathcal{W}_{B,\nu_2}\frac{D\nu_2}{Dt}+\mathcal{W}_{B,\nu_3}\frac{D\nu_3}{Dt}=\\
    &=-\mathcal{W}_A[(\nu_2+\nu_3)\nu_{1,x}+(\nu_1+\nu_3)\nu_{2,x}+(\nu_1+\nu_2)\nu_{3,x}],\\
    &\mathcal{W}_{V,\nu_1}\frac{D\nu_1}{Dt}+\mathcal{W}_{V,\nu_2}\frac{D\nu_2}{Dt}+\mathcal{W}_{V,\nu_3}\frac{D\nu_3}{Dt}=\\
    &=-\mathcal{W}_A[\nu_2\nu_3\cdot\nu_{1,x}+\nu_1\nu_3\cdot\nu_{2,x}+\nu_1\nu_2\cdot\nu_{3,x}],
    \end{split}
\end{equation}
where all the derivatives of$\mathcal{W}$ are represented by integrals
similar to  (\ref{eq3.128}) and (\ref{eq3.130}).

As a clue to further transformations, we note that the
right-hand sides of Eqs.~(\ref{eq3.134}) contain the same factor $\mathcal{W}_A$.
Therefore, their linear combinations can be found such that
the coefficient in front of one of the derivatives vanishes and
the other two coefficients become equal. Indeed, we multiply
the first equation in (\ref{eq3.134}) by $p$, the second by $q$,
and the third by $r$, add them, and choose the parameters $p, q$, and $r$ such that
the coefficient in front of $\nu_{1,x}$ vanishes and the coefficients in
front of $\nu_{2,x}$ and $\nu_{3,x}$ become equal:
\begin{equation}\nonumber
  \begin{split}
  &2p-q(\nu_2+\nu_3)-r\nu_2\nu_3=0,\\
  &2p-q(\nu_1+\nu_3)-r\nu_1\nu_3=2p-q(\nu_1+\nu_2)-r\nu_1\nu_2.
  \end{split}
\end{equation}
It immediately follows from these conditions that
\begin{equation}\nonumber
  q=-r\nu_1,\qquad p=-\frac12r(\nu_1\nu_2+\nu_1\nu_3-\nu_2\nu_3)
\end{equation}
and we can hence set $r=-2$, to obtain $p=\nu_1\nu_2+\nu_1\nu_3-\nu_2\nu_3$,
$q=2\nu_1$, and $r=-2$. The right-hand side of this linear
combination of Eqs.~(\ref{eq3.134}) then takes the form
\begin{equation}\label{eq3.135}
  -2(\nu_2-\nu_1)(\nu_3-\nu_1)\mathcal{W}_A\frac{\prt(\nu_2+\nu_3)}{\prt x}.
\end{equation}
Hence, it follows that, if in a similar linear combination of the
left-hand sides of Eqs.~(\ref{eq3.134}) the coefficient in front of ${D\nu_1}/{Dt}$
vanishes and the coefficient in front of ${D\nu_2}/{Dt}$ and ${D\nu_3}/{Dt}$
are equal to each other, then the modulation equations take a
very simple `diagonal' form.

With the help of the identity
\begin{equation*}
\begin{split}
  & \frac{d}{du}\left(2\sqrt{\frac{(u-\nu_2)(u-\nu_3)}{-(u-\nu_1)}}\right)=\\
  & =-\frac{u^2-2\nu_1 u+\nu_1\nu_2+\nu_1\nu_3-\nu_2\nu_3}{(u-\nu_1)\sqrt{-\mathcal{R}(u)}}
  \end{split}
\end{equation*}
which is easy to verify, we obtain
\begin{equation}\nonumber
\begin{split}
  & p\mathcal{W}_{A,\nu_1}+q\mathcal{W}_{B,\nu_1}+r\mathcal{W}_{V,\nu_1}=\\
  & =-\frac{1}{\sqrt{8}}\oint\frac{d}{du}
\left(2\sqrt{\frac{(u-\nu_2)(u-\nu_3)}{-(u-\nu_1)}}\right)du=0,
\end{split}
\end{equation}
because the integrand is a total derivative of a periodic
function, and the first condition is thus satisfied.

The coefficients in front of ${D\nu_2}/{Dt}$ and ${D\nu_3}/{Dt}$ have the
respective forms
\begin{equation*}
  \begin{split}
&  K_2=p\mathcal{W}_{A,\nu_2}+q\mathcal{W}_{B,\nu_2}+r\mathcal{W}_{V,\nu_2}=\\
&=\frac{1}{\sqrt{8}} \oint\frac{u^2-2\nu_1 u+\nu_1\nu_2+\nu_1\nu_3-\nu_2\nu_3}
{(u-\nu_2)\sqrt{-\mathcal{R}(u)}}du,\\
&  K_3=p\mathcal{W}_{A,\nu_3}+q\mathcal{W}_{B,\nu_3}+r\mathcal{W}_{V,\nu_3}=\\
&=\frac{1}{\sqrt{8}}\oint\frac{u^2-2\nu_1 u+\nu_1\nu_2+\nu_1\nu_3-\nu_2\nu_3}
{(u-\nu_3)\sqrt{-\mathcal{R}(u)}}du,
  \end{split}
\end{equation*}
and their difference, being an integral of the derivative of a
periodic function over the period, vanishes:
\begin{equation*}
\begin{split}
  & K_2-K_3=\\
  & =\frac{\nu_2-\nu_3}{\sqrt{8}}\oint \frac{d}{du}
  \left(2\sqrt{\frac{-(u-\nu_1)}{(u-\nu_2)(u-\nu_3)}}\right)du=0.
  \end{split}
\end{equation*}
Hence, $K_2=K_3$, and the combination $\nu_2+\nu_3$ is a convenient
modulation variable for which the modulation equations are
dramatically simplified. The emerging coefficient $K_2=K_3$ in
front of $D(\nu_2+\nu_3)/Dt$ can also be expressed in terms of $\mathcal{W}_A$.
Indeed, $K_2$ and $K_3$ can be represented as
\begin{equation*}
  \begin{split}
&  K_2=\frac{\nu_2-\nu_1}{\sqrt{2}}\oint\frac{(u-\nu_3)du}{(u-\nu_2)\sqrt{-\mathcal{R}(u)}}+\\
&+  \frac1{\sqrt{8}}\oint\frac{u^2-2\nu_2 u+\nu_1\nu_2+\nu_2\nu_3-\nu_1\nu_3}{(u-\nu_2)\sqrt{-\mathcal{R}(u)}}du,\\
&  K_3=\frac{\nu_3-\nu_1}{\sqrt{2}}\oint\frac{(u-\nu_2)du}{(u-\nu_3)\sqrt{-\mathcal{R}(u)}}+\\
&+  \frac1{\sqrt{8}}\oint\frac{u^2-2\nu_3 u+\nu_1\nu_3+\nu_2\nu_3-\nu_1\nu_2}{(u-\nu_3)\sqrt{-\mathcal{R}(u)}}du.
  \end{split}
\end{equation*}
But the second terms on the right-hand sides vanish due to
identities quite similar to those used above, and the remaining
non-vanishing terms can be easily brought to the form
\begin{equation}\label{eq3.137}
  \begin{split}
&  K_2=(\nu_2-\nu_1)\mathcal{W}_A-2(\nu_2-\nu_1)(\nu_3-\nu_2)\mathcal{W}_{A,\nu_2},\\
&  K_3=(\nu_3-\nu_1)\mathcal{W}_A+2(\nu_3-\nu_1)(\nu_3-\nu_2)\mathcal{W}_{A,\nu_3}.
  \end{split}
\end{equation}
The equality $K_2=K_3$ then leads to the identity
$$
\mathcal{W}_A=-2[(\nu_2-\nu_1)\mathcal{W}_{A,\nu_2}+(\nu_3-\nu_1)\mathcal{W}_{A,\nu_3}],
$$
substituting which in any of the equations in (\ref{eq3.137}) gives
\begin{equation*}
  \begin{split}
  K_2=K_3& =-2(\nu_2-\nu_1)(\nu_3-\nu_1)(\mathcal{W}_{A,\nu_2}+\mathcal{W}_{A,\nu_3})=\\
& = 2(\nu_2-\nu_1)(\nu_3-\nu_1)\mathcal{W}_{A,\nu_1},
  \end{split}
\end{equation*}
because
$$
\mathcal{W}_{A,\nu_1}+\mathcal{W}_{A,\nu_2}+\mathcal{W}_{A,\nu_3}=
\frac{1}{\sqrt{8}}\oint\frac{-\mathcal{R}'(u)du}{(-\mathcal{R})^{3/2}(u)}=0.
$$
We now equate the left-hand side of our linear combination
$$
2(\nu_1-\nu_2)(\nu_1-\nu_3)\mathcal{W}_{A,\nu_1}\frac{D(\nu_2+\nu_3)}{Dt},
$$
to its right-hand side in (\ref{eq3.135}) to obtain the equation
\begin{equation}\label{eq3.138}
\frac{D(\nu_2+\nu_3)}{Dt}+\frac{\mathcal{W}_A}{\mathcal{W}_{A,\nu_1}}\frac{\prt(\nu_2+\nu_3)}{\prt x}=0.
\end{equation}
Cyclic permutations of $\nu_1,\,\nu_2$, and $\nu_3$ give two other
Whitham modulation equations:
\begin{equation}\label{eq3.139}
\begin{split}
  &\frac{D(\nu_3+\nu_1)}{Dt}+\frac{\mathcal{W}_A}{\mathcal{W}_{A,\nu_2}}\frac{\prt(\nu_3+\nu_1)}{\prt x}=0,\\
  &\frac{D(\nu_1+\nu_2)}{Dt}+\frac{\mathcal{W}_A}{\mathcal{W}_{A,\nu_3}}\frac{\prt(\nu_1+\nu_2)}{\prt x}=0.
  \end{split}
\end{equation}

Each of the equations obtained by Whitham involve
derivatives of only one of the quantities $\nu_2+\nu_3$, $\nu_3+\nu_1$,
and $\nu_1+\nu_2$, which means that the equations have acquired a
diagonal form. Therefore, the above transformation is
similar to the transition from the standard form of gas-dynamic equations
to their diagonal form in terms of
different variables, called Riemann invariants (see, e.g., \cite{LL6}).
We therefore define the new modulation variables, the
Riemann invariants $r_1\le r_2\le r_3$ of Whitham modulation equations, as
\begin{equation}\label{eq3.140}
\begin{split}
  & r_1=\tfrac12(\nu_1+\nu_2),\quad r_2=\tfrac12(\nu_1+\nu_3),\\
  & r_3=\tfrac12(\nu_2+\nu_3);\\
  &\nu_1=r_1+r_2-r_3,\quad \nu_2=r_1+r_3-r_2,\\
  & \nu_3=r_2+r_3-r_1,
  \end{split}
\end{equation}
and express the other variables through them. In particular,
we find $\mathcal{W}_{A,r_1}=\mathcal{W}_{A,\nu_1}+\mathcal{W}_{A,\nu_2}-\mathcal{W}_{A,\nu_3}
=-2\mathcal{W}_{A,\nu_3}$, $\mathcal{W}_{A,r_2}=-2\mathcal{W}_{A,\nu_2}$,
and $\mathcal{W}_{A,r_3}=-2\mathcal{W}_{A,\nu_1}$. With $\mathcal{W}_A=L$, we obtain
$$
\frac{\mathcal{W}_A}{\mathcal{W}_{A,\nu_1}}=
-\frac{2\mathcal{W}_A}{\mathcal{W}_{A,r_3}}=-\frac{2L}{\prt L/\prt r_3},
$$
and similar formulas for $\mathcal{W}_A/\mathcal{W}_{A,\nu_2}$ and
$\mathcal{W}_A/\mathcal{W}_{A,\nu_3}$. Finally, because
\begin{equation}\label{eq3.142}
  V=2(\nu_1+\nu_2+\nu_3)=2(r_1+r_2+r_3),
\end{equation}
we can represent Whitham equations as
\begin{equation}\label{eq3.143}
  \frac{\prt r_i}{\prt t}+v_i(r_1,r_2,r_3)\frac{\prt r_i}{\prt x}=0,\quad i=1,2,3,
\end{equation}
with the characteristic velocities
\begin{equation}\label{eq3.144}
\begin{split}
  v_i&=2(r_1+r_2+r_3)-\frac{2L}{\prt L/\prt r_i}=\\
  & =  \left(1-\frac{L}{\prt_i  L}\prt_i\right)V,
    \quad i=1,2,3,
  \end{split}
\end{equation}
where $\prt_i\equiv\prt/\prt r_i$. Because formula (\ref{eq1.287}) for
the wavelength becomes
\begin{equation}\label{eq3.145}
  L=\frac{2K(m)}{\sqrt{r_3-r_1}},\quad m=\frac{r_2-r_1}{r_3-r_1},
\end{equation}
substitution of (\ref{eq3.145}) into (\ref{eq3.144}) using the known expression for
the derivative of the elliptic integral $K(m)$ (see, e.g., \cite{AS-2})
allows expressing the velocities $v_i$ as
\begin{equation}\label{eq6.12}
\begin{split}
&v_1=2(r_1+r_2+r_3)+\frac{4(r_2-r_1)K(m)}{E(m)-K(m)},   \\
&v_2=2(r_1+r_2+r_3)-\frac{4(r_2-r_1)(1-m)K(m)}{E(m)-(1-m)K(m)},   \\
&v_3=2(r_1+r_2+r_3)+\frac{4(r_3-r_1)(1-m)K(m)}{E(m)},
\end{split}
\end{equation}
where $E(m)$ is the full elliptic integral of the second kind. This
is just the form of modulation equations for cnoidal KdV
waves arrived at by Whitham in \cite{whitham-65}.

The possibility of transforming a system of three first-order
equations to diagonal form is a highly nontrivial fact.
Fortunately, Whitham was unaware of a theorem stating that
such a transformation is in general impossible in systems of
more than two equations (see, e.g., \cite{rozhd-yan}). In \cite{whitham-74},
Whitham himself refers to the possibility of such a transformation as
miraculous. It turned out later that, in this case, such a
transformation is made possible by the remarkable mathematical property of
`complete integrability' of the KdV
equation, discovered two years later \cite{ggkm-67}.

If a solution $r_i=r_i(x,t),\,i=1,2,3$, of Whitham equations
for some specific problem is found, then the DSW
profile can be determined by substituting this solution into
the periodic solution, which in the new variables (Riemann
invariants for the system of Whitham modulation equations) takes the form
\begin{equation}\label{eq3.148}
  u=r_2+r_3-r_1-2(r_2-r_1)\sn^2(\sqrt{r_3-r_1}\,(x-Vt),m)],
\end{equation}
with wavelength (\ref{eq3.145}). As $r_2\to r_3$, with $L\to\infty$,
we obtain the soliton limit:
\begin{equation}\label{eq6.14'}
\begin{split}
  & u(x,t)|_{r_2=r_3}=r_1+\frac{2(r_3-r_1)}{\cosh^2[\sqrt{r_3-r_1}(x-V_st)]}, \\
  & V_s=2(r_1+2r_3),
  \end{split}
\end{equation}
and in the small-amplitude limit $r_2-r_1\ll r_2$, the cnoidal
wave becomes harmonic:
\begin{equation}\label{eq6.14''}
\begin{split}
  &u(x,t)=r_3+(r_2-r_1)\cos[2\sqrt{r_3-r_1}\,(x-Vt)],\\
  & V=2(2r_1+r_3)
  \end{split}
\end{equation}
with the wavelength $\pi/\sqrt{r_3-r_1}$, which coincides with the
$m\to0$ limit of (\ref{eq3.145}), as it should.

Whitham equations, even if used alone, allow substantial progress
in the description of the DSW formation in
specific problems, and investigations of this kind were
initiated in Gurevich and Pitaevskii's work~\cite{gp-73}. But, before
discussing these problems, in Section~5 we describe the general
method for solving Whitham equations, developed later
largely by Gurevich and his collaborators~\cite{gkm-89,gke-91,gkme-92,gke-92,ek-93}
(also see \cite{ks-90,ks-91,kud-92,kud-92b,wright-93,tian-93}).

\section{Generalized hodograph method}\label{hodograph}

It was Riemann who made the following observation
regarding the equations of gas dynamics. For arbitrary one-dimensional flows
with the gas density $\rho=\rho(x,t)$ and the flow velocity $u=u(x,t)$ being
functions of the coordinate $x$ and time $t$, the so-called hodograph transformation
making $x$ and $t$ functions of Riemann invariants expressed through $\rho$
and $u$ linearizes the equations for $x$ and $t$; they then allow
solutions in a form quite convenient in applications.
Whitham modulation equations (\ref{eq3.143}) are similar in form
to the equations of gas dynamics after the transformation to
the diagonal form, and it is therefore natural to try to apply a
similar method to solve Whitham equations. Such a `generalized hodograph method'
was proposed in a very general form by Tsarev \cite{tsarev-85} as a strategy
to solve hydrodynamic-type equations with more than two dependent
variables. We give some elementary prolegomena to this
method, which were used by Gurevich and collaborators to
solve Whitham's equations (\ref{eq3.143}) in the Gurevich-Pitaevskii
problem.

In the simplest case of Hopf equation (\ref{rev3.6}), which is the
dispersionless limit of the KdV equation, it is easy to express
solution (\ref{rev3.7}) through the initial distribution of $u$. We now
have three equations (\ref{eq3.143}) of a similar form, and we can seek
their solution in a similar form:
\begin{equation}\label{eq6.24}
  x-v_i(r)t=w_i(r), \quad i=1,2,3,
\end{equation}
where the $w_i(r)$ are the functions to be determined. Differentiating
these relations with respect to $r_j$, we obtain
$ -(\prt v_i/\prt r_j)t=\prt w_i/\prt r_j, \, i\neq j,$, where we can
eliminate $t$ using (\ref{eq6.24}), $ t=-(w_i-w_j)/(v_i-v_j)$. As a result,
we see that the functions $w_i$ must satisfy the Tsarev equations
\begin{equation}\label{eq6.25}
  \frac{1}{w_i-w_j}\frac{\prt w_i}{\prt r_j}=\frac{1}{v_i-v_j}\frac{\prt v_i}{\prt r_j},
  \quad i\neq j.
\end{equation}
Therefore, if we find the general solution $w_i(r)$ of these
equations for the given $v_i(r)$, we obtain the general solution
(\ref{eq6.24}) of Whitham equations (\ref{eq3.143}), which can then be specified
for any particular problem.

We can find a way to solve Eqs.~(\ref{eq6.25}) if we note that these
equation can be represented as compatibility conditions for
Whitham's equations (\ref{eq3.143}) and some auxiliary equations,
\begin{equation}\label{eq6.26}
  \prt r_i/\prt\tau+w_i(r_j)\prt r_i/\prt x=0, \quad i,j=1,2,3,
\end{equation}
for the evolution of Riemann invariants depending on a
fictitious `time' $\tau$ with formal `velocities' $w_i(r_j)$. After simple
transformations, the condition $\prt^2r_i/\prt\tau\prt t=\prt^2r_i/\prt t\prt\tau$
then gives the equation
$w_j\prt v_i/\prt r_j + v_i\prt w_i/\prt r_j = v_j\prt w_i/\prt r_j + w_i\prt v_i/\prt r_j,$
which is equivalent to (\ref{eq6.25}). Regarding $w_i(r)$ as an
analogue of the Whitham velocities, it is natural to seek the
solution $w_i$ of Tsarev equations in a form similar to (\ref{eq3.144}), \cite{gke-91}
\begin{equation}\label{eq6.28}
  w_i=\left(1-(\prt_i\ln{L})^{-1}\prt_i\right)W,\quad \prt_i\equiv\prt/\prt r_i,
\end{equation}
Using the expressions
$v_i=2\sigma_1-2(\prt_i\ln{L})^{-1}$, $\sigma_1=r_1+r_2+r_3$,
 we represent Eq.~(\ref{eq6.28}) as
\begin{equation}\label{eq6.29b}
 w_i=W+\left(\tfrac{1}{2}v_i-\sigma_1\right)\prt_i W,
\end{equation}
and after a simple calculation arrive at
\begin{equation*}
\begin{split}
w_i-&w_j=\left(\tfrac{1}{2}v_j-\sigma_1\right)(\prt_i W-\prt_j W) + \tfrac{1}{2}(v_i-v_j)\prt_i W,\\
\prt_j w_i&=\prt_j W-\prt_i W + \left(\tfrac{1}{2}v_i-\sigma_1\right)\prt_{ij}W + \tfrac{1}{2}\prt_j
v_i \cdot \prt_i W,
\end{split}
\end{equation*}
where $\prt_{ij}=\prt^2/\prt r_i \prt r_j$.  Substituting these expressions into
Eqs.~(\ref{eq6.25}) yields equations for $W$:
\begin{equation}\label{eq6.30}
\begin{split}
  &\prt_j W-\prt_i W+\left(\tfrac{1}{2}v_i-\sigma_1\right)\prt_{ij}W = \\
  &=\left(\tfrac{1}{2}v_j-\sigma_1\right)
  (\prt_i W-\prt_j W) \tfrac{\prt_j v_i}{v_i-v_j}.
  \end{split}
\end{equation}
To simplify, we define the polynomial
\begin{equation}\label{eq6.31b}
\begin{split}
  Q(r)&=(r-r_1)(r-r_2)(r-r_3)=\\
  &=r^3-\sigma_1r^2+\sigma_2r-\sigma_3,\\
  \sigma_1&=\sum_i r_i,\quad \sigma_2=\sum_{i<j}r_ir_j,\quad \sigma_3=r_1r_2r_3,
  \end{split}
\end{equation}
where $r$  is an arbitrary parameter, and use the easily verified identity
\begin{equation}\label{eq6.31}
  \prt_{ij}\frac{1}{\sqrt{Q(r)}} = \frac{1}{2(r_i-r_j)}\left(\prt_i\frac{1}{\sqrt{Q(r)}}
  - \prt_j \frac{1}{\sqrt{Q(r)}}\right),
\end{equation}
It follows from (\ref{eq1.287}) that, up to an inessential factor, the
wavelength is $L=\oint dr/\sqrt{Q(r)}$, where the integral is taken
along a closed contour encircling the interval between two
zeros $r_1$ and $r_2$ of $Q(r)$. Therefore, integrating Eq.~(\ref{eq6.31}) along
the same contour, we obtain the relation
\begin{equation}\label{eq6.32}
  \frac{\prt_{ij}L}{\prt_iL-\prt_jL}=\frac{1}{2(r_i-r_j)}.
\end{equation}
Substituting (\ref{eq3.144}) on the right-hand side of (\ref{eq6.25}), after simple
transformations using the established identities, we obtain a
system of equations for the potential $W$:
\begin{equation}\label{eq6.33}
  \frac{\prt^2 W}{\prt r_i\prt r_j}-
  \frac1{2(r_i-r_j)}\left(\frac{\prt W}{\prt r_i}-\frac{\prt W}{\prt r_j}\right)=0,
  \quad i\neq j.
\end{equation}
These equations are called the Euler-Poisson equations, and
they are the subject of a vast mathematical literature. We here
restrict ourselves to the simplest facts that allow us to solve
several interesting problems from the Gurevich-Pitaevskii
theory for the DSW dynamics.

We first note that comparing Eq.~(\ref{eq6.33}) with identity (\ref{eq6.31})
implies that
\begin{equation}\label{eq6.27}
\begin{split}
  W(r,r_1,r_2,r_3)&=\frac{r^{3/2}}{\sqrt{Q(r)}}=
  \sum_{k=0}^{\infty}\frac{W^{(k)}(r_1,r_2,r_3)}{r^k}
  \end{split}
\end{equation}
is a solution of Eqs.~(\ref{eq6.33}) dependent on an arbitrary
parameter $r$. We hence immediately conclude that (\ref{eq6.27}) can
be considered the generating function of particular solutions
$W^{(k)}(r_1,r_2,r_3)$ given by the coefficients of the expansion of $W$
in inverse powers of $r$. When these are substituted into (\ref{eq6.29b}),
we obtain particular solutions (\ref{eq6.24}) of Whitham's equations in
implicit form. These simplest solutions now allow describing
the behavior of DSWs in several characteristic instances of
the Gurevich-Pitaevskii problem, to which we restrict
ourselves in this paper.

\section{Gurevich-Pitaevskii problem setup}

\begin{figure}[t]
\begin{center}
\includegraphics[width=8cm]{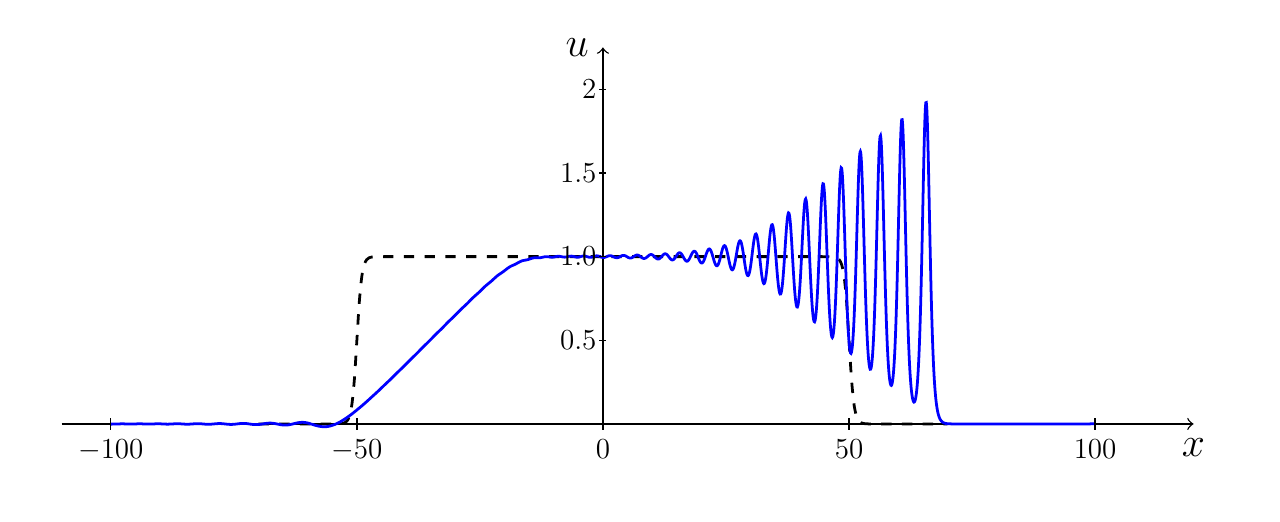}
\caption{ Evolution of a pulse with initial distribution (\ref{rev5.1}) (dashed curve)
over the time $t=5$ in accordance with KdV equation (\ref{eq1.270}).
 }
\label{fig5}
\end{center}
\end{figure}

To present the general physical ideas regarding the problem
setup within the Gurevich-Pitaevskii approach to the DSW
theory, we consider results of a numerical solution of the KdV
equation with the initial distribution given by a `tabletop'
with somewhat rounded edges:
\begin{equation}\label{rev5.1}
  u_0(x)=\left\{
  \begin{array}{ll}
  1,\qquad & |x|\leq l_0,\\
  0,\qquad & |x|> l_0,\\
  \end{array}
  \right.
\end{equation}
In our dimensionless variables, the dispersive size is equal to
unity, and we have therefore chosen the initial tabletop of
a sufficiently large width $2l_0$, such that the width of the
forming DSW could also grow large, and the applicability
condition of Whitham averaging method would safely
hold for $t\gg1$. As can be seen from Fig.~\ref{fig5}, as a result of the
evolution of an initial distribution close to the one in (\ref{rev5.1}),
two structures form on its edges. At the trailing edge, a
rarefaction wave forms, which, ignoring the dispersion,
would be described by the hydrodynamic solution $u(x,t)=(x+l_0)/(6t)$
for $-l_0\leq x\leq -l_0+6t$. The leading edge of
distribution (\ref{rev5.1}) forms the domain of oscillations, i.e., the
DSW, and we must find a suitable way to describe it in the
hydrodynamic limit of vanishing dispersion.

It is useful to briefly discuss here how a similar problem is
solved in the theory of viscous shock waves (see, e.g., \cite{LL6}). As
is known, in media with weak dissipation, the wave breaking
shown in Fig.~\ref{fig3} is eliminated due to the formation of a very
thin transition domain between two states of the medium
flow. Inside this domain, strong irreversible processes occur
that are determined, for example, by the viscosity and heat
conductance of the gas, but, farther away from this transition
domain, the flow rapidly becomes an ideal gas flow, where
any irreversible processes can be disregarded. In the limit of
vanishing viscosity, heat conductance, and other characteristics
of dissipative processes, the thickness of the transition
domain in our macroscopic description tends to zero and we
can replace it with a discontinuity surface of the hydrodynamic
variables, with the flow considered dissipation-free
on both sides of the surface. The characteristics of the flow
and of the thermodynamic state of the gas must satisfy the
conditions of mass, momentum, and energy conservation in
the transition across the discontinuity, which determine the
law of motion of the discontinuity.

In our case of interest, DSWs, we must make a similar
transition to the hydrodynamic limit of vanishing dispersion.
Instead of a discontinuity surface, we now have a domain of
oscillations with a vanishing wavelength inside it, and the
dynamics of this domain are described by Whitham
modulation equations, which on `macroscopic' scales also
have the form of hydrodynamic first-order partial differential
equations. Similarly to the case of a usual shock wave, we
must incorporate a solution of these equations into the
solution of the dispersionless Hopf equation, such that the
smooth dispersionless solution continuously matches the
averaged characteristics of the modulated oscillating solution.

It is obvious that, on the soliton edge of a DSW, this
implies that the leading soliton must propagate over the
background described by a smooth solution at the matching
point. The situation is more delicate at the low-amplitude
edge, where we should apparently expect matching with the
solution of linear modulation equations (\ref{125.12}) and
(\ref{l-19.58}). But in the limit of vanishing dispersion,
the wave amplitude tends to zero at the matching point and
Eq.~(\ref{l-19.58}) is satisfied in that
limit automatically. Still, the conservation law for the number
of waves in Eq.~(\ref{rev4-1}), which we used in deriving Whitham
equations, turns into its linear limit (\ref{125.12}) at the matching
point. Therefore, the small-amplitude edge of the DSW moves
over a smooth background with some group velocity, which
in Whitham's modulation theory becomes a hydrodynamic
variable characterizing the DSW.

Indeed, taking the limit of vanishing dispersion can be
formally regarded as a rescaling, i.e., a transition to `slow'
variables $X=\eps x$ and $T=\eps t$, such that the KdV equation
becomes $u_T+6uu_X+\eps^2u_{XXX}=0$, the wavelength acquires
the order of magnitude $L\sim\eps$, and in the limit $\eps\to0$ the last
equation passes into the Hopf equation. In that same limit,
the parameter $\eps$ drops from the expression for the group
velocity $v_g=-3\eps^2k^2\sim (\eps/L)^2\sim1$, and hence the velocity of
the small-amplitude DSW edge is determined only by the values
of modulation parameters characterizing the DSW envelope.
We emphasize that the DSW picture described here,
as proposed by Gurevich and Pitaevskii, is substantially
different from the earlier proposals by Benjamin-Lighthill
and Sagdeev, according to which the DSW had a stationary
character and its overall characteristics were determined by
the mandatory existence of weak dissipation, which competed
with dispersion. We return to that picture of the transition to
the stationary DSW with dissipation taken into account in
Section~\ref{kdv-perturb}.

We thus assume that the breaking nonlinear solution of
the dispersionless Hopf equation, Eq.~(\ref{rev3.7}), is modified by
dispersion effects, such that, instead of a multi-valuedness
domain, the domain $x_L<x<x_R$ of wave oscillations occurs
in the distribution $u(x,t)$, with its evolution governed by
Whitham modulation equations. Outside the domain
$x_L<x<x_R$, the wave can be described by the smooth
solution of the Hopf equation in Eq.~(\ref{rev3.7}), and inside it, the
DSW is described by expression (\ref{eq3.148}) with good accuracy, with
the parameters $r_1,r_2$, and $r_3$ being a solution of Whitham
equations (\ref{eq3.143}). This solution must satisfy boundary conditions
that ensure matching with the smooth solution. To
clarify the matching conditions, we note that, at these limit
points, the average of $u(x,t)$ over wavelengths,
\begin{equation}\label{eq6.14}
  \langle u\rangle=2(r_3-r_1)\frac{E(m)}{K(m)} + r_1+r_2-r_3,
\end{equation}
can be expressed as
\begin{equation}\label{eq6.15}
  \langle u\rangle_{r_1=r_2}=r_3, \quad  \langle u\rangle_{r_2=r_3}=r_1.
\end{equation}
In other words, on the right edge, the value $r_1$ of the
background over which soliton (\ref{eq6.14'}) is moving is equal to the
value of the dispersionless solution $u(x_R,t)$ at that point; on
the left edge, the background value $r_3$ of small-amplitude limit
(\ref{eq6.14''}) equals the $u(x_L,t)$ value of the same dispersionless
solution. In accordance with the foregoing assumptions, on
the right edge $x_R(t)$, the DSW turns into a sequence of
solitons, and we have $r_2=r_3$, $(m=1)$ in that case. On the
left edge $x_L(t)$, with small amplitude of oscillations, we set
$r_2=r_1$, $(m=0)$.

The coincidence of two Riemann invariants leads to the
equality of the corresponding Whitham velocities (\ref{eq6.12}) at the
DSW edges. We obtain
\begin{equation}\label{eq6.16}
  v_1|_{r_2=r_1}=v_2|_{r_2=r_1}=12r_1-6r_3, \quad v_3|_{r_2=r_1}=6r_3,
\end{equation}
and
\begin{equation}\label{eq6.17}
  v_1|_{r_2=r_3}=6r_1, \quad v_2|_{r_2=r_3}=v_3|_{r_2=r_3}=2r_1+4r_3.
\end{equation}
It then follows that, on the trailing edge $x=x_L(t)$, where the
wave $u(x,t)$ and its averaged value coincide with the Riemann
invariant $r_3$, its evolution is determined by the limit of
Whitham equation
\begin{equation}\label{eq6.18}
  \prt r_3/\prt t+6r_3 \prt r_3/\prt x=0, \quad r_2=r_1, \quad x=x_L(t),
\end{equation}
which coincides with Hopf equation (\ref{rev3.6}) for $u(x,t)$ in the
dispersionless limit. Similarly, on the leading front $x=x_R(t)$,
where the averaged value $\langle u(x,t)\rangle$ coincides with the Riemann
invariant $r_1$, its evolution is determined by the same Hopf
equation:
\begin{equation}\label{eq6.19}
  \prt r_1/\prt t+6r_1 \prt r_1/\prt x=0, \quad r_2=r_3, \quad x=x_R(t).
\end{equation}
We can thus conclude that the boundary condition
\begin{equation}\label{eq6.20}
  v_1|_{r_1=r_2}=v_2|_{r_1=r_2}, \quad v_3|_{r_1=r_2}=6r_L,
\end{equation}
is satisfied at the trailing edge of the DSW, and the condition
\begin{equation}\label{eq6.21}
  v_1|_{r_2=r_3}=6r_R, \quad v_2|_{r_2=r_3}=v_3|_{r_2=r_3}.
\end{equation}
is satisfied at the leading edge. Here, $r_L$ and $r_R$ are the values
that solution (\ref{rev3.7}) of the Hopf equation, which corresponds to
the initial profile $r=u_0(x)$, takes at the DSW matching
points. For the solution of form (\ref{eq6.24}), the DSW endpoints
must match solution (\ref{rev3.7}) of the Hopf equation, and boundary
conditions (\ref{eq6.20}) and (\ref{eq6.21}) can be represented as
\begin{equation}\label{eq6.34}
  w_1|_{r_1=r_2}=w_2|_{r_1=r_2}, \quad w_3|_{r_1=r_2}=\bar{x}(r_3)
\end{equation}
\begin{equation}\label{eq6.35}
  w_1|_{r_2=r_3}=\bar{x}(r_1), \quad w_2|_{r_2=r_3}=w_3|_{r_2=r_3}.
\end{equation}
If we manage to find a solution of Whitham equations (\ref{eq3.143})
satisfying the stated conditions, then we obtain the functions
$r_1$, $r_2$, and $r_3$ in the entire domain $x_L(t)<x<x_R(t)$ and
therefore describe the oscillating wave envelope for the entire
DSW.

Before proceeding to solutions of specific problems, we
note that Whitham equations, as follows from their homogeneity,
have self-similar solutions of the form
\begin{equation}\label{eq6.38}
  r_i(x,t)=t^\gamma R_i(xt^{-1-\gamma}),
\end{equation}
where $\gamma$ is an arbitrary self-similarity exponent and $R_i(z)$ is a
solution of the system of ordinary differential equations
\begin{equation}\label{eq6.39}
  [(1+\gamma)z-v_i(R)]R'_i = \gamma R_i, \quad i=1,2,3,
\end{equation}
where $z=xt^{-1-\gamma}$, $R'_i\equiv dR_i/dz$, and $v_i(R)=t^{-\gamma}v_i(r)$,  i.e.,
$v_i(R)$ is expressed through $R_i$ by the same formulas that
express $v_i(r)$ through $r_i$. This remark allows finding useful
classes of solutions describing DSWs for some especially
chosen initial conditions.

\section{Evolution of the initial discontinuity
in the Korteweg-de Vries theory}\label{step}

We begin with the simplest example \cite{gp-73}, similar to the problem
of the evolution of step-like profile (\ref{eq1.47k}) in the theory of the
linear KdV equation. To simplify formulas, we use the fact
that the KdV equation is invariant under the Galilei
transformations $x\ra x+6At, t\ra t, u\ra u+A$ and the
scale transformations $x\ra x/B^{1/2}, t\ra t/B^{3/2}, u\ra Bu$,
where $A$ and $B$ are constant parameters. Using these
transformations, the initial step-like profile of an arbitrary
amplitude can be represented as
\begin{equation}\label{eq6.48}
 u_0(x)=u(x,0)=
  \begin{cases}
    1, & \text{  \ $x<0$}, \\
    0, & \text{  \ $x>0$}.
  \end{cases}
\end{equation}
In the dispersionless approximation, we obtain the formal
solution of the Hopf equation,
$$
  u(x,t)=
  \begin{cases}
    1, & \text{  \ $x<6t$}, \\
    x/(6t), & \text{  \ $0\le x\le 6t$}, \\
    0,  & \text{  \ $x>6t$},
  \end{cases}
$$
which is multi-valued in the domain $0<x<6t$. According to
Gurevich and Pitaevskii, a DSW emerges instead of this
domain when taking dispersion into account, with the DSW
evolution governed by Whitham's equations.

In Whitham's hydrodynamic approximation, initial conditions
contain no parameters of the dimension of length, and
hence the solution of modulation equations must be self-similar
(see~(\ref{eq6.38} with $\gamma=0$), i.e., $r_i=r_i(z), z=x/t$, where
$r_i(z)$ satisfy the differential equations $(v_i-z)\cdot{dr_i}/{d z}=0$
(see~(\ref{eq6.39})). On the trailing edge $z=z_L$, where the oscillation
amplitude tends to zero, we have $r_1=r_2$, and the averaged
value $\langle{u}\rangle$ coincides with $u=1$ (see~(\ref{eq6.15})), the boundary
condition $ r_1(z_L)=r_2(z_L), r_3(z_L)=1$ must hold. On the
leading soliton front $z=z_R$, where $r_2=r_3$ and the averaged
value $\langle{u}\rangle=r_1$ vanishes, we have another boundary condition:
$r_2(z_R)=r_3(z_R), r_1(z_R)=0$. It is easy to see that we obtain
a solution satisfying both boundary conditions if we set
\begin{equation}\label{eq6.53}
  r_1\equiv 0, \quad   r_3\equiv 1, \quad   v_2=z.
\end{equation}
Then, $m=(r_2-r_1)/(r_3-r_1)=r_2$ and the last equation in
(\ref{eq6.53}) determines the dependence of the self-similar variable
$z=x/t$ on $r_2$,
\begin{equation}\label{eq6.55}
  z=\frac{x}{t}=2(1+r_2)-\frac{4r_2(1-r_2)K(r_2)}{E(r_2)-(1-r_2)K(r_2)}.
\end{equation}
Taking the limit $r_2\ra 0$, we find the value of the self-similar
variable on the trailing edge:
\begin{equation}\label{eq6.56}
  z_L=-6 \quad \text{или} \quad x_L=-6t,
\end{equation}
which means that the oscillation domain expands into the
unperturbed domain of the pulse with the speed $s_L=v_g=-6$
equal to the group velocity of linear waves on the constant
background $u=1$ with the dispersion law $\omega=6k-k^3$.
Indeed, the group velocity $d\omega/dk=6-3k^2$ is $v_g=-6$ for
the wavelength equal to $L(0)=\pi$ in accordance with (\ref{eq3.145}),
and hence for $k=2\pi/L=2$.

\begin{figure}[t]
\begin{center}
\includegraphics[width=6cm]{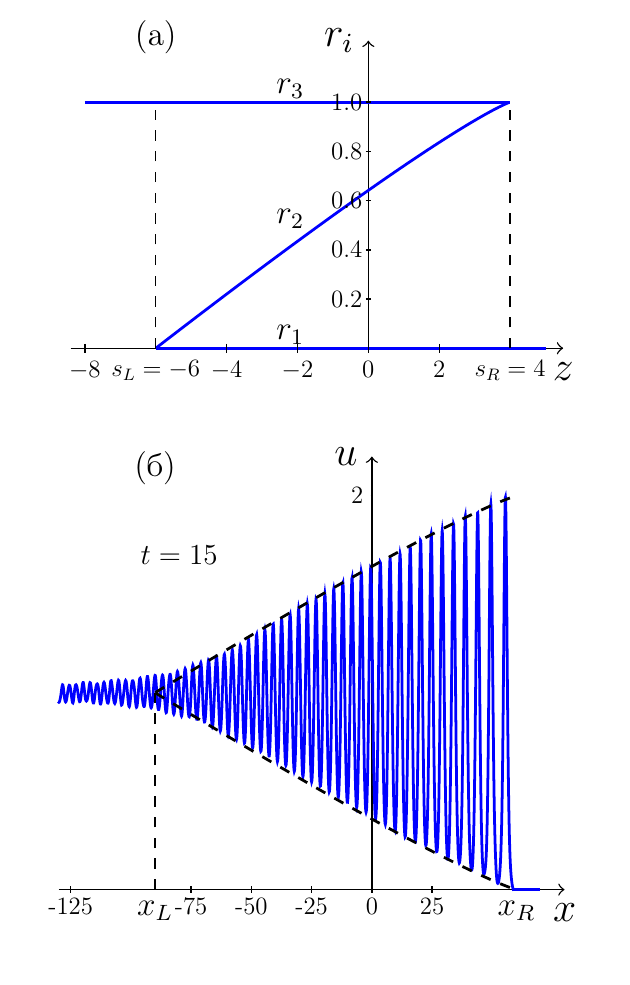}
\caption{  (a) Riemann invariants in the problem of a step-like profile. The
dependence of $r_2$ on $z$ is defined by Eq.~(\ref{eq6.55}).
(b)Evolution of a pulse with a
step-like initial distribution in Eq.~(\ref{eq6.48}) driven by the KdV equation (solid
line). Envelopes of the DSW amplitude are shown with dashed lines.
 }
\label{fig6}
\end{center}
\end{figure}

On the leading front, we haver $r_2 \ra 1$ and Eq.~(\ref{eq6.55}) implies
that
\begin{equation}\label{eq6.57}
  z_R=4 \quad \text{или} \quad x_R=4t,
\end{equation}
and hence this DSW edge moves with the soliton speed
$s_R=V_s=4r_3=4$. The amplitude of the leading soliton is
twice the amplitude of the step-like profile. The dependence of
$r_2=m$ on the variable $z''=4-z$, $|z''|\ll 1$ near the leading
front is determined by the equation $z''\cong 2(1-m)\ln(16/(1-m))$,
which gives $1-m\cong z''/2\ln(1/z'')$ with
logarithmic accuracy. Therefore, the distance between solitons near
the leading front (where $4t-x\sim 1$ or $4-z=z''\sim 1/t$) increases
with time as
\begin{equation}\label{eq6.58}
  L=2K(m)/\sqrt{r_3-r_1} \cong \pi\ln(1/z'')=\pi\ln t.
\end{equation}
Overall, the dependence of $r_2=m$ on $z$ is shown in Fig.~\ref{fig6}(a).
Substituting the values of Riemann invariants into formula
(\ref{eq3.148}) gives an expression for $u(x,t)$ in a DSW:
\begin{equation}\label{eq6.59}
  u(x,t)=1+r_2-2r_2\,\sn^2 (x(r_2)-2(1+r_2)t, r_2),
\end{equation}
with the dependence $x(r_2)$ at a fixed instant $t$ determined by
Eq.~(\ref{eq6.55}). Therefore, the envelope of the maxima is given by
the function $u_{max}=1+r_2$, and the envelope of the minima,
by the function $u_{min}=1-r_2$. In Fig.~\ref{fig6}(b), they are shown with
dashed lines. As we can see, Whitham's theory is quite good at
describing the DSW at a moderate value $t=15$, and it can be
verified that the accuracy increases as $t$ increases. Whitham's
theory correctly predicts the wave number value corresponding to the
small-amplitude edge of a DSW.

\section{Breaking of the wave with a parabolic profile}

In Section~\ref{step}, we considered the simplest Gurevich-Pitaevskii
problem of the formation of a DSW from a very particular
initial profile, a jump-like discontinuity. Although some
interesting problems can be reduced to this idealized case,
including the problem of DSW generation in a flow past an
obstacle \cite{gs-86,smyth-87}, it is rather remote from the typical wave
breaking patterns. As is known (see, e.g., \S101 in \cite{LL6}), there are
two main breaking scenarios for a simple wave. In the first
scenario, the wave propagates into a quiescent medium and at
the instant of breaking the distribution of the wave perturbation
acquires a vertical tangent on the interface with the
quiescent medium. In the most typical case, the wave
amplitude then vanishes in accordance with a square-root
law. In the second, more common, scenario, the breaking
occurs as a result of the evolution of the distribution with an
inflection point: at the instant of breaking, in the dispersionless
approximation, this profile also acquires a vertical tangent at
the inflection point, and in typical situations can be represented
by a cubic parabola. In this section, we consider the first
wave breaking scenario, and in Section~\ref{cubic-kdv} turn to the second.

We thus assume that at the instant of breaking $t=0$, the
pulse amplitude vanishes in accordance with a square-root law,
\begin{equation}\label{eq6.70}
 u_0(x)=u(x,0)=
  \begin{cases}
    \sqrt{-x}, & \text{  \ $x<0$}, \\
    0, & \text{  \ $x>0$}.
  \end{cases}
\end{equation}
Using Galilei and scaling transformations, we can bring
$x|_{t=0}\propto -u^2$ to this simple dimensionless form. The solution
of the Hopf equation with initial condition (\ref{eq6.70}) is (see~(\ref{rev3.7}))
\begin{equation}\label{eq6.71}
  x-6ut=-u^2,
\end{equation}
showing that this solution has a domain of multi-valuedness
for $0<x<9t^2$ after the instant of breaking $t>0$. According
to the Gurevich-Pitaevskii theory, when dispersion effects are
taken into account, this multi-valuedness domain is superseded
by a DSW that occupies the domain $x_L\leq x\leq x_R$. On
its small-amplitude trailing edge $x_L$, the DSW matches solution
(\ref{eq6.71}) (see~(\ref{eq6.34})
\begin{equation}\label{eq6.72}
  w_3|_{r_1=r_2}=-u^2, \quad u=r^L_3.
\end{equation}
It hence follows that we must seek solution (\ref{eq6.24}) with the
functions $w_i$ that are quadratic in the Riemann invariants in
the limit $m\ra 0$. Velocities of this type with power-law
dependences on the Riemann invariants as $m\ra 0$ occur in
studying generating function (\ref{eq6.27}), and the required quadratic
dependence corresponds to the coefficient  $W^{(2)}(r_1,r_2,r_3)$ at
$r^{-2}$. Thus, we take $w_i(r)$ in form (\ref{eq6.29b}) with $W=W^{(2)}$, which,
in view of the linearity of the Euler-Poisson equations, can be
multiplied by an arbitrary constant factor $C$:
\begin{equation}\label{eq6.73}
\begin{split}
  &w_i=C\left(1-({L}/{\prt_iL})\prt_i\right)W^{(2)}(r_1,r_2,r_3), \\
  &W^{(2)}(r_1,r_2,r_3)=2\sigma_2-\tfrac{3}{2}\sigma_1^2, \\
  &\sigma_2=r_1r_2+r_2r_3+r_3r_1, \quad \sigma_1=r_1+r_2+r_3.
\end{split}
\end{equation}
A specific value of $C$ is determined by the condition of
matching with a smooth solution on the small-amplitude
DSW edge, where $r_3=u_L$. On the leading soliton edge $x_R$,
the averaged amplitude then vanishes, and this condition
yields $r_1=0$ and $r_2=r_3$. Hence, we can satisfy the boundary
conditions by taking $r_1\equiv0$ and choosing the constant $C$ such
that condition (\ref{eq6.72}) holds. Calculating $w_3$ at $m\to0$, we obtain
$w_3=-\frac{15}2Cr_3^2$, and it therefore follows from the matching
condition that $C=2/15$. Finally, we obtain formulas for a
solution of Whitham's equations \cite{gkm-89,ks-90}
\begin{equation}\label{eq6.79}
\begin{split}
  &x-v_2t=\tfrac{2}{15}\left[W+\left(\tfrac{1}{2}v_2-\sigma_1\right){\prt W}/{\prt r_2}\right],\\
  &x-v_3t=\tfrac{2}{15}\left[W+\left(\tfrac{1}{2}v_3-\sigma_1\right){\prt W}/{\prt r_3}\right],
\end{split}
\end{equation}
where $W=2r_2r_3-\frac{3}{2}(r_2+r_3)^2$, $\sigma_1=r_2+r_3$.

\begin{figure}[t]
\begin{center}
\includegraphics[width=6cm]{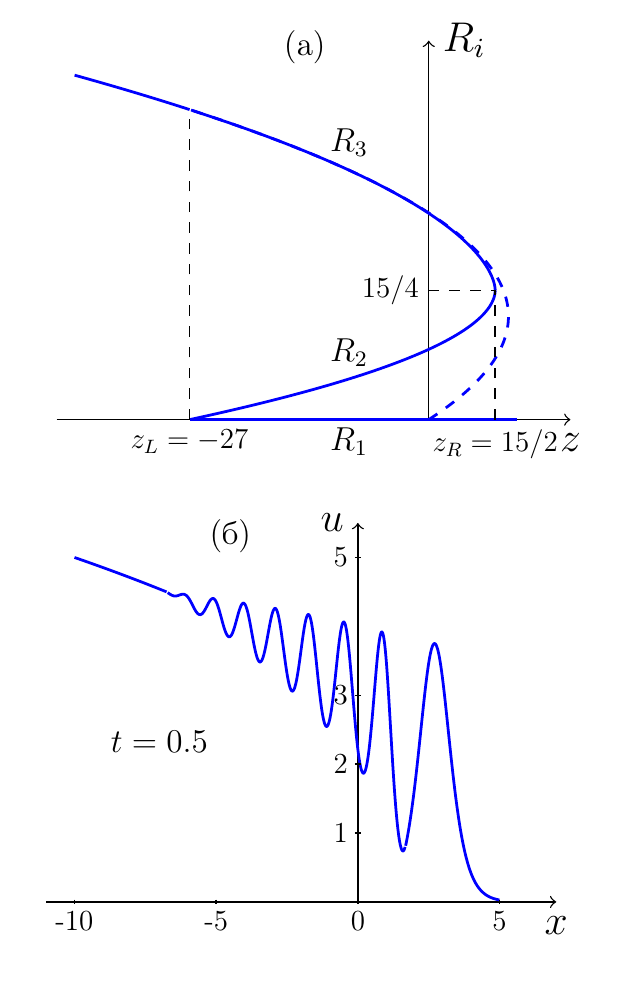}
\caption{(a)Riemann invariants in the problem of breaking of a parabolic-profile pulse.
The dashed line shows the corresponding solution
$z-6R=-R^2$ of the Hopf equation. (b) Evolution of the pulse with
initial profile (\ref{eq6.70}) in accordance with the Gurevich-Pitaevskii theory for
the KdV equation.
 }
\label{fig7}
\end{center}
\end{figure}

On the small-amplitude edge, these equations reduce to
$
 x_L+6r_3^Lt=\tfrac{1}{3}(r_3^L)^2, x_L - 6r_3^Lt=-(r_3^L)^2,
$
which immediately implies the parametric representation
$
 x_L=-\tfrac{1}{3}(r^L)^2,  t=\tfrac{1}{9}r^L,
$
of the law of motion of this
edge, and hence eliminating $r^L$ leads to
\begin{equation}\label{eq6.80}
 x_L=-27t^2.
\end{equation}

On the soliton edge at $r_2=r_3$, both equations (\ref{eq6.79}) tend to
the same limit $ x_R-4r_3t=-\tfrac{8}{15}r_3^2$, and the value of $r^R_3$
is determined by the maximum value of $x$ in the DSW domain,
whence $r_3^R=15t/4$ and
\begin{equation}\label{eq6.81}
  x_R=\frac{15}{2}t^2.
\end{equation}
This is the law of motion of the leading soliton edge.

It follows from the obtained formulas that we have
arrived at a self-similar solution of Whitham's equations
(see~(\ref{eq6.38})) with $\gamma=1$, where the Riemann invariants are
\begin{equation}\label{t3-137.5}
  r_1=R_1\equiv0,\quad r_2=tR_2(z),\quad r_3=tR_3(z)
\end{equation}
with the self-similarity variable $z=x/t^2$. The dependence of
the Riemann invariants $R_i$ on $z$ is shown in Fig.~\ref{fig7}(a). It is clear
that $R_3$ matches the solution of the Hopf equation shown in
the figure with a dashed line. Substituting the found values of
$r_2$ and $r_3$, together with $r_1=0$, into Eq.~(\ref{eq3.148}), we obtain a
parametric form of $u(x,t)$ as a function of the coordinate and
time in the DSW domain. An example of such a dependence
$u(x,t)$ at a fixed instant $t$ is shown in Fig.~\ref{fig7}(b).

\section{ Breaking of a cubic profile}\label{cubic-kdv}

As we have noted, typical wave breaking occurs when the
initial wave profile has an inflection point and in the
dispersionless limit of the solution of the Hopf equation
acquires a vertical tangent at some instant. Because this
breaking point remains an inflection point, the second
derivative of the profile also vanishes at that point. Assuming
that the third derivative of the profile does not vanish at that
point, and also choosing the origin at the breaking point and
the instant of breaking as zero time, we can approximate the
profile near the inflection point with a cubic parabola. As a
result, we obtain a solution of the dispersionless Hopf
equation corresponding to the initial condition $\bar{x}(u)=-u^3$
at $t=0$ in the form
\begin{equation}\label{eq6.86}
 x-6ut=-u^3.
\end{equation}
It is obvious from the foregoing that this is the most typical
distribution at the instant of breaking, and we here discuss the
evolution of the corresponding DSW. The main features of
the solution were investigated in \cite{gp-73}, and an exact analytic
solution was obtained in \cite{potemin-88}.

To solve the problem, we note that the velocities $w_i(r)$ in
(\ref{eq6.29b}) that correspond to the third term $W=W^{(3)}$ in the
expansion of generating function (\ref{eq6.27}) have a cubic dependence
on $r_i$ at the endpoints with $m=0$ and $m=1$. Using the
formula (see~(\ref{eq6.28}))
\begin{equation}\label{eq6.88}
 w_i=\left(1-({L}/{\prt_i L})\prt_i\right)W^{(3)}(r_1,r_2,r_3),
\end{equation}
where
\begin{equation}\label{eq6.89}
  W^{(3)}(r_1,r_2,r_3)=-\frac{5}{4}\sigma_1^3 + 3\sigma_1\sigma_2 - 2\sigma_3,
\end{equation}
and $\sigma_i$ are coefficients of polynomial (\ref{eq6.31b}), it is easy to evaluate
\begin{equation}\label{eq6.90}
\begin{split}
 & w_3=-\frac{35}{4}r_3^3 \quad \text{при} \quad m\ra 0,\\
 & w_1=-\frac{35}{4}r_1^3 \quad \text{при} \quad m\ra 1.
 \end{split}
\end{equation}
Multiplying $w_i$ by $-4/35$, we satisfy the boundary conditions
of DSW matching on the edges with a smooth dispersionless
solution in Eq.~(\ref{eq6.86}), and we find a solution of Whitham's
equations (\ref{eq3.143}) in the form
\begin{equation}\label{eq6.93}
 x-6v_i(r_1,r_2,r_3)t=\frac{4}{35}w_i(r_1,r_2,r_3), \quad i=1,2,3,
\end{equation}
where the functions $w_i$, $i=1,2,3$, are defined by Eqs.~(\ref{eq6.88}
and \ref{eq6.89}). The expressions for $v_i$ and $w_i$ , even if somewhat
bulky, can be given in terms of elliptic integrals as functions of
the Riemann invariants (explicit formulas are presented
below in a self-similar form; see Eqs.~(\ref{eq6.127})-(\ref{eq6.129})).
Therefore, system (\ref{eq6.93}) allows finding $r_i$ as functions of $x$ and $t$.
Before passing to the self-similar form, we consider characteristic properties
of the obtained solution.

On the small-amplitude edge, we have $r_1=r_2$ ($m=0$), and
Eq.~(\ref{eq6.93}) with $i=3$ becomes
\begin{equation}\label{eq6.94}
 x-6r_3t=-r_3^3 \quad \text{(at  $r_1=r_2$)}.
\end{equation}
Similarly, on the soliton edge, we have $r_2=r_3$, and Eq.~(\ref{eq6.93})
with $i=1$ becomes
\begin{equation}\label{eq6.95}
 x-6r_1t=-r_1^3 \quad \text{(at  $r_2=r_3$)}.
\end{equation}
Therefore, these Riemann invariants match the smooth
solution on the DSW edges, as they should:
\begin{equation}\label{eq6.96}
\begin{split}
 & r_3=u \quad \text{at} \quad r_1=r_2, \\
 & r_1=u \quad \text{at} \quad  r_2=r_3,
 \end{split}
\end{equation}
where $u$ is the solution (\ref{eq6.86}) of the Hopf equation. In the
neighborhood of the trailing small-amplitude edge, we introduce a local
coordinate $x'$,
\begin{equation}\label{eq6.97}
 x=x_L + x',
\end{equation}
and small deviations $r'_i$ of the Riemann invariants from their
limit values,
\begin{equation}\label{eq6.98}
 r_1=r_1^L + r_1', \quad r_2=r_1^L +r_2', \quad r_3=r_3^L +r_3'.
\end{equation}
Expanding Eqs.~(\ref{eq6.93}) in powers of $r'_i$ at a fixed instant $t$,
we obtain
\begin{equation}\label{eq6.99}
\begin{split}
  x^L&+x'-(12r_1-6r_3)t-(9r_1'+3r_2'-6r_3')t\\
 &= \tfrac{1}{5}(-16r_1^3+8r_1^2r_3+2r_1r_3^2+r_3^3)-\\
 & -\tfrac{3}{10}(24r_1^2-8r_1r_3-r_3^2)r_1' \\
 & - \tfrac{1}{10}(24r_1^2-8r_1r_3-r_3^2)r_2' +\\
 & + \tfrac{1}{5}(8r_1^2+4r_1r_3+3r_3^2)r_3', \\
  x^L&+x'-(12r_1-6r_3)t-(3r_1'+9r_2'-6r_3')t=\\
&  =\tfrac{1}{5}(-16r_1^3+8r_1^2r_3+2r_1r_3^2+r_3^3)-\\
 & -\tfrac{1}{10}(24r_1^2-8r_1r_3-r_3^2)r_1'-\\
&   - \tfrac{3}{10}(24r_1^2-8r_1r_3-r_3^2)r_2' +\\
 & + \tfrac{1}{5}(8r_1^2+4r_1r_3+3r_3^2)r_3', \\
  x^L&+x'-6r_3t-6r'_3t = -r^3_3-3r_3^2r_3',
 \end{split}
\end{equation}
where we introduce the temporary notation $r_1\equiv r_1^L$ and $r_3\equiv r_3^L$.
Subtracting the second equation from the first, we obtain the relation
\begin{equation}\label{eq6.100}
  t=\frac{1}{30}(24r_1^2-8r_1r_3-r_3^2).
\end{equation}
It hence follows that the coefficients in front of $r'_1$ and $r'_2$ in
the first two equations in (\ref{eq6.99}) vanish, and therefore $x'$ is a
quadratic function of $r'_1$ and $r'_2$:
$$
 x' \propto {r_1'}^2, {r_2'}^2, r_3'.
$$
At the point $x^L$, these two equations give
\begin{equation}\label{eq6.101}
  x^L -(12r_1-6r_3)t=\frac{1}{5}(-16r_1^3+8r_1^2r_3+2r_1r_3^2+r_3^3),
\end{equation}
and the third equation in (\ref{eq6.99}), as we have already noted,
reduces to the solution $x^L -6r_3t=-r_3^3$ of the Hopf equation.
We can hence find the law of motion of the trailing edge.
Subtracting Eq.~(\ref{eq6.94}) with $x=x^L$ from (\ref{eq6.101}) and dividing
the result by $(r_1-r_3)$, we obtain the relation
$$
 t=\frac{1}{30}(8r_1^2+4r_1r_3+3r_3^2),
$$
Comparing this with (\ref{eq6.100}), we find the relation between
values of Riemann invariants on the trailing edge:
\begin{equation}\label{eq6.103}
  r_1^{L} = r_2^{L} =-\frac{1}{4}r_3^{L}.
\end{equation}
It then follows from Eqs.~(\ref{eq6.100}) and (\ref{eq6.94}) that
\begin{equation}\label{eq6.104}
  t=\frac{1}{12}(r_3^L)^2, \quad x^L=-\frac{1}{2}(r_3^L)^3,
\end{equation}
and hence the small-amplitude edge moves according to the law
\begin{equation}\label{eq6.130'}
 x^L=-12\sqrt{3}\,t^{3/2}.
\end{equation}
The amplitude of oscillations here tends to zero as
\begin{equation}\label{eq6.131}
 a=r_2-r_1\simeq 2r_2'\propto\sqrt{x'},
\end{equation}

Near the leading soliton front, we introduce small
variables:
\begin{equation}\label{eq6.113}
 x=x^R-x'',\quad x''>0,
\end{equation}
\begin{equation}\label{eq6.114}
 r_1=r_1^R+r_1'',\quad
 r_2=r_3^R+r_2'',\quad
 r_3=r_3^R+r_3''.
\end{equation}
The expansions of Eqs.~(\ref{eq6.33}) with only the leading corrections
retained have the form
\begin{equation}\label{eq6.115}
 \begin{split}
 &x^R-x''-6r_1t+\left[\frac{8(r_3-r_1)}{\ln(16/(1-m))}\right]t=\\
 &=-r_1^3+\frac{4}{35}(15r_1^2+12r_1r_3+8r_3^2)
 \left[\frac{r_3-r_1}{\ln(16/(1-m))}\right],\\
 &x^R-x''-(2r_1+4r_3)t+2\ln(16/(1-m))(r_3''-r_2'')t=\\
 &=-\tfrac{1}{35}(5r_1^3+6r_1^2r_3+8r_1r_3^2+16r_3^2)\\
 &+ \tfrac{1}{35}(3r_1^2+8r_1r_3+24r_3^2)\ln(16/(1-m))(r_3''-r_2''),\\
 &x^R-x''-(2r_1+4r_3)t-2\ln(16/(1-m))(r_3''-r_2'')t=\\
 &=-\tfrac{1}{35}(5r_1^3+6r_1^2r_3+8r_1r_3^2+16r_3^2)\\
 &- \tfrac{1}{35}(3r_1^2+8r_1r_3+24r_3^2)\ln(16/(1-m))(r_3''-r_2''),
 \end{split}
\end{equation}
where $ 1-m=(r_3''-r_2'')/(r_3-r_1)$,  and we revert to the
temporary notation  $r_1\equiv r_1^R$ and $r_3\equiv r_3^R$.
Subtracting the third equation in (\ref{eq6.115}) from the second,
we obtain the relation
\begin{equation}\label{eq6.117}
 t=\frac{1}{70}(3r_1^2+8r_1r_3+24r_3^2),
\end{equation}
which together with the leading approximation in Eqs.(\ref{eq6.115}),
\begin{equation}\label{eq6.118}
 \begin{split}
 x^R&-6r_1t=-r_1^3,\\
 x^R&-(2r_1+4r_3)t=\\
&=-\frac{1}{35}(5r_1^3+6r_1^2r_3+8r_1r_3^2+16r_3^3),
 \end{split}
\end{equation}
defines the law of motion of the leading edge. Indeed, the
difference between Eqs.~(\ref{eq6.118}) gives another relation,
\begin{equation}\label{eq6.119}
 t=\frac{1}{70}(15r_1^2+12r_1r_3+8r_3^2),
\end{equation}
which, when compared with (\ref{eq6.117}), yields
\begin{equation}\label{eq6.120}
 r_3^R=-\frac34r_1^R,\quad (r_1^R<0),
\end{equation}
whence
\begin{equation}\label{eq6.121}
 t=\frac{3}{20}(r_1^R)^2,\quad x^R=\frac{1}{10}|r_1^+|^3
\end{equation}
and therefore the soliton edge moves in accordance with the law
\begin{equation}\label{eq6.130b}
  x^R=\frac49\sqrt{15}\,t^{3/2}.
\end{equation}
The distance between solitons on the leading edge depends on $x''$ x 00 as
\begin{equation}\label{eq6.132}
 L\propto\ln(1/|x''|).
\end{equation}

\begin{figure}[t]
\begin{center}
\includegraphics[width=5.5cm]{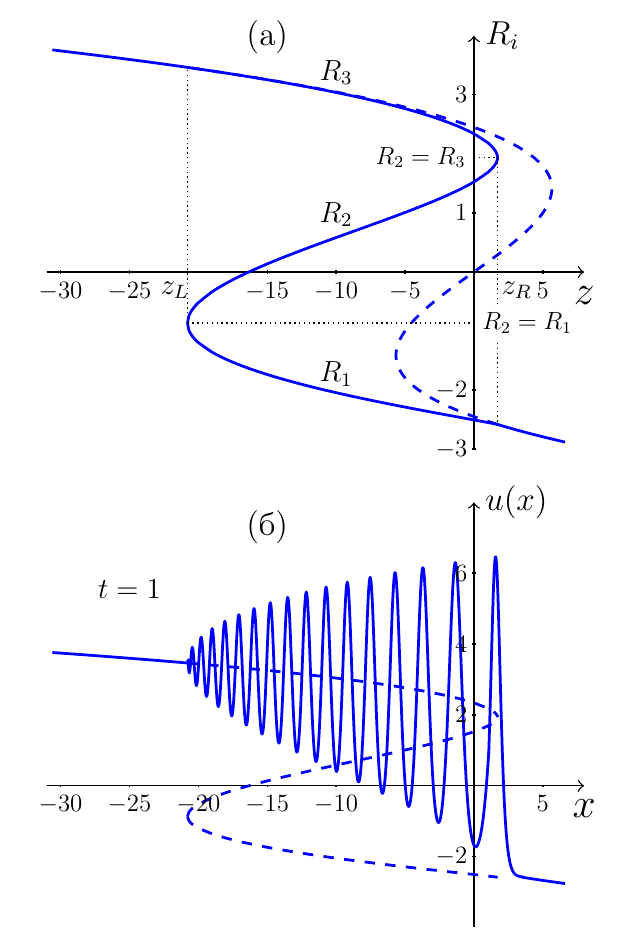}
\caption{ (a) Riemann invariants in the problem of breaking of a cubic-profile
pulse. The dependence of $R_i=r_i/t^{1/2}$ on $z=x/t^{3/2}$ is determined
by Eqs.~(\ref{eq6.127}). The dashed line shows the corresponding solution
$z-6R=-R^3$ of the Hopf equation. (b) Evolution of the pulse with the
initial cubic profile in accordance with Whitham's approximation for the
KdV equation. The dashed line shows the dependence of Riemann
invariants on the coordinate $x$. The evolution time is $t=1$.
 }
\label{fig8}
\end{center}
\end{figure}

The obtained solution, which can be written in the self-similar form
\begin{equation}\label{eq6.93'}
 r_i=t^{1/2}R_i\left(x/t^{3/2}\right),
\end{equation}
is a solution of Eqs.~(\ref{eq6.39}) with $\gamma=1/2$:
\begin{equation}\label{eq6.93''}
 \frac{dR_i}{d z}=\frac{R_i}{3z-v_i(R)}\quad z=x/t^{3/2}.
\end{equation}
The above relations allow easily finding boundary values of
$R_i$. On the trailing small-amplitude edge of the DSW, we have
$z^L=x^L/t^{3/2}=-12\sqrt{3}$ and
\begin{equation}\label{eq6.107}
  R_1^L=R_2^L=-\frac{1}{2}\sqrt{3},\quad R_3^L=2\sqrt{3},
\end{equation}
and on the leading soliton edge, $z^R=4\sqrt{15}/9$ and
\begin{equation}\label{eq6.123}
 R_1^R=-\frac23\sqrt{15},\quad R_2^R=-R_3^R=-\frac12\sqrt{15}.
\end{equation}

The global dependence of $R_i$ on z $z$ defined implicitly by
the expressions
\begin{equation}\label{eq6.127}
 z=6v_1-w_1,\quad z=6v_2-w_2,\quad z=6v_3-w_3,
\end{equation}
where
\begin{equation}\label{eq6.128}
 \begin{split}
 v_1=&2(R_1+R_2+R_3)+\frac{4(R_2-R_1)K(m)}{E(m)-K(m)},\\
 v_2=&2(R_1+R_2+R_3)-\\
-&\frac{4(R_2-R_1)(1-m)K(m)}{E(m)-(1-m)K(m)},\\
 v_3=&2(R_1+R_2+R_3)+\\
&+\frac{4(R_3-R_1)(1-m)K(m)}{E(m)};
 \end{split}
\end{equation}
with $m=(R_2-R_1)/(R_3-R_1)$;  the functions $w_i(R_1,R_2,R_3)$
have the form
\begin{equation}\label{eq6.129}
  w_i=W+\left(\frac12v_i-R_1-R_2-R_3\right)\frac{\prt W}{\prt R_i},
\end{equation}
where
\begin{equation*}
  \begin{split}
   W=&\frac4{35}\Big[-\frac54(R_1+R_2+R_3)^3+3(R_1+R_2+R_3)\times\\
   &\times(R_1R_2+R_2R_3+R_3R_1)-2R_1R_2R_3\Big].
   \end{split}
\end{equation*}
Thus, system of algebraic equations (\ref{eq6.127}) allows finding the
dependence of the invariants $R_i$ on $z$ \cite{potemin-88}. This dependence is
shown in Fig.~\ref{fig8}(a), where the dashed line shows the cubic curve
$ z=6R-R^3$ matching the Riemann invariants $R_3$ and $R_1$ at
the respective points $z^L$ and $z^R$. With the dependence of
the invariants $r_i=t^{1/2}R_i(x/t^{3/2})$ on the self-similar variable
found, their substitution in (\ref{eq3.148}) gives a description of the
DSW forming in the neighborhood of the breaking point due
to dispersion effects. This DSW is plotted in Fig.~\ref{fig8}(b). The
self-similar solution considered here is valid for as long as the
smooth part of the solution is described by cubic curve (\ref{eq6.86})
with sufficient accuracy.

\section{Motion of edges of dispersive shock waves}\label{KdV-edges}

The solutions found in Sections 8 and 9 give an idea of the
nature of the DSW evolution at a stage not too distant in time
from the wave breaking instant, when the smooth part of the
solution remains a monotonic function of the coordinate and
is sufficiently close to a parabola or a cubic parabola. But in
practice the pulses typically have a finite duration, which
raises a question about the DSW shape at the stage when its
full length is comparable to or much greater than the initial
length of the pulse. The hodograph method outlined in
Section~5 allows obtaining a solution to such a problem in
the form of a solution to the system of Euler-Poisson equations (\ref{eq6.33})
\cite{gkm-89,gke-91,gkme-92,gke-92,ek-93,kud-92,wright-93,tian-93}.
However, this form of the
solution is rather complicated, and even a very detailed
quantitative description of the process does not give an
intuitively clear picture of the effect. We therefore do not go
into the details of that theory and discuss a simpler approach
\cite{gkm-89,kamch-19}, which readily yields simple formulas for the principal
parameters of the DSW and, in addition, allows a generalization to a
rather broad class of other nonlinear wave equations.

\begin{figure}[t]
\begin{center}
\includegraphics[width=7cm]{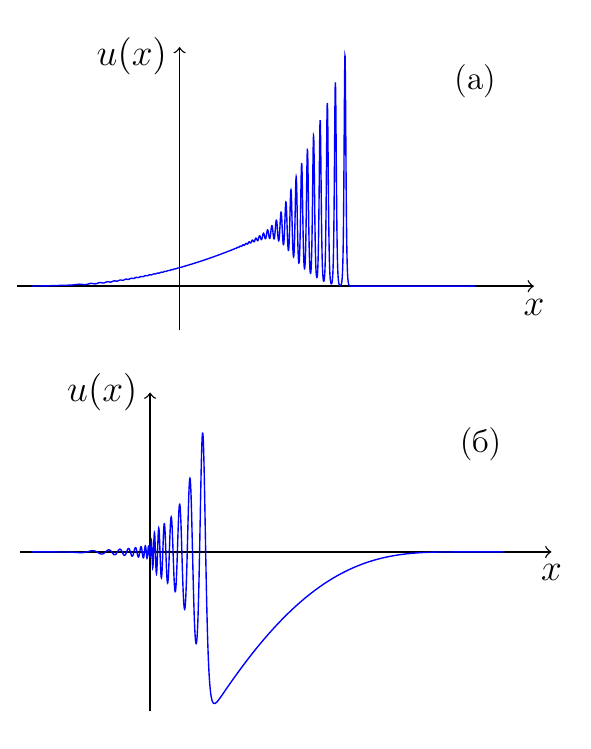}
\caption{  (a) Evolution of a `positive' initial pulse. (b) Evolution of a
`negative' initial pulse.
 }
\label{fig9}
\end{center}
\end{figure}

We first note that `positive' and `negative' pulses with the
respective initial distributions $u_0(x)>0$ and $u_0(x)<0$ must
be distinguished: they exhibit qualitatively different behaviors and must
be considered separately. An idea of how they
evolve can be gleaned from Fig.~\ref{fig9}, where we show the results
of a numerical solution of the KdV equation with appropriate
initial data.

\begin{figure}[t]
\begin{center}
\includegraphics[width=6.5cm]{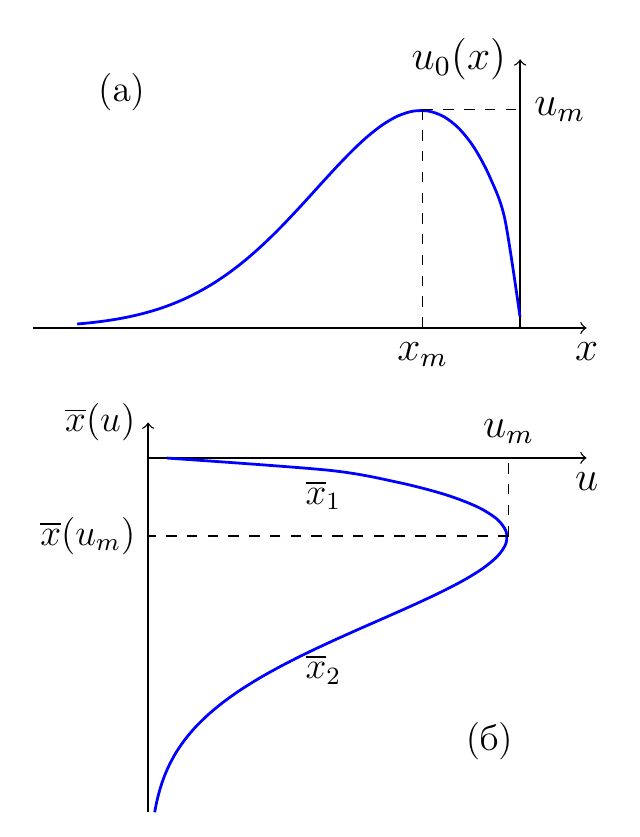}
\caption{ (a) Initial profile of a `positive' pulse. (b) The inverse function
$\ox(u)$ consisting of two branches, $\ox_1(u)$ and $\ox_2(u)$.
 }
\label{fig10}
\end{center}
\end{figure}

For a positive pulse, breaking occurs on the leading front,
and the leading part of the DSW consists of a sequence of
solitons (\ref{eq6.14'}), moving over the zero background, whereas the
trailing small-amplitude edge matches the smooth solution and
propagates over an inhomogeneous background. It must be
recalled here that, in the case of a localized initial pulse $u_0(x)$
with a single maximum $u_m$ of the distribution at $x=x_m$
(Fig.~\ref{fig10}(a)), the inverse function consists of two branches,
$\ox_1(u)$ and $\ox_2(u)$ (Fig.~\ref{fig10}(b)), and hence the dispersionless
solution is given by two formulas (\ref{rev3.7}), one for each branch.

\begin{figure}[t]
\begin{center}
\includegraphics[width=6.5cm]{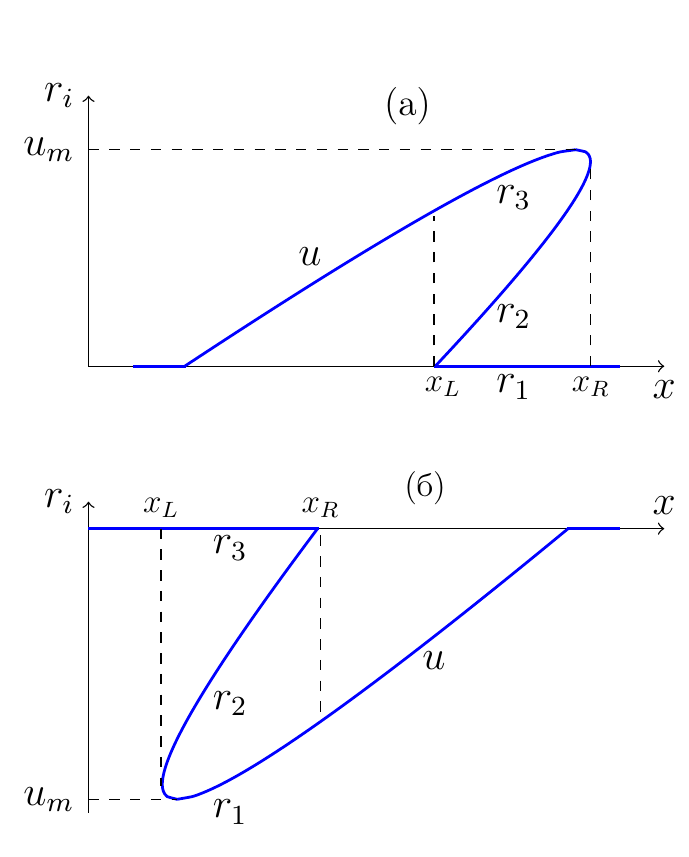}
\caption{ (a) Diagram of Riemann invariants for a `positive' pulse.
(b) Diagram of Riemann invariants for a `negative' pulse.
 }
\label{fig11}
\end{center}
\end{figure}

At the initial stage of the DSW evolution, its small-amplitude edge matches
the solution corresponding to the
branch $\ox_1(u)$, and at the matching point $x_L$ we have
\begin{equation}\label{rev7.1}
  x_L-6ut=\ox_1(u).
\end{equation}
On the other hand, at that point the Riemann invariants $r_1,r_2$
are equal to zero and $r_3=u$ (Fig.~\ref{fig11}(a)), wavelength (\ref{eq3.145})
becomes $L=\pi/\sqrt{u}$, with the corresponding wave number
$k=2\sqrt{u}$, and the velocity of motion of this point, determined
by the group velocity of the linear wave on the background $u$,
is equal to $v_g=6u-3k^2=-6u$. Hence, $dx_L+6udt=0$
along the path of the small-amplitude edge, and the compatibility
condition between Eq.~(\ref{rev7.1}) and the equation
\begin{equation}\label{rev7.2}
  \frac{dx_L}{du}+6u\frac{dt}{du}=0
\end{equation}
leads to the differential equation
\begin{equation}\label{rev7.3}
  2u\frac{dt}{du}+t=-\frac16\frac{d\ox_1}{du},
\end{equation}
which can be easily solved with the initial condition $t(0)=0$,
assuming that the breaking occurs at the zero instant on the
interface with the medium `at rest', where $u=0$. We hence obtain
\begin{equation}\label{rev7.4}
  t(u)=\frac1{12\sqrt{u}}\int_{\ox_1(u)}^0\frac{dx}{\sqrt{u_0(x)}},
\end{equation}
and substituting this into (\ref{rev7.1}) gives the law of motion of the
small-amplitude edge in parametric form:
\begin{equation}\label{rev7.5}
  x_L(u)=\ox_1(u)+\frac{\sqrt{u}}2\int_{\ox_1(u)}^0\frac{dx}{\sqrt{u_0(x)}}.
\end{equation}
It is easy to verify that these formulas reproduce law (\ref{eq6.80}) for
the parabolic initial profile $u_0(x)=\sqrt{-x}$ with a single branch
of the inverse function $\ox_1(u)=-u^2$.

For a localized initial pulse, the obtained solution is valid
until the instant
\begin{equation}\label{rev7.6}
  t_m=\frac1{12\sqrt{u_m}}\int_{x_m}^{0}\frac{dx}{\sqrt{u_0(x)}},
\end{equation}
when the small-amplitude edge reaches the point corresponding to the
maximum amplitude $u_m$. After that, we must solve
Eq.~(\ref{rev7.3}) with the replacement $\ox_1(u)\rightarrow\ox_2(u)$ and with the
initial condition $t(u_m)=t_m$. As a result, we obtain the law of
motion of the small-amplitude edge in parametric form:
\begin{equation}\label{rev7.7}
  \begin{split}
  & t(u)=\frac1{12\sqrt{u}}\int_{\ox_2(u)}^0\frac{dx}{\sqrt{u_0(x)}},\\
  & x_L(u)=\ox_2(u)+\frac{\sqrt{u}}2\int_{\ox_2(u)}^0\frac{dx}{\sqrt{u_0(x)}},
  \end{split}
\end{equation}
where $u_0(x)$ is understood as the full initial profile of the
pulse, vanishing at $x=0$ and tending to zero as $x\to-\infty$.
If the initial pulse vanishes on the trailing edge at
$x=-l\equiv\ox_2(0)$, then, as $t\to\infty$, it is obvious that
 $t\approx \mathcal{A}/(12\sqrt{u})$, where $\mathcal{A}=\int_{-l}^0dx/\sqrt{u_0(x)}$,
 and the law of motion of the trailing edge takes the asymptotic form
\begin{equation}\label{rev7.8}
  x_L\approx-l+\frac{\mathcal{A}^2}{24t},\quad t\to\infty.
\end{equation}

The asymptotic form of the law of motion can also be easily
found for the leading soliton edge of the DSW. We see from
Fig.~\ref{fig11}(a) for Riemann invariants that, as $t\to\infty$, the plots of
$r_2(x)$ and $r_3(x)$ elongate into an extended `tongue', with
$r_1=0$ and $r_2\approx r_3\approx u_m$ near the leading edge. Therefore,
the leading edge moves with the soliton velocity $V_s\approx 4u_m$ and
\begin{equation}\label{rev7.9}
  x_R\approx4u_mt.
\end{equation}

\begin{figure}[t]
\begin{center}
\includegraphics[width=6.5cm]{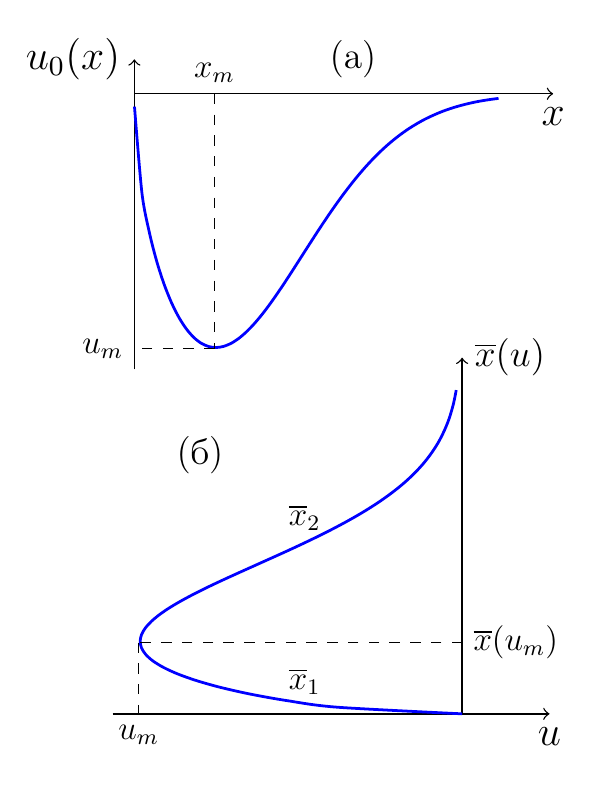}
\caption{ (a) Initial profile of a `negative' pulse. (b) The inverse function
$\ox(u)$ consisting of two branches, $\ox_1(u)$ and $\ox_2(u)$.
 }
\label{fig12}
\end{center}
\end{figure}

Turning now to the question of the evolution of a negative
initial pulse, we see from Fig.~\ref{fig9}(b) that the smooth
dispersionless solution is adjacent to the soliton edge of the DSW, which
therefore propagates over an inhomogeneous background.
On that boundary, the Riemann invariants are $r_1=u$ and $r_2=r_3=0$ and
(Fig.~\ref{fig11}(b)), and hence the soliton edge velocity is
$V_s=2u$ or $dx_R=2udt$, in accordance with (\ref{eq6.17})
\begin{equation}\label{ref8.10}
  \frac{d x_R}{d u}-2u\frac{d t}{d u}=0
\end{equation}
must again be made compatible with the dispersionless solution
\begin{equation}\label{rev8.11}
  x_R-6ut=\ox_i(u),\quad i=1,2,
\end{equation}
if the edge borders the $i$th branch of that solution. Eliminating
$x_R$, we obtain a differential equation for $t=t(u)$:
\begin{equation}\label{rev8.12}
  2u\frac{dt}{du}+3t=-\frac12\ox_i'(u),
\end{equation}
where $\ox_i(u)$  is the corresponding branch of the inverse
function of the initial distribution (Fig.~\ref{fig12}). For the branch
$i=1$, a solution is sought with the initial condition $t(0)=0$,
which defines a parametric form of the law of motion of the
soliton edge:
\begin{equation}\label{rev8.13}
\begin{split}
  &t(u)=\frac1{4(-u)^{3/2}}\int_0^u\sqrt{-u}\,\ox_1'(u)du,\\
  &x_R=-\frac3{2\sqrt{-u}}\int_0^u\sqrt{-u}\,\ox_1'(u)du+\ox_1(u).
  \end{split}
\end{equation}
For example, for a parabolic initial pulse $u_0(x)=-\sqrt{x}$,
$\ox_1(u)=u^2$, $x>0$, we hence find the law of motion
$x_R=-5t^2$.

Solving Eq.~(\ref{rev8.12}) with the initial condition
$$
t(u_m)=\frac1{4(-u_m)^{3/2}}\int_0^{u_m}\sqrt{-u}\,\ox_1'(u)du
$$
for a localized initial pulse with a minimum $u=u_m$ at $x=x_m$,
we obtain the law of motion
\begin{equation}\label{rev9.15}
  \begin{split}
  &t(u)=\frac1{4(-u)^{3/2}}\int_0^{x_2(u)}\sqrt{-u_0(x)}\,dx,\\
  &x_R=x_2(u)-\frac3{2\sqrt{-u}}\int_0^{x_2(u)}\sqrt{-u_0(x)}\,dx.
  \end{split}
\end{equation}

Negative solitons are nonexistent for the KdV equation,
and therefore a negative pulse cannot decay into a sequence of
solitons at asymptotically large times. Instead, it transforms
into a nonlinear wave packet whose soliton edge moves at
$t\to\infty$ in accordance with the law
\begin{equation}\label{ref9.16}
  x_R\approx-\frac{3\mathcal{A}^{2/3}}{2^{1/3}}t^{1/3},\quad
   \mathcal{A}=\int_0^{\infty}\sqrt{-u_0(x)}\,dx,
\end{equation}
matching a virtually rectilinear asymptotic dispersionless
solution $u\approx x/(6t)$ for $x_R<x<0$. Accordingly, the leading
soliton amplitude in the DSW decreases with time as
\begin{equation}\label{rev9.16b}
  a=2|r_1|\approx\frac{\mathcal{A}^{2/3}}{2^{4/3}}t^{-2/3}.
\end{equation}
Near this edge, solutions of Whitham's equations are self-similar
and depend on the variable $z=x/t^{1/3}$. Although this
solution can be obtained in analytic form \cite{dvz-94,ikp-19},
the self-similarity domain is relatively small, and we do not discuss
this theory here. The solution of Whitham's equations in the
entire DSW domain was obtained in \cite{ek-93,ikp-19}. In approaching
the small-amplitude edge, the DSW evolution again becomes
self-similar, with the modulation parameters depending on
 $z=x/t$. We can obtain the asymptotic law of motion of the
small-amplitude edge by noting that, according to Fig.~\ref{fig11}(b),
$r_1\approx r_2\approx u_m$ and $r_3=0$ on that edge, and hence from (\ref{eq3.145}) we
can find the wave number $k=2\pi/L\approx2\sqrt{-u_m}$. Therefore, at
the matching point, the group velocity of the linear wave is
$v_g=-3k^2=-12u_m$ and
\begin{equation}\label{rev9.17}
  x_L\approx-12u_mt.
\end{equation}

\section{Theorem on the number of oscillations
in dispersive shock waves}

An important theorem given in \cite{gp-87} states that, due to the
difference between the velocity of the small-amplitude edge $v_g$
and the phase velocity of the wave $V$, the DSW length
increases on that edge by $(v_g-V)dt$ in a time $dt$, and
therefore the number of wave periods in the domain of
oscillations increases with the rate
\begin{equation}\label{rem1}
  \frac{dN}{dt}=\frac1{2\pi}k(v_g-V),
\end{equation}
where all the variables are evaluated at the DSW wave
number on the small-amplitude edge. The right-hand side of
\cite{rem1} can also be interpreted as the flux of the wave number
$\om=kV$ into the DSW domain with a Doppler shift due to the
motion of the boundary, with the speed $v_g$ taken into account.
Therefore, the total number of oscillations entering the DSW
over all of its evolution time is up to a sign given by
\begin{equation}\label{rem2}
  N=\frac1{2\pi}\int_{-\infty}^{\infty}k(v_g-V)dt=
  \frac1{2\pi}\int_{-\infty}^{\infty}\left(k\frac{d\om}{dk}-\om\right)dt.
\end{equation}
The integrand can be interpreted as a Lagrangian of a
classical particle with the momentum $k$ and the Hamiltonian
$\om$, which is associated with the wave packet co-moving with
the small-amplitude edge of the DSW. The integral is then
equal to the classical action $S$ of such a particle and the
number of oscillations is
\begin{equation}\label{rem3}
  N=\frac{S}{2\pi}.
\end{equation}
It is clear that these formulas are of a general nature and their
validity is not limited to the KdV equation.

For actual calculations, we must know the main characteristics of
the DSW at least on its small-amplitude edge. For
example, in the case of the KdV equation, it is easy to find that
$|k(v_g-V)|=2k^3$; for the evolution of a unit-height step, as
shown in Section~\S\ref{step}, the wave number on the small-amplitude
edge is $k=2$. We hence find the number of oscillations
formed in the DSW over time $t$: $N=(8/\pi)t$. For the time
$t=15$, this formula predicts $N\approx 38$, whereas counting the
oscillations in Fig.~~\ref{fig6}(b), which shows the results of a numerical
solution of the KdV equation, gives approximately $N\approx 39$, in
good agreement with the theory. However, the agreement
with this asymptotic calculation worsens at smaller times. For
example, in the case of breaking of a cubic profile, the values
of Riemann invariants on the small-amplitude edge, according
to formula (\ref{eq6.103}), are $r_3=u$ and $r_1=r_2=-u/4$, where $u$ is
the wave amplitude at the matching point, depending on time
as $u=\sqrt{12t}$ (see (\ref{eq6.103})). Substitution into (\ref{eq3.145}) gives the
wavelength $L=2\pi/\sqrt{5u}$ and the wave number $k=\sqrt{5u}=\sqrt{10}\cdot3^{1/4}t^{1/4}$.
Hence, for the number of oscillations formed by the instant $t$, we obtain
$$
N=\frac{40\sqrt{10}\cdot3^{3/4}}{7\pi}t^{7/4}\approx 13,1\cdot t^{7/4}.
$$
For $t=1$, the number of oscillations $N\approx13$ is somewhat
different from the number of oscillations $N\approx15\div16$ discernible in
Fig.~\ref{fig8}(b), but can still be considered satisfactory for
such a short evolution time.

As regards a positive pulse of finite duration, it eventually
evolves mainly into a sequence of solitons propagating over
the zero background $u=r_1=0$. The group velocity of the
small-amplitude edge, which is a hydrodynamic variable in
Whitham's theory, then has the meaning of the velocity of the
interface between the oscillations that turn into solitons as
$t\to\infty$ and the linear wave packet. The number of solitons
formed from a localized pulse is determined by the initial
profile $u_0(x)$ and can be evaluated as follows.

On the low-amplitude edge, $k=2\sqrt{u}$ and $k(v_g-V)=-2k^3=-16u^{3/2}$.
Integration over $t$ from $0$ to $t_m$ can be replaced using (\ref{rev7.3})
and (\ref{rev7.4}) with integration over $u$ from $0$ to $u_m$,
and similarly integration from $t_m$ to $+\infty$ transforms
with the help of (\ref{rev7.7}) into integration over the same interval
of $u$. As a result, we obtain
\begin{equation}\label{rev21.1}
  N=\frac4{\pi}\int_0^{u_m}\left[t_2-t_1+\frac16(\ox_2'-\ox_1')\right]du,
\end{equation}
where
\begin{equation}\label{rev21.2}
  t_2-t_1=\frac1{12\sqrt{u}}\int_u^{u_m}\frac{\ox_2'-\ox_1'}{\sqrt{u_1}}\,du_1.
\end{equation}
The double integral that occurs in substituting (\ref{rev21.2}) into
(\ref{rev21.1}) can easily be made single-fold by integration by parts,
which leads to the formula
\begin{equation}\label{rev21.3}
  N=\frac1{\pi}\int_0^{u_m}\sqrt{u}(\ox_2'-\ox_1')du
  =\frac1{\pi}\int_{-\infty}^0\sqrt{u_0(x)}\,dx,
\end{equation}
where, as usual, $u_0(x)$ is the initial profile of the wave. This
formula was first derived in \cite{karpman-67} using profound mathematical
properties of the KdV equation associated with its complete
integrability \cite{ggkm-67}. In our presentation, it is a simple corollary
of the Gurevich-Pitaevskii approach to the DSW theory.

\section{ Theory of dispersive shock waves for
the Korteweg-de Vries equation with dissipation}\label{kdv-perturb}

In the Introduction, we discussed the development of the
DSW concept, starting with Sagdeev's idea that dispersion
effects transform the transition layer of a viscous shock wave
into a stationary oscillatory structure, and on to Gurevich and
Pitaevskii's idea of the formation of non-stationary DSWs as a
result of wave breaking, with the evolution of the DSW
modulation parameters governed by Whitham's equations.
It must be clear, however, that the existence of small
dissipation or other perturbing terms in the KdV equation
also leads to the evolution of modulation parameters, which
means that Whitham's modulation equations must then be
modified accordingly. The picture proposed by Sagdeev must
then be described by stationary solutions of modified
Whitham's equations that take small dissipation effects into
account, in addition to dispersion. In this section, we discuss
such a modified Whitham's theory and the simplest corollaries.

We assume that the perturbed KdV equation has the form
\begin{equation}\label{t5-26.1}
    u_t+6uu_x+u_{xxx}=R[u],
\end{equation}
where the perturbing term is small, $R\sim\eps\ll1$, and depends
on both the field $u$ and its spatial derivatives. Generally
speaking, two types of perturbation must be distinguished.
For one type, Whitham's equations acquire right-hand sides
with the old form of Riemann invariants, and perturbations
of the other type lead to a non-diagonal form of the averaged
equations
$$
\frac{\prt r_i}{\prt t}+\sum_jv_{ij}\frac{\prt r_j}{\prt x}=0,
$$
diagonalizing which, as noted in Section 4, is typically
impossible. We discuss only the first case, which includes
physically important problems with small dissipation. We
again derive perturbed Whitham's equations by averaging the
conservation laws. We then take into account that the
conservation law for the number of waves, Eq.~(\ref{rev4-1}),
preserves its form, while conservation laws (\ref{eq3.127})
acquire right-hand sides:
\begin{equation}\label{t5-26.3}
  \begin{split}
  & u_t+(3u^2+u_{xx})_x=R,\\
  & (\tfrac12u^2)_t+(2u^3+uu_{xx}-\tfrac12u_x^2)_x=uR.
  \end{split}
\end{equation}
The averaged equations
\begin{equation}\label{t5-27.5}
  \begin{split}
  & \langle u\rangle_t+\langle 3u^2+u_{xx}\rangle_x=\langle R \rangle,\\
  & \langle \tfrac12u^2\rangle_t+\langle 2u^3+uu_{xx}-\tfrac12u_x^2\rangle_x=\langle uR \rangle
  \end{split}
\end{equation}
can be transformed just as we did previously, and instead of
(\ref{eq3.133}) we now obtain the equations
\begin{equation}\label{t5-27.7}
\begin{split}
  & \frac{D\mathcal{W}_A}{Dt}=\mathcal{W}_A\frac{\prt V}{\prt x},\quad
  \frac{D\mathcal{W}_B}{Dt}=\mathcal{W}_A\frac{\prt B}{\prt x}-\mathcal{W}_A\langle R \rangle,\\
  & \frac{D\mathcal{W}_V}{Dt}=\mathcal{W}_A\frac{\prt A}{\prt x}-\mathcal{W}_A\langle uR \rangle.
  \end{split}
\end{equation}
which differ from the preceding equations only by additional
termsdependingontheperturbation.Movingtothevariables
$\nu_1,\nu_2,$ and $\nu_3$ and introducing Riemann invariants (\ref{eq3.140}) for
unperturbed Whitham's equations as the modulation parameters, we find the
desired Whitham's equations accounting for the perturbation:
\begin{equation}\label{t5-27.8}
\begin{split}
 & \frac{\prt r_i}{\prt t}+v_i\frac{\prt r_i}{\prt x}=\frac{L}{\prt L/\prt r_i}\times\\
 & \times\frac{\langle(\sigma_1-2r_i-u) R\rangle }{4\prod_{j\neq i}(r_i-r_j)},\quad
   i=1,2,3,
  \end{split}
\end{equation}
where $v_i$ are Whitham's velocities (\ref{eq6.12}) of the unperturbed
equations and $\sigma_1=r_1+r_2+r_3$. In the particular case of
Burgers viscosity, the perturbed Whitham equations were
derived in \cite{gp-87,akn-87}, and for nonlocal viscosity, in \cite{gp-91}. In the
general case, they are derived in form (\ref{t5-27.8}) in \cite{mg-95,kamch-04,kamch-16}.

To obtain an insight into the role of small dissipation, we
turn to the Gurevich-Pitaevskii problem of the decay of an
initial discontinuity. We recall from Section~\ref{step} that, at the
initial stage of the evolution, dissipation is inessential and the
DSW expands in a self-similar fashion. But when its length
reaches a size $\sim\eps^{-1}$, all terms in Whitham's equations (\ref{t5-27.8})
become equally significant, and the transition to the stationary regime
of propagation is to be expected, with the full
size of the DSW determined by the balance of terms with
derivatives with respect to coordinates and dissipative
corrections. We therefore seek the solution of Whitham's
equations (\ref{t5-27.8}) with the invariants $r_i$ depending only on the
variable $\xi=x-Vt$. It is a simple observation that this system reduces to
\begin{equation}\label{t5-29.22}
    \frac{dr_i}{d\xi}=-\frac{\langle(\sigma_1-2r_i-u)R\rangle}{8\prod_{j\neq i}(r_i-r_j)},
    \quad i=1,2,3,
\end{equation}
if we take $V$ to be the wave velocity $V=2\sigma_1$. Because the
profile is stationary, this system must have the integral
\begin{equation}\label{t5-29.23}
  \sigma_1=\mathrm{const}.
\end{equation}
It is easy to verify that $\sigma_1$ is indeed an integral, and the
other two symmetric functions $\sigma_2=r_1r_2+r_1r_3+r_2r_3$ and
$\sigma_3=r_1r_2r_3$ satisfy the equations
\begin{equation}\label{t5-29.24}
     \frac{d\sigma_2}{d\xi}=\frac14\langle R\rangle,\quad
     \frac{d\sigma_3}{d\xi}=\frac18[\sigma_1\langle R\rangle-\langle uR\rangle].
\end{equation}
We have thus reduced the problem to solving a system of two
ordinary differential equations for $\sigma_2$ and $\sigma_3$, with $r_i$ being the
functions of $\sigma_2$ and $\sigma_3$ to be found from the cubic equation
\begin{equation}\label{t5-29.25}
  Q(r)=r^3-\sigma_1 r^2+\sigma_2 r-\sigma_3=0.
\end{equation}
The problem can be simplified even more if $\langle R\rangle=0$, in which
case we have another integral $\sigma_2=\mathrm{const}$, and it remains to
solve a single differential equation,
\begin{equation}\label{t5-29.26}
  \frac{d\sigma_3}{d\xi}=-\frac18\langle uR\rangle.
\end{equation}
It is now convenient to return from the symmetric functions
to the variables $r_i$ and, for example, regard $r_1$ and $r_2$ as
functions of $r_3$, where $r_3=r_3(\xi)$. From (\ref{t5-29.22}), we then find
\begin{equation}\label{t5-29.27}
    \frac{dr_1}{dr_3}=\frac{r_3-r_2}{r_2-r_1},\quad \frac{dr_1}{dr_3}=-\frac{r_3-r_1}{r_2-r_1}.
\end{equation}
This system has two integrals: $\sigma_1=\mathrm{const}$ and $\sigma_2=\mathrm{const}$.
Therefore, $r_1$ and $r_2$ as functions of $r_3$ are the roots of the
quadratic equation
\begin{equation}\label{t5-29.29}
    r^2-(\sigma_1-r_3)r+\sigma_2-(\sigma_1-r_3)r_3=0.
\end{equation}
Its roots must be ordered as $r_1\leq r_2$; the constants $\sigma_1$ and $\sigma_2$
are determined by the boundary conditions.
We let $u_L$ denote the limit value of the wave amplitude as
$x\to-\infty$ and assume that the wave propagates in a medium
with $u=0$ at $x\to+\infty$. On the small-amplitude edge, where
$m\to0$, $r_2\to r_1$, we have $u_L=r_3=r_3^L$ and
\begin{equation}\label{t5-31.30}
    \sigma_1=2r_1^L+u_L,\quad \sigma_2=(r_1^L)^2+2r_1^Lu_L.
\end{equation}
On the soliton edge, $r_1^R=0$ and $r_2^R=r_3^R$, and substituting
these into the definition of $\sigma_1$ and $\sigma_2$ yields the relation
\begin{equation}\label{t5-31.32}
    \sigma_1^2-4\sigma_2=0
\end{equation}
between the integrals. Substituting formulas (\ref{t5-31.30}) into (\ref{t5-31.32}),
we obtain an equation for $r_1^L$, whose solution gives $r_1^L=u_L/4$,
and hence
\begin{equation}\label{t5-31.33}
    \sigma_1=\frac32u_L,\quad \sigma_2=\frac9{16}u_L^2.
\end{equation}
on the small-amplitude edge. The integrals take the same
values as on the soliton edge, where $r_1=0$ and $\sigma_3=0$, and
hence Eq.~(\ref{t5-29.25}) has a double root $r_2^R=r_3^R=\frac34u_L$. As a
result, the amplitude $a_s=2r_3^R$ of the leading soliton and its
velocity $V_s=4r_3^R$, coincident with the shock wave velocity, are
\begin{equation}\label{t5-31.34}
    a_s=\frac32u_L,\qquad V=3u_L.
\end{equation}
Thus, the speed of a stationary DSW is determined only by
the magnitude of the discontinuity, in accordance with the
general theory of viscous small-amplitude shock waves \cite{LL6}.
Interestingly, not only the speed but also the amplitude of the
leading soliton is expressed by universal formulas (\ref{t5-31.34}) in
terms of the initial discontinuity and is independent of the
form of the dissipative term. In the particular case of Burgers-type
dissipation, formulas (\ref{t5-31.34}) were derived in \cite{johnson-70} directly
from the perturbation theory without using Whitham's theory.

\begin{figure}[t]
\begin{center}
\includegraphics[width=6.5cm]{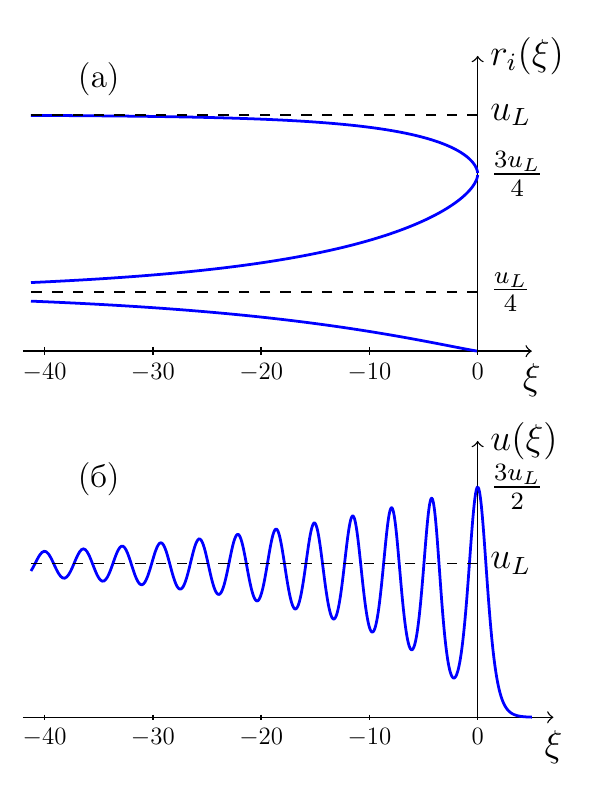}
\caption{  (a) Plots of Riemann invariants for the stationary solution of
Whitham's equations corresponding to a DSW with Burgers viscosity
(\ref{t6-4.1}) at $\eps=0.1$, $u_L=1.0$. (b) Profile of a stationary shock wave $u(\xi)$
with Burgers viscosity at the same parameter values.
 }
\label{fig13}
\end{center}
\end{figure}

To find a global solution along all of the DSW, we note
that, after substituting integrals (\ref{t5-31.33}) into (\ref{t5-29.29}) and solving
this quadratic equation, we obtain $r_1$ and $r_2$ as functions of
$r_3$. Their substitution into expression (\ref{eq3.145}) for $m$ gives
an equation whose solution for $r_3$ allows expressing this
Riemann invariant in terms of $m$, and then $r_1$ and $r_2$ can also
be represented as functions of $m$. As a result of these
elementary calculations, we obtain
\begin{equation}\label{rev9.18}
  \begin{split}
  & r_1=\frac{u_L}2\left(1-\frac{1+m}{2\sqrt{1-m+m^2}}\right),\\
  & r_2=\frac{u_L}2\left(1-\frac{1-2m}{2\sqrt{1-m+m^2}}\right),\\
  & r_3=\frac{u_L}2\left(1+\frac{1-m/2}{\sqrt{1-m+m^2}}\right).
  \end{split}
\end{equation}
The problem is solved when we obtain the dependence of the
parameter $m$ on the coordinate $\xi$. Evaluating the derivative
$dm/dr_3$ with the help of formulas (\ref{t5-29.27}) and multiplying the
result by $dr_3/d\xi$ in (\ref{t5-29.22}), we obtain the desirable equation,
\begin{equation}\label{t6-8.20}
  \frac{dm}{d\xi}=-\frac{1-m+m^2}{4(r_2-r_1)(r_3-r_1)(r_3-r_2)}\langle uR\rangle,
\end{equation}
where the right-hand side can be expressed in terms of $m$ for a
perturbation $R$ of a given form.

We specify this theory by choosing the perturbation as
Burgers friction \cite{gp-87,akn-87}:
\begin{equation}\label{t6-4.1}
  R=\eps u_{xx}.
\end{equation}
To actually take the averages, it is convenient to pass to
the variable $\upsilon=(\sigma_1-u)/2$ that satisfies the equation
$\upsilon_x^2=4Q(\upsilon)$, whence $u_{xx}=-2\upsilon_{xx}=-4dQ/d\upsilon$.
As a result, we find
$$
-\langle uR\rangle=\frac{8\eps}L\oint\sqrt{Q(\upsilon)}\,d\upsilon.
$$
This elliptic integral is readily reduced to tabulated ones, and
we hence obtain the equation
\begin{equation}\label{t6-8.22}
\begin{split}
  &\frac{dm}{d\xi}  =\Phi(m)\equiv \frac{8\eps}{15}\frac{1-m+m^2}{m(1-m)}\times\\
  &\times   \left[(1-m+m^2)\frac{E(m)}{K(m)}-(1-m)\left(1-\frac{m}2\right)\right].
  \end{split}
\end{equation}
The problem solution has thus been reduced to the quadrature
\begin{equation}\label{t6-8.23}
  \xi=-\int_m^1\frac{dm}{\Phi(m)}.
\end{equation}
This formula, together with (\ref{rev9.18}), parametrically defines the
dependence of the modulation parameters, i.e., the Riemann
invariants $r_i$ of the system of Whitham's equations, on the
coordinate $\xi$, referenced to the DSW front. An example of
such a dependence is shown in Fig.~\ref{fig13}(a), and the corresponding
DSW profile is shown in Fig.~\ref{fig13}(b).

\section{ Gross-Pitaevskii equation}

Besides the KdV equation, which has a universal character,
another very important equation, also occurring in very
diverse circumstances, is the Gross-Pitaevskii equation,
which in particular describes the dynamics of a weakly
non-ideal Bose gas at zero temperature \cite{gross,pit-1} in the mean
field approximation, when the coherent state of the macroscopic
Bose gas is described by a classical wave function,
similar to the Maxwell field in classical electrodynamics. This
theory came to the forefront after the experimental realization of
Bose-Einstein condensation of atoms, and the main
ideas underlying the theory are available in reviews \cite{dps-99,pit-2}.
Here, we restrict ourselves to writing the Gross-Pitaevskii
equation for the wave function $\psi(\mathbf{r})$ in the standard notation:
\begin{equation}\label{rev9.1}
   i\hbar\frac{\prt \psi}{\prt t}=-\frac{\hbar^2}{2m}\Delta\psi+
   U(\mathbf{r})\psi+g|\psi|^2\psi,
\end{equation}
where $m$  is the atom mass, $\Delta$ is the Laplace operator, $U(\mathbf{r})$ is
the potential of an external field acting on the atoms, and the
parameter $g$, expressed in terms of the atom-atom scattering
length $a$,
$$
g=\frac{4\pi\hbar^2 a}m
$$
characterizes the strength of interatomic interaction; it
is repulsive for $g>0$ and attracting for $g<0$. We are
interested in the first case, where the homogeneous state of
the condensate is stable and waves can propagate over it.

We note that the mathematically equivalent equation
occurred in describing self-focusing of light beams in non-linear
media \cite{talanov-65,kelley-65}, where the role of time is played by the
coordinate along the beam and diffraction replaces dispersion,
but the papers just cited discussed only the focusing
nonlinearity, for which the state with a homogeneous
distribution of light intensity is unstable. Another interpretation
of Eq.~(\ref{rev9.1}) occurs when describing the evolution of the
envelope of a wave packet propagating in a medium with low
dispersion and weak nonlinearity \cite{bn-67}. In that case, the first
term on the right-hand side corresponds to second-order
dispersive effects, which, besides the packet motion with the
group velocity, takes its slow spreading into account, and the
last term corresponds to the dependence of the medium
response on the wave intensity. This situation occurs rather
frequently in physics, from the description of deep-water
waves to the theory of propagation of light pulses in non-linear
optical fibers. In this context, the resultant equation is
often called the nonlinear Schr\"{o}dinger (NLS) equation, but
we here use the physical interpretation due to Gross-Pitaevskii,
which allows addressing more transparent representations and notions
of gas dynamics. In particular, the condensate density is $\rho=|\psi|^2$,
and its flow speed is expressed in terms of the gradient of the wave
function phase \cite{dps-99,pit-2}. If we represent the wave function as
\begin{equation}\label{rev9.2}
  \psi=\sqrt{\rho}\,e^{i\phi},\qquad \mathbf{u}=\frac{\hbar}m\nabla\phi,
\end{equation}
then, substituting this into (\ref{rev9.1}), after simple transformations,
leads to the system of equations (with $U(\mathbf{r})=0$)
\begin{equation}\label{rev9.3}
  \begin{split}
  & \rho_t+\nabla(\rho\mathbf{u})=0,\\
  & \nabla\mathbf{u}+(\mathbf{u}\nabla)\mathbf{u}+\frac{g}m\nabla\rho+
  \frac{\hbar^2}{2m}\nabla\left[\frac{(\nabla\rho)^2}{4\rho^2}-\frac{\Delta\rho}{2\rho}\right]=0.
  \end{split}
\end{equation}
The first equation is the standard continuity equation
corresponding to the conservation of the number of particles
in the condensate, and the second equation has the form of a
modified Euler equation for the flow of gas with the equation
of state $p=g\rho^2/(2m)$ and with the last term containing
higher-order spatial derivatives. It is clear that this term
corresponds to dispersive properties of the gas caused by
quantum dispersion of atoms. If we consider extremely long
waves and ignore this term, we arrive at an expression for the
speed of sound in the condensate,
\begin{equation}\label{rev10.4}
  c_s=\sqrt{\frac{dp}{d\rho}}=\sqrt{\frac{g\rho}m}.
\end{equation}
which depends on the local density $\rho$. If we turn to linear
waves in a homogeneous condensate with a constant density
$\rho$, then a standard calculation gives Bogoliubov's dispersion
law \cite{bogol-47}
\begin{equation}\label{rev10.5}
  \om(k)=k\sqrt{c_s^2+\left(\frac{\hbar k^2}{2m}\right)^2}.
\end{equation}
where, as the wave number $k$ increases, the sound dispersion
law $\om=c_sk$ passes into the standard dispersion law of
quantum particles $\eps=\hbar\om=(\hbar k)^2/(2m)$ when the de Broglie
wavelength becomes less than the coherence length
\begin{equation}\label{rev10.6}
  \xi_C=\frac{\hbar}{\sqrt{2}mc_s}=\frac{\hbar}{\sqrt{2mg\rho}}.
\end{equation}
We introduce parameters characterizing the state of the
condensate: the length $\xi_c$ and the speed $c_s$ at the characteristic
density $\rho_0$, which allows us to define convenient
dimensionless variables $\mathbf{r}\to\mathbf{r}/(\sqrt{2}\xi_C)$,
$t\to c_st/(\sqrt{2}\xi_C)$, and $\psi\to\psi/\sqrt{\rho_0}$.
In addition, we restrict ourselves in what follows
to only one-dimensional motions of the condensate, and
therefore, in the new variables, the Gross-Pitaevskii equation takes the form
\begin{equation}\label{rev10.7}
    i\psi_t+\frac12\psi_{xx} -|\psi|^2\psi=0,
\end{equation}
and its `hydrodynamic' representation (\ref{rev9.3}) becomes
\begin{equation}\label{rev10.8}
  \begin{split}
  & \rho_t+(\rho u)_x=0,\\
  & u_t+uu_x+\rho_x+\left(\frac{\rho_x^2}{8\rho^2}-\frac{\rho_{xx}}{4\rho}\right)_x=0.
  \end{split}
\end{equation}
Accordingly, for linear waves, thedispersion law in Eq.~(\ref{rev10.5}) becomes
\begin{equation}\label{rev10.9}
  \om(k)=k\sqrt{1+\frac{k^2}4}.
\end{equation}

\begin{figure}[t]
\begin{center}
\includegraphics[width=7cm]{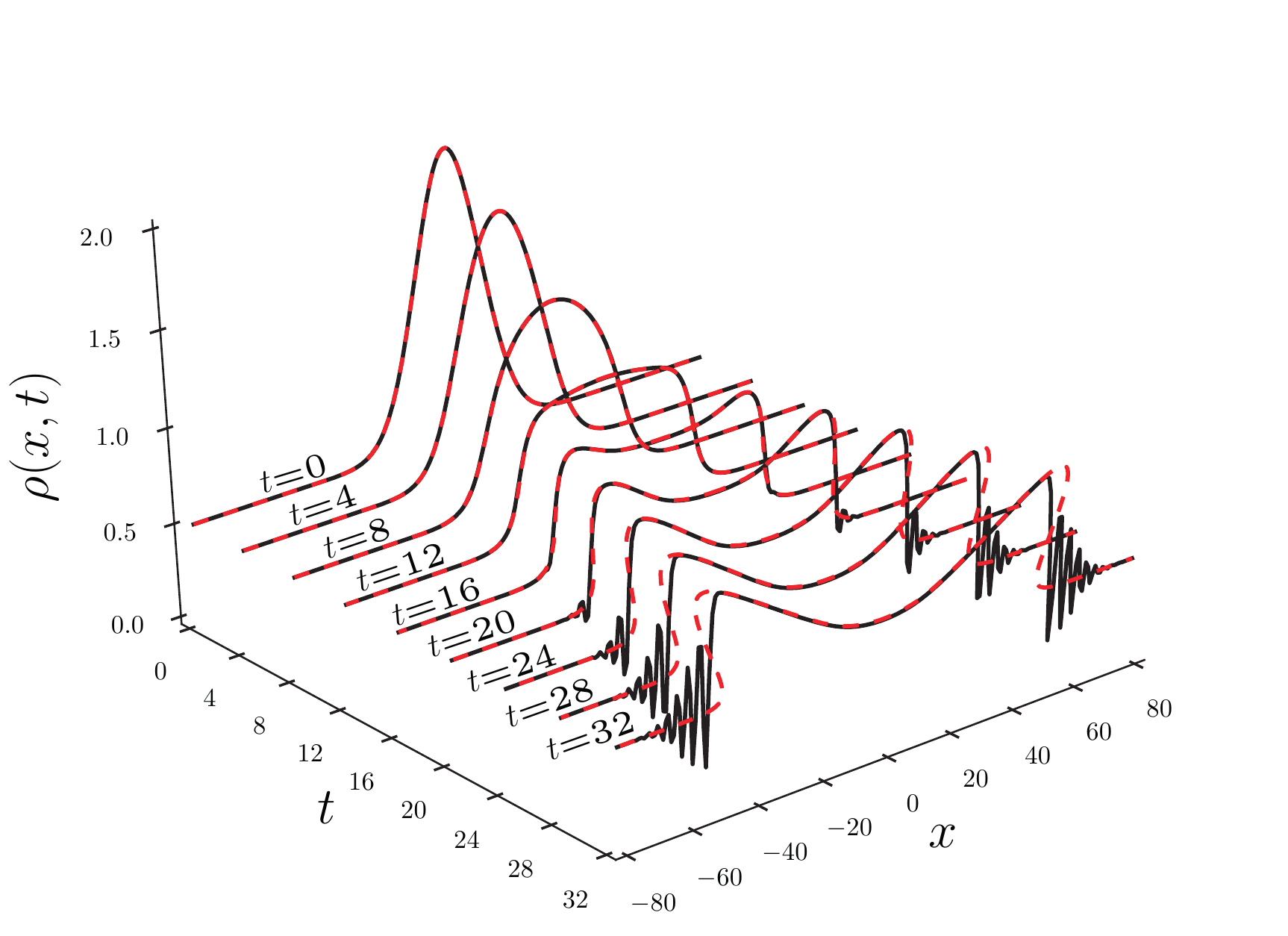}
\caption{ Evolution of an initially localized density pulse in a Bose-Einstein
condensate. After the wave breaking at $t_b\approx20$, DSWs emerge on
the pulse edges.
 }
\label{fig14}
\end{center}
\end{figure}

It is clear that waves can propagate in both directions of
the $x$ axis, and therefore any initial perturbation evolves with
time into two wave pulses propagating in opposite directions.
For example, if the initial pulse has a shape describing a hump
in the condensate density above a homogeneous background,
then the numerical solution of Gross-Pitaevskii equation
(\ref{rev10.7}) or the equivalent system (\ref{rev10.7}) exhibits the wave
evolution shown in Fig.~\ref{fig14}. As we can see, the pulse splits
into two with time, and each of them experiences breaking
with the formation of a DSW. We therefore have the task to
describe the evolution of shock waves satisfying the Gross-Pitaevskii equation.
In accordance with the Gurevich-Pitaevskii approach, each DSW borders a smooth
solution of the dispersionless equations, and we therefore first discuss
this last approximation.

In the dispersionless limit, the last term in Euler equation
(\ref{rev10.8}) can be dropped, and the system takes the simple
hydrodynamic form
\begin{equation}\label{rev10.10}
   \rho_t+(\rho u)_x=0,\quad
   u_t+uu_x+\rho_x=0.
\end{equation}
As is standard in the theory of linear waves, local changes in
the density $\delta\rho$ and velocity $\delta u$ of the flow are related as
$\delta\rho/\rho\approx\pm\delta u/c$, where the choice of sign corresponds to the
wave propagation direction. Therefore, for example, in a
wave propagating to the right, the differential relation
$du=cd\rho/\rho=d\rho/\sqrt{\rho}$ is satisfied, integrating which shows
that, in such a simple wave, the flow velocity $u$ and the density
$\rho$ are related as $u/2-\sqrt{\rho}=\mathrm{const}$, and a similar relation with
the other sign in front of the square root holds for a wave
propagating to the left. This argument shows that the so-called Riemann invariants,
related to the density and velocity of the flow as
\begin{equation}\label{rev10.11}
    r_+=\frac{u}2+\sqrt{\rho},\quad  r_-=\frac{u}2-\sqrt{\rho}.
\end{equation}
are natural variables in the physics of waves. Equations (\ref{rev10.10}),
when written in these variables, take a simple diagonal form,
\begin{equation}\label{rev10.12}
\begin{split}
    \frac{\prt r_+}{\prt t}+v_+(r_+,r_-)\frac{\prt r_+}{\prt x}=0,\\
    \frac{\prt r_-}{\prt t}+v_-(r_+,r_-)\frac{\prt r_-}{\prt x}=0,
    \end{split}
\end{equation}
where the velocities $v_{\pm}=u\pm c$ have a clear physical meaning
of the signal propagation speed, equal to the sum of and the
difference between the flow velocity and the speed of sound
propagating downstream or upstream. In our case of the
Bose-Einstein condensate, they are especially simply
expressed in terms of the Riemann invariants:
\begin{equation}\label{rev10.13}
    v_+=\frac32r_++\frac12r_-,\quad  v_-=\frac12r_++\frac32r_-.
\end{equation}
Simple waves are characterized by the constancy of one of the
Riemann invariants. For example, for a wave propagating to
the right, the invariant $r_-=r_-^{(0)}=\mathrm{const}$ is constant, the
second equation in (\ref{rev10.12}) is then satisfied automatically, and
the first equation becomes the Hopf equation, which we
already discussed in the case of ion-sound waves in plasma.
Obviously, because of the relation between $\rho$ and $u$, this Hopf
equation can also be written for only one of these variables,
which would then give a dispersionless approximation for
unidirectional propagation of waves in the condensate.
Additionally taking dispersion (\ref{rev10.9}) into account in the
leading approximation, $\om\approx k+k^3/8$, leads to the KdV
equation for nonlinear waves in the limit of a large
wavelength and a small amplitude. It is easy to see that the
nonlinear and dispersion terms have opposite signs in this
equation, and therefore soliton solutions correspond to
troughs in the density distribution, and the KdV equation
describes `shallow' solitons on a homogeneous background.
Naturally, the DSW theory for KdV is entirely applicable to
the description of shock waves in a condensate under the
condition of their small amplitude and unidirectional propagation.
But for deep solitons and large-amplitude DSWs,
development of the Gurevich-Pitaevskii theory is required.

With the dispersionless approximation equations conveniently written
in form (\ref{rev10.12}), we can now turn to the theory of
periodic solutions of the Gross-Pitaevskii equation, whose
modulations describe the DSWs. If we seek a solution to
system (\ref{rev10.8}) in the form of a traveling wave $\rho=\rho(\xi)$, $u=u(\xi)$,
$\xi=x-Vt$, then the first equation is readily
integrated, and the second, after eliminating the variable $u$
and some transformations, reduces to the equation
\begin{equation}\label{t3-62.2}
  \rho_\xi=2\sqrt{\mathcal{R}(\rho)},\quad \mathcal{R}(\rho)=\prod_{i=1}^3(\rho-\nu_i).
\end{equation}
Evidently, the density $\rho$ oscillates in the range $\nu_1\leq\rho\leq\nu_2$
where the polynomial $\mathcal{R}(\rho)$ is positive, and a standard
calculation similar to the derivation of the cnoidal wave
solution of the KdV equation leads to a periodic solution of
the Gross-Pitaevskii equation in the form
\begin{equation}\label{t3-63.3}
  \rho=\nu_1+(\nu_2-\nu_1)\,\sn^2\left(\sqrt{\nu_3-\nu_1}\,(x-Vt),m\right),
\end{equation}
where $m=(\nu_2-\nu_1)/(\nu_3-\nu_1)$ and the velocity $V$, unlike the
one in the KdV theory, is now an independent parameter. The
condensate flow velocity is
\begin{equation}\label{t3-63.4}
  u=V\pm\frac{\sqrt{\nu_1\nu_2\nu_3}}{\rho}.
\end{equation}
In the soliton limit, as $\nu_3\to\nu_2$ and $m\to0$, we obtain the
solution \cite{tsuzuki-71}
\begin{equation}\label{t3-63.7}
\begin{split}
  &\rho=\rho_0\left(1-\frac{1-V^2/\rho_0}{\ch^2(\sqrt{\rho_0-V^2}\,(x-Vt))}\right),\\
  &u=V\left(1-\frac{\rho_0}{\rho}\right)
  \end{split}
\end{equation}
for a soliton moving over a condensate that has the density
$\nu_2=\rho_0$ and is at rest at infinity. As the depth of the soliton
tends to zero, its velocity tends to the speed of sound
$c_0=\sqrt{\rho_0}$, never exceeding it. If the soliton velocity is zero,
the density $\rho$ at its center also vanishes; such a soliton is called
`black.' In view of the relation $u=\phi_x$, the wave function phase jumps by
\begin{equation}\label{t3-64.2}
  \Delta\phi\equiv \phi(\infty)-\phi(-\infty)=
  -2\arccos\frac{V}{\sqrt{\rho_0}},\quad V>0,
\end{equation}
when crossing the domain occupied by the soliton. For the
black soliton, with $V\to+0$, this jump is $\Delta\phi=-\pi$. Because
the phase is defined up to $2\pi$, this state of the condensate is not
different from the state having the velocity $V\to-0$ and the
jump $\Delta\phi=\pi$. Due to this property, a dark soliton moving in
an inhomogeneous condensate confined by a trap can change
the direction of motion at the points where the density in its
center vanishes. Formulas (\ref{t3-63.7}) can be combined into the
expression
\begin{equation}\label{t3-64.4}
  \psi=\left\{\sqrt{\rho_0-V^2}\th[\sqrt{\rho_0-V^2}(x-Vt)]+iV\right\}e^{-i\rho_0t}
\end{equation}
for the soliton solution of Gross-Pitaevskii equation (\ref{rev10.7}). In
the low-amplitude limit $\nu_2-\nu_1\ll\nu_3-\nu_1$, $m\ll1$ wave (\ref{t3-63.3})
degenerates into a trigonometric one,
\begin{equation}\label{t3-64.5}
  \rho=\nu_1+\frac{a}2\cos[2\sqrt{\nu_3-\nu_1}(x-Vt)],
\end{equation}
with the wave number $k=2\sqrt{\nu_3-\nu_1}$ and the phase velocity
$V=\pm\sqrt{\nu_3}$ related with each other as $V^2=\nu_3=\nu_1+k^2/4=\rho_0+k^2/4$,
in accordance with dispersion law (\ref{rev10.9}).

The obtained periodic solution depends on four parameters $V,\nu_1,\nu_2,\nu_3$,
and describing the DSWs requires deriving the corresponding modulation equations.
Evidently, the conservation law for the number of waves, Eq.~(\ref{125.12}),
extends to nonlinear waves (\ref{t3-63.3}) with the corresponding
expression for the wave number in terms of the modulation
parameters, and it is easy to find three more conservation
laws for Gross-Pitaevskii equation (\ref{rev10.7}), whose averages in
principle give a full set of modulation equations. But their
transformation into the diagonal form by Whitham's direct
method turns out to be technically complicated, and these
equations were first derived in diagonal form in \cite{FL-86,pavlov-87} only
after the complete integrability of the Gross-Pitaevskii
equation was discovered in \cite{zs-73} and relations between the
complete integrability and diagonalization of Whitham's
equations were revealed in \cite{ffm-80}. We do not go into the details
of this theory and give Whitham's equations for the Gross-Pitaevskii
equation in the final form, especially because they
are quite similar to the already familiar Whitham's equations
for modulation of periodic KdV waves and can be investigated by similar methods.

In the KdV case, the transition from the parameters $\nu_i$ to
the Riemann invariants $r_i$ of Whitham's system is effected by
very simple formulas (\ref{eq3.140}), but in the case of the Gross-Pitaevskii
equation, the parameters $V$ and $\nu_i$ are related to the
Riemann invariants $r_i$, $r_1\leq r_2\leq r_3\leq r_4,$, by the more
complicated expressions
\begin{equation}\label{rev11.21}
  \begin{split}
  & \nu_1=\frac14(r_1-r_2-r_3+r_4)^2,\\
  & \nu_2=\frac14(r_1-r_2+r_3-r_4)^2,\\
  & \nu_3=\frac14(r_1+r_2-r_3-r_4)^2,\\
  & V=\frac12(r_1+r_2+r_3+r_4).
  \end{split}
\end{equation}
It is worth noting that the polynomial $\mathcal{R}(\nu)=\prod_{i=1}^3(\nu-\nu_i)$ is
Ferrari's resolvent for the polynomial $Q(r)=\prod_{i=1}^4(r-r_i)$,
allowing the roots of the equation $Q(r)=0$ to be expressed in
radicals in terms of its coefficients. The polynomial $Q(r)$ and
symmetric functions of its roots play an important role in the
theory of periodic solutions and their modulation for a wide
class of integrable equations. The periodic solution of the
Gross-Pitaevskii equation can be expressed in terms of the
Riemann invariant as
\begin{equation}\label{rev11.22}
\begin{split}
&\rho =\frac14(r_4-r_3-r_2+r_1)^2+ (r_4-r_3)\times\\
&\times(r_2-r_1)\,{\rm sn}^2\left(\sqrt{(r_4-r_2)(r_3-r_1)}\,
\xi,m\right)  ,
\end{split}
\end{equation}
where
\begin{equation}\label{rev12.23}
  m=\frac{(r_2-r_1)(r_4-r_3)}{(r_4-r_2)(r_3-r_1)}.
\end{equation}
Whitham's modulation equations have the diagonal form
\begin{equation}\label{rev12.24}
\frac{\partial r_i}{\partial
t}+v_i(r)\frac{\partial r_i}{\partial x}=0, \quad i=1,2,3,4,
\end{equation}
where the characteristic velocities are expressed through the
wavelength
\begin{equation}\label{rev12.25}
L= \frac{2{K}(m)}{\sqrt{(r_4-r_2)(r_3-r_1)}}
\end{equation}
by the formula
\begin{equation}\label{rev12.26}
\begin{split}
v_i(r)&=\left(1-\frac{{L}}{\partial_i{L}}
\partial_i\right)V , \quad i=1,2,3,4 \, ,\\
\end{split}
\end{equation}
which is similar to (\ref{eq3.144}). Substituting (\ref{rev12.25})
into (\ref{rev12.26}), we obtain
\begin{equation}\label{rev12.27}
\begin{split}
v_1&=\frac12 \sum r_i
-\frac{(r_4-r_1)(r_2-r_1)K}{(r_4-r_1)K-(r_4-r_2)E},\\
v_2&=\frac12 \sum r_i
+\frac{(r_3-r_2)(r_2-r_1)K}{(r_3-r_2)K-(r_3-r_1)E},\\
v_3&=\frac12 \sum r_i
-\frac{(r_4-r_3)(r_3-r_2)K}{(r_3-r_2)K-(r_4-r_2)E},\\
v_4&=\frac12 \sum r_i
+\frac{(r_4-r_3)(r_4-r_1)K}{(r_4-r_1)K-(r_3-r_1)E}.
\end{split}
\end{equation}
On the soliton edge of a DSW with $r_2=r_3$ $(m=1)$, these
expressions become
\begin{equation} \label{rev12.28}
\begin{split}
         &v_1= \frac{3}{2} r_{1} + \frac12{r_{4}}\, , \quad
         v_4= \frac{3}{2} r_{4} + \frac12{r_{1}}, \\
         &v_2=v_3=\frac{1}{2}(r_1+ 2r_2 + r_4),
\end{split}
\end{equation}
and on the small-amplitude edge with $r_3=r_4$ and $m=0$, we have
\begin{equation}\label{rev12.29}
\begin{split}
& v_1= \frac{3}{2} r_{1} + \frac12{r_{2}}\, , \quad
v_2= \frac{3}{2} r_{2} + \frac12{r_{1}} , \\
   & v_3=v_4=2r_4 +
\frac{(r_2-r_1)^2}{2(r_1+r_2-2r_4)}  ,
\end{split}
\end{equation}
Similar formulas can be derived in the limit $r_1=r_2$ $(m=0)$.

On the DSW edges, as we can see, one pair of velocities
merges into a single expression and the other pair takes the
form of expressions (\ref{rev10.13}) for dispersionless velocities if
Whitham's Riemann invariants are properly identified with
the dispersionless Riemann invariants $r_{\pm}$ (see (\ref{rev10.11})). This
allows incorporating the solution of Whitham's equations
describing the DSW into a smooth solution of dispersionless
equations (\ref{rev10.12}). These dispersionless equations, as well as
Whitham's equations, can be solved by the hodograph
method. For Whitham's system, the solution has the form
\begin{equation}\label{rev12.30}
    x-v_i(r_j)t=w_i(r_j),\quad i,j=1,2,3,4,
\end{equation}
where
\begin{equation}\label{rev12.31}
w_i(r_j)=\left(1-\frac{L}{\partial_i{L}}
\partial_i\right)W(r_j) , \quad i,j=1,2,3,4 ,
\end{equation}
and the function $W(r_1,r_2,r_3,r_4)$ is a solution to the system of
Euler-Poisson equations (\ref{eq6.33}). In particular, as in the case of
the KdV equation, an important class of self-similar solutions
is represented by the generating function
\begin{equation}\label{rev12.32}
    W=\frac{r^2}{\sqrt{Q(r)}}
    =\sum_{k=0}^{\infty}\frac{W^{(k)}(r_j)}{r^k},
\end{equation}
which depends on an arbitrary parameter $r$ and satisfies
Euler-Poisson equation (\ref{eq6.33}). The coefficients of its expansion
in inverse powers of $r$ give particular solutions of the
Euler-Poisson equation, for which the functions $w_i(r_j)$ take
the particular form
\begin{equation}\label{rev12.33}
    w_i^{(k)}(r_j)=W^{(k)}(r_j)+2(v_i-V)\prt_iW^{(k)}(r_j).
\end{equation}
In view of the linearity of the Euler-Poisson equations, any
linear combination $w_i=\sum_kA_kw_i^{(k)}$ of functions (\ref{rev12.33}) also
gives a solution (\ref{rev12.30}). Here, the $W^{(k)}$ are expressed in terms
of $\sigma_i$, symmetric functions of the roots of the polynomial
$Q(r)=\prod_{i=1}^4(r-r_i)$ (the coefficients of the polynomial). In particular,
\begin{equation}\label{rev17.5}
  \begin{split}
  & W^{(1)}=\frac12\sigma_1,\quad W^{(2)}=\frac3{16}\sigma_1^2-\frac14\sigma_2,\\
  & W^{(3)}=\frac5{32}\sigma_1^3-\frac38\sigma_1\sigma_2+\frac14\sigma_3.
  \end{split}
\end{equation}
This elementary treatment suffices for solving the Gurevich-Pitaevskii problem
in several characteristic cases.

\section{Evolution of the initial discontinuity
in the Gross-Pitaevskii theory}

\begin{figure}[t]
\begin{center}
\includegraphics[width=7cm]{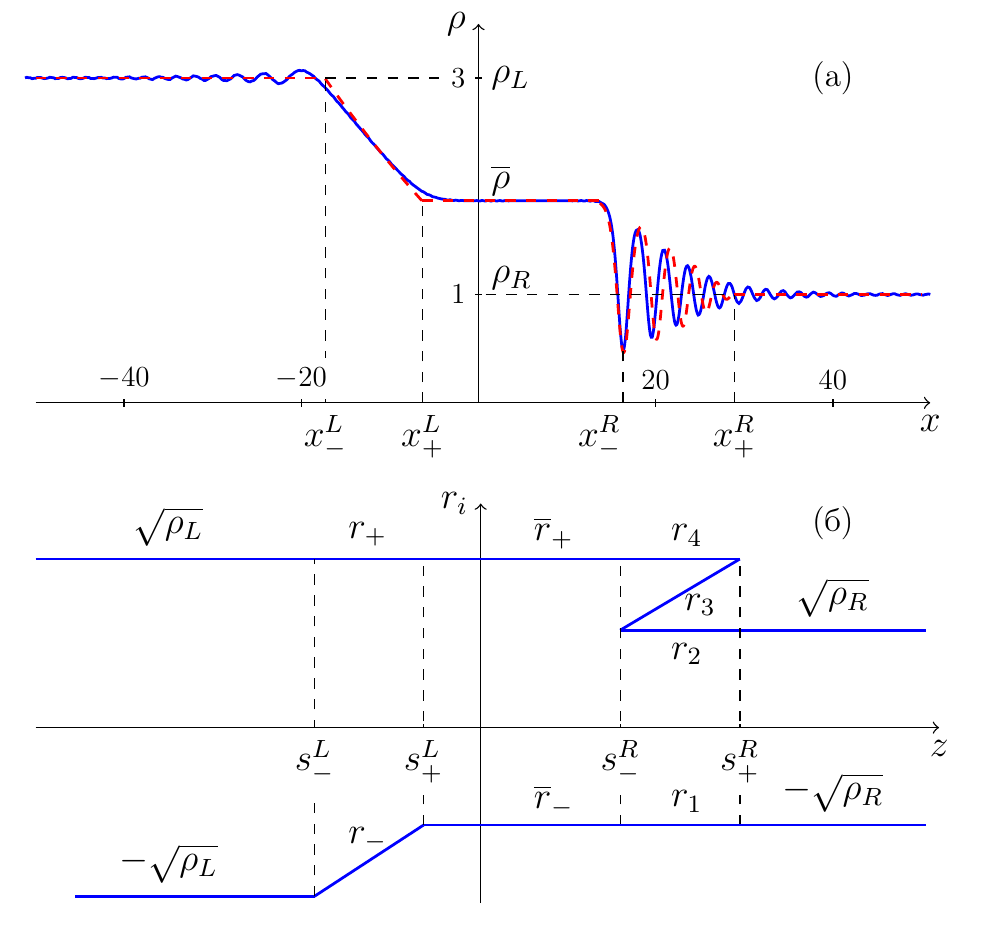}
\caption{(a) Wave structure formed in the evolution of an initial discontinuity
in the density distribution of a Bose condensate with $\rho_L=3$ and $\rho_R=1$,
the evolution time being $t=10$. Solid line shows the numerical
solution and the dashed line shows the analytic solution to the Gurevich-Pitaevskii
problem. Theoretical values $\overline{\rho}=1.87$, $x_-^L/t=-1.73$, $x_+^L/t=-0.63$, $x_-^R/t=1.37$,
$x_+^R/t=2.89$ agree well with the numerical
solution. (b) Diagram of Riemann invariants corresponding to the wave
structure formed in the evolution of the initial density discontinuity.
 }
\label{fig15}
\end{center}
\end{figure}

Just as in case of the KdV theory discussed in Section~7, we
begin with the simplest problem of the evolution of the initial
discontinuity, with the condensate state having different
densities and different flow velocities, $\rho_L,u_L$ and $\rho_R,u_R$ , on
the respective half-lines $x<0$ and $x>0$. The values of
Riemann invariants are to be matched in the emerging wave
structure, and we therefore specify the condensate state by
their values on both sides of the discontinuity:
\begin{equation}\label{rev13.1}
  r_{\pm}(x,t)=
  \left\{
  \begin{array}{l}
   r_{\pm}^L=u_L/2\pm\sqrt{\rho_L},\quad x<0,\\
   r_{\pm}^R=u_R/2\pm\sqrt{\rho_R},\quad x>0.
  \end{array}
  \right.
\end{equation}
As an example, we consider the evolution of an initial
discontinuity in the density distribution with the initial state
$u_L=u_R=0$, and assume for definiteness that $\rho_L>\rho_R$,
whence $r_+^L=-r_-^L>r_+^R=-r_-^R$.

The numerical solution of the Gross-Pitaevskii equation
for this initial condition gives the wave structure shown with a
solid line in Fig.~\ref{fig15}(a). As we see, this structure consists of two
waves joined by the domain of homogeneous flow (`plateau').
Because parameters with the dimension of length are absent
in the initial distribution, solutions of both dispersionless
equations (\ref{rev10.12}) and Whitham's equations (\ref{rev12.24}) must be
self-similar and depend only on the variable $z=x/t$. Therefore, as
can be easily verified, only one of the Riemann invariants can
change along these waves. On the left, there is a rarefaction
wave, along which the Riemann invariant $r_+$ is constant, i.e.,
$\sqrt{\rho_L}=\overline{u}/2+\sqrt{\overline{\rho}}$, where the bar
over a variable denotes its
value on the plateau. In the solution of Whitham's equations,
too, only one of the Riemann invariants $r_i$ varies, and we
conclude that they can be matched continuously only if the
Riemann invariant $r_3$ varies. The resultant wave structure can
be represented by the diagram of the Riemann invariant
shown in Fig.~\ref{fig15}(b), which schematically shows the
dependences of all the invariants on the self-similarity variable $z$.
Because the invariant $r_1$ is constant along the DSW and
matches the invariants $\overline{r}_-$ and $r_-^R$ on the DSW edges, we
obtain one more equation $\overline{u}/2-\sqrt{\overline{\rho}}=-\sqrt{\rho_R}$
for the parameters of the flow along the plateau. The obtained equations
determine the values of flow parameters on the plateau
\begin{equation}\label{rev13.2}
   \overline{\rho}=\frac14(\sqrt{\rho_L}+\sqrt{\rho_R})^2,\quad
   \overline{u}=\sqrt{\rho_L}-\sqrt{\rho_R}.
\end{equation}
which are in excellent agreement with the numerical solution.

\begin{figure}[t]
\begin{center}
\includegraphics[width=7cm]{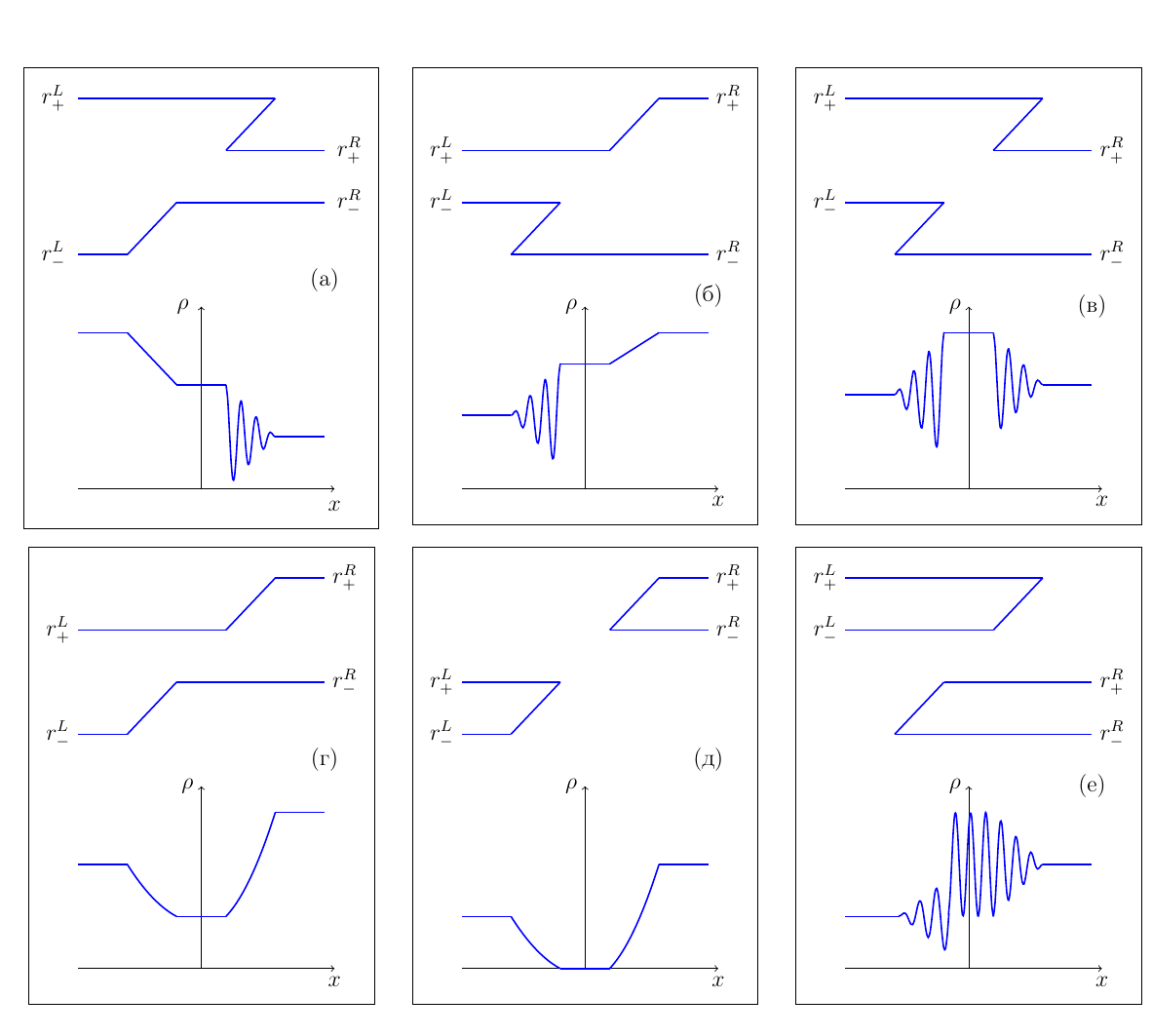}
\caption{Wave structures formed in the evolution of the initial
discontinuity in the theory of the Gross-Pitaevskii equation and the
corresponding diagrams of Riemann invariants.
 }
\label{fig16}
\end{center}
\end{figure}

The above example shows that the shape of the wave
structure resulting from the evolution of the initial discontinuity
can be determined by joining pairs of Riemann
invariant values corresponding to wave edges with lines
having a positive slope and corresponding to self-similar
solutions of the form $v_i=z$ (for the rarefaction wave, the
positivity of the slope is obvious from expression (\ref{rev10.13}) for
characteristic dispersionless velocities, and for the DSW it
follows from a more detailed investigation of expressions
(\ref{rev12.27})). If there are only two Riemann invariants in the
resultant domain, this domain corresponds to the rarefaction
wave. If four invariants are defined in that domain, then it
corresponds to the DSW.

It can be easily verified \cite{gk-87,eggk-95} that only six possible
diagrams exist, which we present in Fig.~\ref{fig16} together with the
corresponding wave structure types. In the cases shown in
Fig.~\ref{fig16}(a,b), one rarefaction wave and one DSW emerge, and
these differ only in the wave propagation directions. In the
case shown in Fig.~\ref{fig16}(c) (`collision of condensates'), two DSWs
emerge on different sides of the plateau. In the cases in
Fig.~\ref{fig16}(d,e), the condensates on different sides of the
discontinuity have opposite velocities and, as the condensates
recede, a lower-density plateau appears between them; in
Fig.~\ref{fig16}(e), the initial velocities are so high that this density
decreases to zero. Finally, in the case shown in Fig.~\ref{fig16}(f),
conversely, the head-on motion of the colliding condensates is
so fast that, instead of a plateau, as in Fig.~\ref{fig16}(c), a nonlinear
periodic wave appears between the DSWs, with the $m$
parameter determined by the boundary values:
\begin{equation}\label{rev14.3}
  m=m^*=\frac{(r_+^R-r_-^R)(r_+^L-r_-^L)}{(r_+^L-r_+^R)(r_-^L-r_-^R)}.
\end{equation}
So that just this combination of wave structures is realized, we
must verify that the velocities of the rarefaction wave and
DSW edges are ordered in a proper manner. This requires
exploring the corresponding solutions of hydrodynamic and
modulation equations.

A self-similar solution of Eqs.~(\ref{rev10.12}) with the required
boundary conditions is not difficult to find. For example, for
the rarefaction wave in Figs.~\ref{fig15} or \ref{fig16}(a), the Riemann invariant
$r_+=u/2+\sqrt{\rho}=\sqrt{\rho_L}$ is constant, which defines the relation
between $u$ and $\rho$ and in the simple wave. The first equation in (\ref{rev10.12})
is satisfied, and the self-similar solution of the second
equation has the form
$$
v_-=\frac12r_++\frac32r_-=\frac32u-\sqrt{\rho_L}=z=\frac{x}t.
$$
It readily follows from the obtained relations that
\begin{equation}\label{rev14.4}
  \rho=\frac19\left(\sqrt{\rho_L}-\frac{2x}t\right)^2,\quad
  u=\frac23\left(\sqrt{\rho_L}+\frac{x}t\right).
\end{equation}
The left edge of the rarefaction wave moves to the left with the
speed of sound $s_-^L$, equal in modulus to
$\sqrt{\rho_L}$, and the speed $s_+^L$ of the right edge can be
found by equating one of the variables
in (\ref{rev14.4}) to its value (\ref{rev13.2}) on the plateau, whence
\begin{equation}\label{rev14.5}
  s_-^L=-\sqrt{\rho_L},\quad s_+^L=\frac12\sqrt{\rho_L}-\frac32\sqrt{\rho_R}.
\end{equation}

In the DSW in Fig.~\ref{fig15}, the values of three Riemann invariants are known,
\begin{equation}\label{rev14.6}
  r_1=-\sqrt{\rho_R},\quad r_2=\sqrt{\rho_R},\quad r_4=\sqrt{\rho_L},
\end{equation}
and the dependence of $r_3$ on $z=x/t$ is determined by the self-similar
solution of Whitham's equations:
\begin{equation}\label{rev14.7}
  v_3(-\sqrt{\rho_R},\sqrt{\rho_R},r_3,\sqrt{\rho_L})=z=\frac{x}t.
\end{equation}
Substituting all these values and the functions $r_r=r_3(z)$ into
(\ref{rev11.22}) gives the density profile in the DSW, which is shown
with a dashed line in Fig.~\ref{fig15}(a), in good agreement with the
numerical solution. The velocities of the DSW edges can be
found by substituting values (\ref{rev14.6}) in the limit expressions
(\ref{rev12.28}) and (\ref{rev12.29}) for $v_3$:
\begin{equation}\label{rev14.8}
  s_-^{R}=\frac12(\sqrt{\rho_L}+\sqrt{\rho_R}),\quad
  s_+^{R}=\frac{2\rho_L-\rho_R}{\sqrt{\rho_L}}.
\end{equation}
It is easy to verify that, for $\rho_L>\rho_R$, the velocities of the
rarefaction wave and DSW edges are ordered in accordance
with the inequalities $s_-^L<s_+^L<s_-^R<s_+^R$, in agreement with
the diagram in Fig.~\ref{fig15}(b).

\begin{figure}[t]
\begin{center}
\includegraphics[width=7cm]{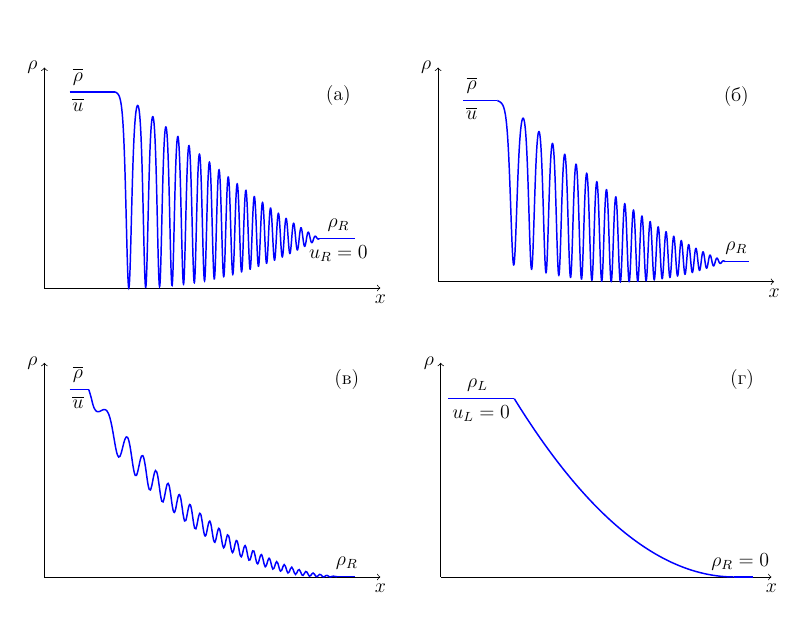}
\caption{ Transformation of a DSW as the density on the right boundary decreases.
(a) Occurrence of avacuum point. (b) Vacuum point inside theDSW.
(c) DSW with a small amplitude on the background of a rarefaction wave.
(d) Transformation of the DSW into a rarefaction wave at zero density on the
right boundary.
 }
\label{fig17}
\end{center}
\end{figure}

The soliton amplitude on the border with the plateau is
\begin{equation}\label{rev15.9}
  a_s=(r_4-r_2)(r_2-r_1)=2(\sqrt{\rho_L\rho_R}-\rho_R).
\end{equation}
If we fix $\rho_L$ and decrease $\rho_R$ from its maximum value $\rho_L$ , we
see that at $\rho_R=\rho_L/9$ the soliton depth $a_s$ becomes equal to
the background density $\overline{\rho}$ defined on the plateau by expression
(\ref{rev13.2}). This means that this soliton becomes black, and
the condensate density distribution acquires a `vacuum point'
\cite{gk-87,eggk-95}. As $\rho_R$ decreases further, the leading soliton
amplitude becomes less than the background density, and
the vacuum moves inwards the DSW. For the vanishing density
$\rho_R$, the amplitude of oscillations in the DSW tends to zero
together with soliton amplitude (\ref{rev15.9}), the plateau disappears
together with the left rarefaction wave, but the entire DSW
domain becomes a rarefaction wave, Eq.~(\ref{rev14.4}), corresponding
to the expansion of the condensate into the vacuum. This
transformation of the DSW depending on the boundary
conditions is illustrated in Fig.~\ref{fig17}.

Other configurations shown in Fig.~\ref{fig16} can be considered
similarly. It must only be kept in mind that, in Fig.~\ref{fig16}(f), the
modulated waves are matched not with the homogeneous
flow on the plateau but with a non-modulated periodic
solution with a known value (\ref{rev14.3}) of the $m$ parameter.

\begin{figure}[t]
\begin{center}
\includegraphics[width=7cm]{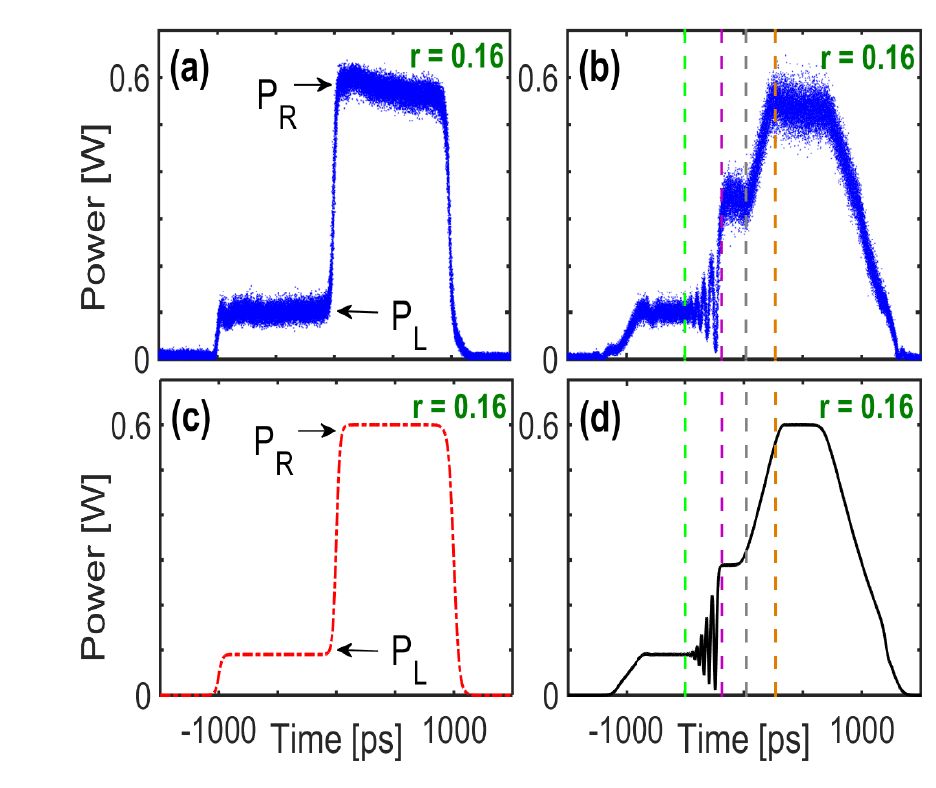}
\caption{Intensity profiles measured (a) at the entrance to the optical fiber and
(b) at the exit from it. (c) Initial condition and (d) the result of solving the
NLS equation numerically with that initial condition. (From \cite{xckmt-17}.)
 }
\label{fiber2}
\end{center}
\end{figure}

The theory expounded here was confirmed quantitatively
in a dedicated experiment \cite{xckmt-17}, in which an optical pulse had
an artificially produced discontinuity in the light intensity
distribution and the evolution of the pulse was governed by
the NLS equation, equivalent to the Gross-Pitaevskii equation.
Figure~\ref{fiber2}(a), which is borrowed from that paper, shows
the intensity profile of the pulse entering the optical fiber, and
Figs.~\ref{fiber2}(b,d) show the pulse profile at the exit. Figures~\ref{fiber2}(a,b)
show the results of measurements, and Figs.~\ref{fiber2}(c,e), the results
of a numerical solution of the NLS equation. The initial pulse
has the shape of two table tops with different heights placed
next to each other without a gap, such that a discontinuity in
intensity occurs in the center. Its evolution is the main subject
of interest here, whereas the rarefaction waves emerging on
the outer edges of the structure can be ignored. As we can see,
the wave emerging in the center corresponds to the case in
Fig.~\ref{fig16}(b), and the velocities of the rarefaction wave and DSW
edges agree well the theoretical values.

The problem of the evolution of a discontinuity, despite its
simplicity, is being used in more realistic applications, such as
DSW formation in a condensate flowing past an obstacle \cite{hakim-07,legk-09},
which allows explaining the result of the experiments
in \cite{ea-07}, at least qualitatively. We also note that experiments
with a nonlinear evolution of pulses in a more complicated
geometry, both in the physics of condensates \cite{kgk-04,hoefer-06} and in
nonlinear optics \cite{wjf-07}, also allow interpretations within that
scheme. In Section~15, we illustrate the method with the
solution to a simple problem on condensate motion under the
action of a steadily moving piston~\cite{hae-08}.

\section{Piston problem}

\begin{figure}[t]
\begin{center}
\includegraphics[width=6.5cm]{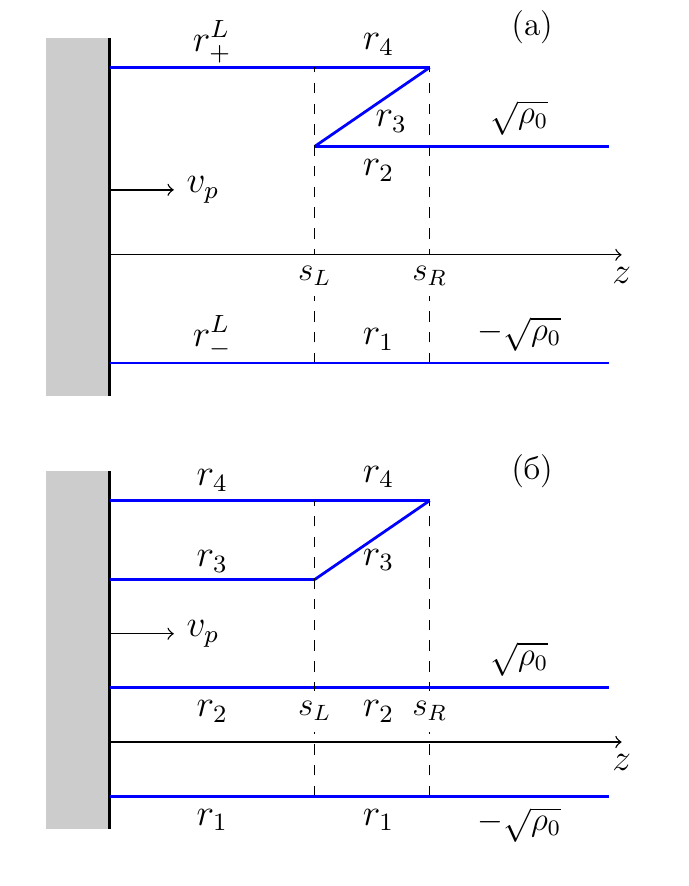}
\caption{ (a) Diagram of Riemann invariants in the problem of a piston
for $v_p<2\sqrt{\rho_0}$. (b) Diagram of Riemann invariants for $v_p>2\sqrt{\rho_0}$.
 }
\label{fig18}
\end{center}
\end{figure}

We consider the problem of the flow of a condensate under
the action of a piston \cite{hae-08}. We assume that the piston started
moving at the instant $t=0$ with a constant velocity $v_p$ and
that, prior to the motion of the piston, the condensate with a
constant density $\rho_0$ was at rest to the right of the piston. It is
clear that, as a result of that motion, a wave starts
propagating from the piston; if the piston speed is not too
high, it is natural to assume that adjacent to it is a
homogeneous flow of the condensate with the same speed $v_p$
and with some increased density $\rho_L$. Between this homogeneous flow
and the condensate at rest far from the piston,
there is a DSW, and the values of Riemann invariants on the
left and on the right of it can be expressed as
\begin{equation}\label{rev15.1}
  r_{\pm}^L=\frac12v_p\pm\sqrt{\rho_L},\quad r_{\pm}^R=\pm\sqrt{\rho_0}.
\end{equation}
The DSW originates instantaneously as the piston starts
moving, and hence the solution of Whitham's equations
must be self-similar, and the diagram of Riemann invariants
must have the form shown in Fig.~\ref{fig18}(a). We use the equality
$r_-^L=r_1=r_-^R$ to find the density $\rho_L$ of the flow adjacent to the piston:
\begin{equation}\label{rev15.2}
  \rho_L=\left(\frac12v_p+\sqrt{\rho_0}\right)^2,
\end{equation}
This, in turn, determines the value of the Riemann invariant
$r_4=r_+^L$. Hence, the values of three invariants that are
constant along the DSW are known,
\begin{equation}\label{rev15.3}
  r_1=-\sqrt{\rho_0},\quad r_2=\sqrt{\rho_0}, \quad r_4=v_p+\sqrt{\rho_0},
\end{equation}
and the dependence of invariant $r_3$ on the self-similarity
variable $z=x/t$ is defined implicitly by the equation
\begin{equation}\label{rev15.4}
  v_3(-\sqrt{\rho_0},\sqrt{\rho_0},r_3,v_p+\sqrt{\rho_0})=z.
\end{equation}
Using the limit expressions for $v_3$ in (\ref{rev12.28}) and (\ref{rev12.29}),
we find the velocities of the DSW edges as
\begin{equation}\label{rev16.5}
  s_L=\frac12v_p+\sqrt{\rho_0},\quad
  s_R=\frac{2v_p^2+4v_p\sqrt{\rho_0}+\rho_0}{v_p+\sqrt{\rho_0}}.
\end{equation}
At the location of the deepest soliton adjacent to the
homogeneous flow, formulas (\ref{t3-63.3}) and (\ref{t3-63.4}) give the
minimal condensate density and the flow velocity:
\begin{equation}\label{rev16.6}
\begin{split}
 & \rho_{\text{мин}}=\left(\sqrt{\rho_0}-\frac12v_p\right)^2,\\
  & u_{\text{мин}}=-v_p\frac{\sqrt{\rho_0}+v_p/2}{\sqrt{\rho_0}-v_p/2}.
  \end{split}
\end{equation}
For a sufficiently low piston speed, $v_p<2\sqrt{\rho_0}$, the flow
velocity $u_{\text{мин}}$ is negative, and hence the condensate flows
into the domain of increased density $\rho_L>\rho_0$, as expected.

For $v_p=2\sqrt{\rho_0}$, a vacuum point is formed in the DSW,
with the velocity of the left DSW edge becoming equal to the
piston speed, and hence the homogeneous flow domain
adjacent to the piston disappears. For $v_p>2\sqrt{\rho_0}$, similarly
to the case of the collision of condensates with too high
velocities (Fig.~\ref{fig16}(f)), the domain of a non-modulated periodic
solution of the Gross-Pitaevskii equation occurs instead of
the plateau, and this wave structure therefore corresponds to
the diagram of Riemann invariants shown in Fig.~\ref{fig18}(b). In the
periodic wave, the Riemann invariants $r_1,r_2$, $r_4$, preserve
their values (\ref{rev15.3}), and the condition that the wave velocity
coincide with the piston speed $V=(r_3+r_4)/2=v_p$ gives
$r_3=v_p-\sqrt{\rho_0}$. Thus, in the periodic solution domain,
\begin{equation}\label{rev16.7}
  m^*=\frac{4\rho_0}{v_p^2}<1\quad\text{for}\quad v_p>2\sqrt{\rho_0},
\end{equation}
and the condition of matching with the DSW determines the
velocity of this DSW edge:
\begin{equation}\label{rev16.8}
  s_L=v_p+\frac{2\sqrt{\rho_0}(v_p-2\sqrt{\rho_0})K(m^*)}
  {v_pE(m^*)-(v_p-2\sqrt{\rho_0})K(m^*)}.
\end{equation}
The maximum density of the condensate in this structure is
\begin{equation}\label{rev16.9}
  \rho_{\text{max}}=(r_4-r_3)(r_2-r_1)=4\rho_0.
\end{equation}
The density profile in the DSW can be constructed without
difficulty by substituting the Riemann invariants in (\ref{rev11.22}), and
the analytic results agree well with numerical computations \cite{hae-08}

The Gurevich-Pitaevskii method thus allows completely
solving the problem posed in this section.

\section{ Uniformly accelerated piston problem}

As in the case of the KdV equation, there are two scenarios for
a simple wave breaking: the profile of one of dispersionless
Riemann invariants $r_{\pm}$ acquires a vertical tangent either at
the interface with the condensate, which is at rest, or at the
inflection point. We here consider the first case and
assume for definiteness that this profile is produced by a
uniformly accelerated moving piston \cite{kk-10}, such that, at a
time $t$, the coordinate of the condensate-piston boundary
is $X(t)=at^2/2$.

Prior to the instant of breaking, the condensate flow can
be described by dispersionless equations (\ref{rev10.12}) with good
accuracy, and we now give their solution in the form that we
need. Under the action of the piston, the condensate flow is
unidirectional and hence can be described by a simple wave
with a constant Riemann invariant, $r_-=u/2-\sqrt{\rho}=-\sqrt{\rho_0}$,
where $\rho_0$ is the initial density of the condensate in the domain
that has not yet been reached by the wave produced by the
piston. The invariant $r_+$ satisfies the first equation in (\ref{rev10.12});
the general solution $x-(\frac32r_+-\sqrt{\rho_0})t=w(r_+)$ of that
equation must satisfy the boundary condition $u(X(t),t)=\dot{X}(t)$,
which states that the flow velocity on the boundary with
the piston coincides with the piston velocity. Therefore,
$r_+-\sqrt{\rho_0}=at$, and using the general solution for the
condensate flow on the boundary with the piston gives
$w=at^2/2-(\frac32r_+-\sqrt{\rho_0})t$. After eliminating $t=(r_+-\sqrt{\rho_0})/a$,
we obtain the general solution for the condensate flow in the form
\begin{equation}\label{rev17.1}
  x-\left(\frac32r_+-\frac12\sqrt{\rho_0}\right)t=\frac1a\sqrt{\rho_0}r_+-\frac1ar_+^2.
\end{equation}

\begin{figure}[t]
\begin{center}
\includegraphics[width=7cm]{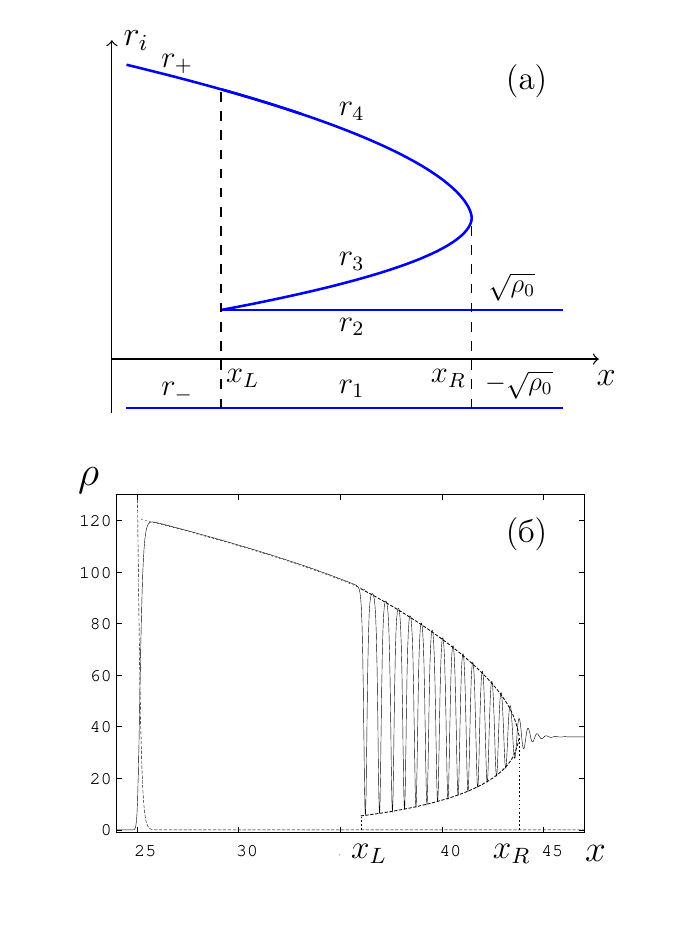}
\caption{ (a) Diagram of Riemann invariants in the problem of a
uniformly accelerated piston. (b) Density profile in the condensate
moving under the action of a uniformly accelerated piston. Solid line
shows the numerical computation result and the dashed line shows the
envelope, in accordance with the analytic theory.
 }
\label{fig19}
\end{center}
\end{figure}

This solution holds in the entire inhomogeneous flow
domain until the instant $t_b={2\sqrt{\rho_0}}/(3a)$ when the $r_+(x)$
profile acquires a vertical tangent at the point $x_b={2\rho_0}/(3a)$
on the boundary with the condensate at rest. After that
instant of breaking, a wave structure involving a DSW
emerges, with the distribution of Riemann invariants represented
by the diagram shown in Fig.~\ref{fig19}(a).
We therefore have to find a solution of Whitham's
equations with the constant Riemann invariants $r_1=-\sqrt{\rho_0}$
and $r_2=\sqrt{\rho_0}$, a solution satisfying the condition that $r_4$
match the invariant $r_+$ of dispersionless solution (\ref{rev17.1}) as
$r_3\to r_2$. The right-hand side of (\ref{rev17.1}) contains linear and
quadratic terms in $r_+$. As in the KdV problems considered
above, it suffices to take a linear combination of the
expressions $w_i^{(1)}\equiv v_i$ and $w_i^{(2)}$
that has just that dependence in the limit as $r_3\to r_2$.
The coefficients of this linear combination are chosen from the condition
of matching $r_4$ with $r_+$, and a straightforward calculation
\cite{kk-10} yields a solution in the form
\begin{equation}\label{rev17.4}
    \begin{split}
    x-v_3(r)t=\frac2{5a}\left(\rho_0+\sqrt{\rho_0}v_3(r)-\frac83w_3^{(2)}(r)\right),\\
    x-v_4(r)t=\frac2{5a}\left(\rho_0+\sqrt{\rho_0}v_4(r)-\frac83w_4^{(2)}(r)\right).
    \end{split}
\end{equation}
These formulas implicitly define the dependences of $r_3$and $r_4$
on $x$ and $t$, and their substitution in (\ref{rev11.22}) gives the DSW
density profile, whose envelope is compared in Fig.~\ref{fig19}(b) with
the results of a numerical solution of the Gross-Pitaevskii
equation.

Importantly, formulas (\ref{rev17.4}) allow finding the main DSW
parameters analytically. For example, in the soliton limit
$r_3=r_2$, the difference between these formulas on the
boundary $x=x_L(t)$ gives the time dependence of $r_4$ in the
form $r_4=5at/4+\sqrt{\rho_0}/6$, substituting which in any of
formulas (\ref{rev17.4}) leads to the law of motion of the soliton edge
of the DSW:
\begin{equation}\label{rev17.7}
    x_L(t)=\frac5{36}\frac{\rho_0}a+\frac7{12}\sqrt{\rho_0}\,t+\frac5{16}at^2.
\end{equation}
In the small-amplitude limit $r_3=r_4$, Eqs.(\ref{rev17.4}) reduce to a
single equation on the boundary $x=x_R(t)$:
\begin{equation}\nonumber 
    x-\frac{2r_4^2-\rho_0}{r_4}t=\frac2{5a}\left(3\rho_0+2\sqrt{\rho_0}r_4-\frac{\rho_0^{3/2}}{r_4}
    -4r_4^2\right),
\end{equation}
with the boundary value $x_R$ corresponding to the maximum
of this function $x(r_4)$ at a fixed value of $t$. This implies the
dependence of $t$ on $y=r_R/\sqrt{\rho_0}$:
\begin{equation}\label{ref18.9}
    t=    \frac{2\sqrt{\rho_0}}{5a}\cdot\frac{8y^3-2y^2-1}{2y^2+1},\quad y\geq1,
\end{equation}
substituting which in the limit expression for (\ref{rev17.4}) gives
\begin{equation}\label{18-7}
    x_R=\frac{2\rho_0}{5a}\cdot\frac{8y^4-6y^2+3}{2y^2+1},
\end{equation}
The obtained formulas define the law of motion of the small-
amplitude DSW edge in parametric form. At $t=t_b$ ($y=1$),
the coordinates of both edges are equal to the breaking point
coordinate $x_b$, in accordance with the fact that in the
asymptotic Gurevich-Pitaevskii approach the DSW has a
vanishing length at the instant of formation. The derived laws
of motion for the DSW edges agree well with numerical
solutions of the Gross-Pitaevskii equation \cite{kk-10}. The solution
to the breaking problem for a simple wave expanding into a
medium at rest and having a power-law profile $r_+\propto(-x)^{1/n}$
at the instant of breaking can be found similarly for any
integer $n$ (see \cite{kamch-18a}).

\section{Motion of edges
of `quasi-simple' dispersive shock waves}

A characteristic feature of a wave formed in the condensate as
a result of the motion of a piston was that it expanded into the
depth of the condensate at rest, and therefore in the DSW
domain two out of the four Riemann invariants of Whitham's
system were constant, and only the other two changed in the
course of evolution. This is similar to the KdV case considered
in Section~\S\ref{KdV-edges}, where one invariant was constant and two
others were variable. In \cite{gkm-89}, DSWs of this type were called
`quasi-simple'. The law of motion of their edges can again be
found in the theory of the Gross-Pitaevskii equation
following a strategy similar to that presented in Section~\S\ref{KdV-edges}.
In view of a close analogy with Section~\S\ref{KdV-edges}, we here give only
the basic facts of the corresponding theory \cite{kamch-19,kamch-20}.

For definiteness, we consider the breaking of a simple
wave for which the invariant $r_-=u/2-c=-c_0$ is constant,
where $c=\sqrt{\rho}$ is the local speed of sound, which takes the
value $c_0=\sqrt{\rho_0}$ in the unperturbed domain of the condensate.
We then have $r_+=u/2+c=2c-c_0$ and $v_+=3c-2c_0$, and
the solution of dispersionless equations (\ref{rev10.12}) can be written as
\begin{equation}\label{rev19.1}
  x-(3c-2c_0)t=\ox(c-c_0),
\end{equation}
where $\ox(c-c_0)$ is a function inverse to the initial distribution
$c-c_0=w(x)$ at the instant of breaking $t=0$. We first
assume that the initial pulse is `positive,' i.e., $c-c_0>0$.
This solution borders the soliton edge of the DSW, which
moves with the soliton velocity $V_s=(r_4+r_2)=c$, where we
used the fact that $r_2=-r_1=c_0$ along the quasi-simple DSW
and $r_4=r_+=2c-c_0$ at the matching point. Therefore,
$dx_L-cdt=0$ and dispersionless solution (\ref{rev19.1}) on the
boundary with the DSW for $x=x_L$ must be compatible with the equation
\begin{equation}\label{rev19.2}
  \frac{dx_L}{dc}-c\frac{dt}{dc}=0,
\end{equation}
where $x_L$ and $t$ are regarded as functions of the local speed of
sound $c$, which varies on the soliton edge as a result of the
DSW evolution. After eliminating $x_L$, we hence obtain the equation
\begin{equation}\label{rev19.3}
  2z\frac{dt}{dc}+3t=-\frac{d\ox}{dc},\quad z=c-c_0,
\end{equation}
solving which with the initial condition $t(0)=0$,
\begin{equation}\label{rev19.4}
  t(z)=-\frac1{2z^{3/2}}\int_0^z\sqrt{z}\,\ox'(z)dz,
\end{equation}
together with the equation
\begin{equation}\label{rev19.5}
  x_L(z)=(3z+c_0)t(z)+\ox(z)
\end{equation}
defines the law of motion of the soliton DSW edge over a
monotonic dispersionless profile in parametric form.

If the profile is not monotonic and has a maximum
$c_m=c_0+z_m$, then, for $t>t_m=t(z_m)$, when the soliton
edge borders the branch $\ox_2(c-c_0)$ of the dispersionless
solution, instead of (\ref{rev19.4}) and (\ref{rev19.5}) we
easily find the relations
\begin{equation}\label{rev19.6}
  \begin{split}
   & t(c)=-\frac1{2(c-c_0)^{3/2}}\int_0^{\ox_2(c-c_0)}\sqrt{\widetilde{c}_0(x)}dx,\\
  & x_L(c)=(3c-2c_0)t(c)+\ox_2(c),
  \end{split}
\end{equation}
where $c_0+\widetilde{c}_0(x)$ is the initial distribution of the local speed of
sound. At asymptotically large times, we hence find
\begin{equation}\label{rev20.7}
   x_L=c_0t+3\left(\frac{\mathcal{A}}2\right)^{2/3}t^{1/3},\quad
   \mathcal{A}=\int_{-\infty}^0\sqrt{\widetilde{c}_0(x)}\,dx.
\end{equation}
In this asymptotic limit, the DSW amplitude becomes much
less than the background density $\rho_0$, and the Gross-Pitaevskii
equation can be approximated for unidirectional wave
propagation with the KdV equation; hence, solution (\ref{rev20.7})
coincides with (\ref{ref9.16}) in the corresponding variables.

On the low-amplitude edge, in the same asymptotic
regime, $r_3\approx r_4\approx r_m=2c_m-c_0$ and $r_2=-r_1=c_0$, and
therefore formula (\ref{rev12.25}) gives the wavelength
$$
L=\frac{\pi}{2\sqrt{c_m(c_m-c_0)}}
$$
and the wave number $k=4\sqrt{c_m(c_m-c_0)}$. Hence, the group
velocity of motion of the small-amplitude edge is
\begin{equation}\label{rev20.9}
  \frac{dx_R}{dt}=\left.\frac{d\om}{dk}\right|_{k=k_m}=2r_m-\frac{c_0^2}{r_m}.
\end{equation}

In the case of a negative initial pulse with $\widetilde{c}_0(x)=c-c_0<0$,
similarly, the small-amplitude edge borders the
dispersionless solution (\ref{rev19.1}), with the Riemann invariants of
Whitham's system given by $r_3=r_4=-r_1=c_0$ and $r_2=2c-c_0$,
where $c$ is the local speed of sound on that edge. Therefore, the wavelength
is here given by $L=\pi/(2\sqrt{c_0(c_0-c)})$, i.e. $k=4\sqrt{c_0(c_0-c)}$,
and this edge moves over the background with the parameters
$\rho=c^2$, $u=2(c-c_0)$ with the group velocity
\begin{equation}\label{rev20.10}
  \frac{dx_R}{dt}=\frac{d\om}{dk}=u+\frac{c^2+k^2/2}{\sqrt{c^2+k^2/4}}=2c_0-\frac{c^2}{2c_0-c}.
\end{equation}
The compatibility condition of Eq.~(\ref{rev19.1}) with the equation
\begin{equation}\label{rev20.11}
  \frac{d x_R}{d c}-\left(2c_0-\frac{c^2}{2c_0-c}\right)\frac{d t}{d c}=0
\end{equation}
leads to the differential equation
\begin{equation}\nonumber
  \frac{(4c_0-c)(c_0-c)}{2c_0-c}\frac{dt}{dc}-\frac32t=\frac12\ox'(c-c_0),
\end{equation}
whose solution gives a parametric law of motion of the right DSW edge
\begin{equation}\label{rev20.12}
\begin{split}
  &t(c)=\frac1{2(4c_0-c)\sqrt{c_0-c}}\int_{c_0}^c\frac{(2c_0-c)\ox'(c-c_0)}{\sqrt{c_0-c}}dc,\\
  &x_R(c)=(3c-2c_0)t(c)+\ox(c).
  \end{split}
\end{equation}
It is easy to rewrite it, with obvious changes, for localized
pulses with a single local minimum.

In the case of a negative initial pulse, the asymptotic state
mainly consists of dark solitons, and it is easy to find the
velocity of the deepest soliton on the left DSW edge. We here
have $r_4=-r_1=c_0$ and $r_2\approx r_3\approx r_m=2c_m-c_0$, whence
\begin{equation}\label{rev2013}
  \frac{dx_L}{dt}=\frac12\sum r_i\approx r_m=2c_m-c_0.
\end{equation}
The number of dark solitons into which the initial negative
pulse eventually decays can be found following the same
strategy that we used to derive Karpman's formula (\ref{rev21.3}) for
the KdV equation. On the small-amplitude edge, we now have
$k(v_g-V)=k^3/(4\sqrt{c^2+k^2/4})$ and $k=4\sqrt{c_0(c_0-c)}$. Substituting
these expressions into the general formula (\ref{rem2}) and
using (\ref{rev20.12}) to replace the integration over $t$ with integration
over $c$, after simple transformations we obtain
\begin{equation}\label{rev22.1}
  N=\frac2{\pi}\int\sqrt{c_0(c_0-c(x))}\,dx,
\end{equation}
where $c(x)$ is the initial distribution of the local speed of
sound in the wave. The Gross-Pitaevskii equation, just like
the KdV equation, is completely integrable, making the
inverse scattering transform method \cite{zs-73} applicable to it,
which allows finding \cite{jlml-99,kku-02} the general expression for the
number of solitons originating from the pulse with the given
initial distributions of dispersionless Riemann invariant $r_{\pm}(x)$:
\begin{equation}\label{rev22.2}
  N=\frac1{\pi}\int\sqrt{(c_0-r_-(x))(c_0-r_+(x))}\,dx.
\end{equation}
In our case of the evolution of the pulse in the form of a simple
wave, $r_-(x)=-c_0$ and $r_+(x)=2c(x)-c_0$, and formula (\ref{rev22.2})
reduces to (\ref{rev22.1}). We must note, however, that both formula
(\ref{rev21.3}) for the KdV equation and formula (\ref{rev22.1}) for the
Gross-Pitaevskii equation can be represented as
\begin{equation}\label{rev22.3}
  N=\frac1{2\pi}\int k_0(x)dx,
\end{equation}
 where $k_0(x)$ is the wave number on the small-amplitude edge
corresponding to the initial distribution of the parameters of
the simple wave. Formula (\ref{rev22.3}) apparently is of a general
nature and can also be applied to equations that are not
completely integrable \cite{egkkk-07,egs-08}, for which the dependence $k_0(x)$
is to be found by solving equation for the conservation of the
number of waves along the trajectory of the small-amplitude
edge \cite{te-99,el-05}.

\section{Breaking of a cubic profile
in the Gross-Pitaevskii theory}

In the general case, a wave governed by the Gross-Pitaevskii
equation breaks in such a way that the profile of one of the
dispersionless Riemann invariants $r_{\pm}$ acquires a vertical
tangent and can be approximately represented by a cubic
curve near the inflection point. We assume for definiteness
that the invariant $r_+$ undergoes breaking, and it hence varies
in the neighborhood of that point very rapidly, which allows
assuming the $r_-$ invariant to be constant. By an appropriate
change of variables, it can be ensured that the condensate flow
is described by the formulas
\begin{equation}\label{rev18.1}
  x-\left(\frac32r_++\frac12r_-^0\right)t=-r_+^3,\quad r_-=r_-^0=\mathrm{const},
\end{equation}
up to the instant of breaking. These formulas give a solution
of hydrodynamic equations (\ref{rev10.12}). Naturally, it is assumed
here that $r_-^0<r_+$ in the domain of interest, including the
solution branch in (\ref{rev18.1}) with $r_+<0$. For $t>0$, solution (\ref{rev18.1})
becomes multi-valued. Taking dispersion into account, i.e.,
solving the full Gross-Pitaevskii equation, eliminates this
multi-valuedness by the formation of a DSW. Following the
Gurevich-Pitaevskii approach, we solve this problem \cite{kku-02,kk-10}
in Whitham's approximation by incorporating the solution
of Whitham's equations in dispersionless solution (\ref{rev18.1}) such
that the equality $r_1=r_-^0$ holds and the boundary conditions
\begin{equation}\label{rev18.2}
  \begin{split}
  & r_4(x_L(t),t)=r_+(x_L(t),t) \quad\text{при}\quad r_3=r_2,\\
  & r_2(x_R(t),t)=r_+(x_R(t),t) \quad\text{при}\quad r_3=r_4.
  \end{split}
\end{equation}
are satisfied. Because the right-hand side of the first equation
in (\ref{rev18.1}) involves a cubic function of $r_3$, we can satisfy all the
conditions by taking solution (\ref{rev12.30}) with $r_1=r_-^0$ and
$w_i=\sum_{k=0}^3A_kw_i^{(k)}$ are given by formulas (\ref{rev12.33}) and (\ref{rev17.5})
and the coefficients $A_k$ are chosen such that the
matching conditions are satisfied. As a result, we obtain
\begin{equation}\label{rev18.3}
    \begin{split}
    & x-v_i(r)t=-\tfrac{32}{35}{w_i^{(3)}}(r)+\tfrac{16}{35}{w_i^{(2)}}(r)r_-^0+\\
    & +\tfrac2{35}{v_i}(r)(r_-^0)^2+\tfrac1{35}(r_-^0)^3,\quad i=2,3,4,
    \end{split}
\end{equation}
These formulas implicitly define the dependence of the
invariants $r_2,r_3$, and $r_4$ on $x$ and $t$. In particular, investigating
the limit $r_3\to r_3$, we can easily find the law of motion of the
soliton edge of the DSW:
\begin{equation}\label{rev18.4}
    x_L(t)=\frac12r_-^0t-\frac16\sqrt{\frac53}\,t^{3/2}.
\end{equation}
The law of motion of the small-amplitude edge
\begin{equation}\label{rev18.5}
  x_R=\left(\frac32r_2+\frac12r_-^0\right)t(r_2,r_4)-r_2^3
\end{equation}
is defined in parametric form, with the time $t$ depending on
the parameters $r_2$ and $r_4$ as
\begin{equation}\label{rev19.6}
    t=\frac{2[8(r_4-7r_-^0)(3r_2^2+4r_2r_4
    +8r_4^2)-15r_2^3]}{35(4r_4-r_2-3r_-^0)},
\end{equation}
and the parameters themselves related as
\begin{equation}\label{21-8}
\begin{split}
    & 21(r_-^0)^2(4r_4+r_2)-10r_-^0(20r_4^2+2r_2r_4+r_2^2)+\\
    & +16r_4(8r_4^2-r_2r_4-r_2^2)+9r_2^3=0.
    \end{split}
\end{equation}
We see that this particular Gurevich-Pitaevskii problem has
also been given a fully analytic solution.

\section{ Conclusions}

We have presented the Gurevich-Pitaevskii theory for DSWs
in some detail following \cite{gp-73} and other closely related papers. It
remains to briefly mention some avenues of further development of this theory.

We first note that, simultaneously with the appearance
and development of the theory of DSWs, other important
events were taking place in nonlinear physics associated with
the discovery of the inverse scattering transform method
for solutions of nonlinear wave equations \cite{ggkm-67,lax-68,zs-73}. A
fundamental fact of that method is the relation between the
so-called completely integrable equations, a class to which
the KdV and Gross-Pitaevskii equations belong, and the
associated linear spectral problems. For example, associated
with the KdV equation is the problem of the spectrum of a
quantum particle moving in the potential $u(x,t)$; the relation
is such that, in particular, the parameters of the soliton
solution are related to the discrete spectrum of that potential.

An extension of this method to periodic solutions of the
KdV equation \cite{nov-74,dmn-76} has shown that the Riemann invariants
of Whitham's system coincide with the endpoints of gaps
where the motion of the quantum particle is forbidden in the
corresponding periodic potential. This allowed, on the one
hand, generalizing the Whitham method to multi-phase
solutions \cite{ffm-80} and, on the other hand, extending it to other
integrable equations. In particular, we have used Whitham's
equations for the Gross-Pitaevskii theory, which were found
in \cite{FL-86,pavlov-87} by methods based on the complete integrability of
that equation.

It turns out as a result that three sets of parameters
characterizing the periodic solutions arise naturally in the
theory: (1) physical parameters $\nu_i$ related to the wave
amplitude and other quantities that bear a clear physical
meaning; (2) the end points $\lambda_i$ of the periodic spectral problem;
(3) the Riemann invariants $r_i$ of Whitham's modulation
system for the considered periodic wave.

In the simplest case of the KdV equation, the relations
among all these parameters are linear, and this is why
Whitham could diagonalize the modulation equations
derived for physical parameters by choosing appropriate
linear combinations. In the case of the Gross-Pitaevskii
equation, the relation between $\lambda_i$ and $r_i$ remains linear, and
that is why we were able to not invoke $\lambda_i$ in our presentation,
but the physical parameters $\nu_i$ are related to $r_i$ (or $\lambda_i$) by more
complicated formulas (\ref{rev11.21}). This complication, technical at
first glance, becomes fundamentally important when the
relation between $\lambda_i$ and $r_i$ becomes multi-valued: one solution
of Whitham's equations corresponds to two different
periodic waves. This situation is characteristic of the so-called
not genuinely nonlinear equations, in which nonlinear terms
can vanish for some amplitude of the wave. This was noted in
\cite{pt-06} for a higher KdV equation, an element of a hierarchy of
equationsas sociated with the samespectral problem, and also
in \cite{marchant-08} for the modified KdV equation $u_t\pm6u^2u_x+u_{xxx}=0$,
where the coefficient in the nonlinear term has a maximum or
a minimum at $u=0$, depending on the sign.

In the problem of the evolution of a step-like profile, this
led to the appearance of more complicated structures than
rarefaction waves and modulated cnoidal waves that we are
familiar with from the theory outlined in the foregoing. A
classification of such structures evolving from the initial
discontinuity in accordance with the Gardner equation
$u_t+6(u\pm \alpha u^2)u_x+u_{xxx}=0$ that occurs in the theory of
internal water waves was given in \cite{kamch-12}. In the theory of
the modified NLS equation $i\psi_t+\tfrac12\psi_{xx}-i(|\psi|^2\psi)_x=0$,
which has applications in nonlinear optics and magnetohydrodynamic waves,
the use of all three sets of parameters
becomes necessary: periodic solutions and Whitham's equations were obtained
in \cite{kamch-90}, and the evolution of the initial
discontinuity was analyzed in \cite{gke-92b,ik-17,kamch-18}. Finally, the most
complicated case of this type, a ferromagnet with `easy plane'
anisotropy and the equivalent limit for two-component
Gross-Pitaevskii equations, was studied in \cite{kamch-92,ikcp-17}.

Besides the development of Whitham's averaging method,
the discovery of the complete integrability of the most
important equations in nonlinear wave physics has allowed
developing other approaches to the theory of DSWs. In
particular, it was shown in \cite{ll-83a,ll-83b,ll-83c,ven-85a,ven-85b,mazur-96}
that the solution to the Gurevich-Pitaevskii problem in Whitham's approximation
can also be obtained as a semiclassical limit of exact multi-soliton solutions
of the KdV equation. Another aspect of a
more exact theory of DSWs is that, similarly to how the linear
problem solution (\ref{eq1.51k}) obtained by the averaging method is an
asymptotic form of the Airy function, Whitham's approximation for breaking
waves is a semiclassical asymptotic form of
some special functions that are `standard' solutions of the
Painlev\'{e} nonlinear differential equations (see, e.g., \cite{sul-94,gs-10,cg-10}).
Solutions expressed in terms of such special functions
are also exact at the small-amplitude edge of the DSW.

Another area of investigations is to generalize the
Gurevich-Pitaevskii approach to equations that are not
completely integrable. Naturally, the Whitham theory
considered above for the perturbed KdV equation can be
generalized to a rather wide class of equations close to
completely integrable ones \cite{kamch-04,lkp-12}. However, a large
number of physically important equations do not fall into
that category and the modulation equations for periodic
solutions of such equations do not have Riemann invariants
in any approximation. Still, the general Gurevich-Pitaevskii
approach is also valid for them and some important
characteristic of DSWs can be calculated with no Riemann
invariants defined.

The first important statement regarding such systems,
made by Gurevich and Meshcherkin \cite{gm-84}, was that only a
DSW is formed in the breaking of a simple wave, and the
constant Riemann invariant of the dispersionless limit
transports its value across the DSW, despite the absence of
Whitham's Riemann invariant conserved along the DSW.
This statement is already sufficient in order to calculate the
parameters of the plateau appearing between two wave
structures in the evolution of a discontinuity.

The next important step was made in \cite{te-99,el-05}, where it
was noted that, on the border with a simple wave,
Whitham's system reduces to an ordinary differential
equation whose solution gives a relation between the DSW
parameters on that edge. Because one of the modulation
equations (the conservation law for the number of waves) is
certainly known on the small-amplitude edge, the solution
of that equation gives a relation between the wave number
and the background amplitude of the wave. On the soliton
edge, such an equation is absent in general. But it can be
verified that, in the case of KdV and Gross-Pitaevskii
equations, the equation $\widetilde{k}_t+\widetilde{\om}_x=0$ holds
for pulse expansion into a medium at rest with two constant Riemann
invariants, with $\widetilde{k}$ being the inverse half-width of the soliton
and $\widetilde{\om}(\widetilde{k})$ obtained from the linear dispersion law
$\om(k)$ by the substitution $\widetilde{\om}(\widetilde{k})=-i\om(i\widetilde{k})$.
According to an old remark by
Stokes quoted in a note to \S252 in \cite{lamb}, $\widetilde{\om}(\widetilde{k})$
determines the soliton velocity: the tails of the soliton propagate with the
same velocity as the soliton itself, and on the tails the
linearized equations have the same form as in the small-amplitude harmonic limit.

Assuming the validity of the equation $\widetilde{k}_t+\widetilde{\om}_x=0$ in the
general case of the breaking of simple waves expanding into a
`quiescent' homogeneous medium with two constant dispersionless Riemann invariants,
we can obtain an ordinary differential equation for the parameters along the soliton
edge of the DSW. These two equations are entirely sufficient
for finding the parameters of the edges of the DSW forming in
the evolution of a discontinuity and satisfying an unintegrable
equation, as was indeed done in series of studies
\cite{el-05,egs-06,egkkk-07,ep-11,lh-13,hoefer-14,ckp-16,hek-17,ams-18}.
Requiring the compatibility of the thus obtained
ordinary differential equation with the solution of the
dispersionless equations on that boundary allows obtaining
the equation of motion for the DSW edge propagating over
the general profile of a simple wave \cite{kamch-19,ik-19,kamch-20}.

A new type of DSW can occur when taking higher-order
dispersion effects into account when the soliton velocity is
equal to the phase velocity of linear waves and these are in
resonance with other. The general Gurevich-Pitaevskii
approach is also applicable in that case \cite{smyth-16,es-16,hss-18,bs-20}.

In this paper, we mentioned applications of the Gurevich-Pitaevskii
problem to water waves, plasmas, Bose-Einstein
condensate, and nonlinear optics. To these, we can add the
observations and the theory of DSWs in internal waves in the
ocean \cite{hm-06} and the atmosphere \cite{ps-02}, and on jets of a liquid in
viscous media \cite{maiden-16,maiden-20}. The Gurevich-Pitaevskii approach
to the DSW theory also extends to waves with several spatial
variables \cite{abr-18} and finds applications in other areas in
physics, including the quantum gravity theory \cite{sul-94}. The
reader can find more examples of DSWs, e.g., in review
\cite{eh-16} and the references therein. In addition, the creation of
the DSW theory was related to the substantial progress in
modern mathematical physics, and the reader can glean some
aspects of the mathematical theory from reviews \cite{dm-80,dn-89}.

To conclude, we can say that in the years that have passed
since the appearance of paper \cite{gp-73}, the Gurevich-Pitaevskii
problem, understood as a general approach to the DSW
theory based on Whitham's modulation equations, has
become an area of vibrant research in nonlinear physics,
with a distinctive problem setting and with profound
mathematical methods for solving problems and clear
physical ideas that enrich the entire physics of nonlinear
waves.

I am grateful to L.~P.~Pitaevskii for discussions of the
problems considered in this paper and for his useful remarks.


\begin{thebibliography}{99}

\bibitem{gp-73} Gurevich A V, Pitaevskii L P {\it Sov. Phys. JETP} {\bf 38} 291 (1974);
{\it Zh. Eksp. Teor. Fiz.} {\bf 65} 590 (1973)

\bibitem{whitham-65} Whitham G B {\it Proc. Roy. Soc. London} {\bf 283} 238 (1965)

\bibitem{russel-1844} Scott Russel J {\it Report on waves} (British Association Reports, 1844).

\bibitem{bouss-1871a} Boussinesq J {\it Comptes Rendus} {\bf 72} 755 (1871)

\bibitem{rayleigh-1876} Lord Rayleigh {\it Phil. Mag.} {\bf 1} 257 (1876)

\bibitem{kdv} Korteweg D J, de~Vries G {\it Phil. Mag.} {\bf 39} 422 (1895)

\bibitem{bl-54} Benjamin T B, Lighthill M J {\it Proc. Roy. Soc. Lond.} A {\bf 224} 448 (1954)

\bibitem{lamb} Lamb H {\it Hydrodynamics} (Cambridge: Cambridge Univ. Press, 1932)

\bibitem{stoker} Stoker J J {\it Water Waves; the Mathematical Theory with Applications}
(New York: Interscience Publ., 1957)

\bibitem{gm-60} Gardner S C, Morikawa G K, ``Similarity in the asymptotic behavior of
collision-free hydromagnetic waves and water waves'', Courant Institute of Mathematical
Sciences Report No. NYO‐9082 (N. Y. 1960)

\bibitem{vvs-61} Vedenov A A, Velikhov E P, Sagdeev R Z {\it Nucl. Fusion} {\bf 1} 82 (1961);
{\it Yad. Sintez} {\bf 1} 82 (1961)

\bibitem{sagdeev}  Sagdeev R Z, in {\it Reviews of Plasma Physics} Vol. 4 (Ed.
M A Leontovich) (New York: Consultants Bureau, 1966) p. 23;
Translated from Russian: in {\it Voprosy Teorii Plazmy} Issue 4 (Ed.
M A Leontovich) (Moscow: Gosatomizdat, 1964) p. 20

\bibitem{ABS-68} Alikhanov S G, Belan V G, Sagdeev R Z {\it JETP Lett.} {\bf 7} 318 (1968);
{\it Pis'ma Zh. Eksp. Teor. Fiz.} {\bf 7} 405 (1968)

\bibitem{TBI-70} Taylor R J, Baker D R, Ikezi H, {\it Phys. Rev. Lett.} {\bf 24} 206 (1970)

\bibitem{gpp-65a} Gurevich A V, Pariiskaya L V, Pitaevskii L P {\it Sov. Phys. JETP} {\bf 22}
449 (1966); {\it Zh. Eksp. Teor. Fiz.} {\bf 49} 647 (1965)

\bibitem{gpp-65b} Gurevich A V, Pariiskaya L V, Pitaevskii L P {\it Sov. Phys. JETP} {\bf 27}
476 (1968); {\it Zh. Eksp. Teor. Fiz.} {\bf 54} 891 (1968)

\bibitem{gp-69} Gurevich A V, Pitaevskii L P {\it Sov. Phys. JETP} {\bf 29} 954 (1969);
{\it Zh. Eksp. Teor. Fiz.} {\bf 56} 1178 (1969)

\bibitem{gp-71} Gurevich A V, Pitaevskii L P {\it Sov. Phys. JETP} {\bf 33} 1159 (1971);
{\it Zh. Eksp. Teor. Fiz.} {\bf 60} 2155 (1971)

\bibitem{LL-10} Lifshitz E M, Pitaevskii L P {\it Physical Kinetics} (Oxford: Pergamon
Press, 1981); Translated from Russian: {\it Fizicheskaya Kinetika}
(Moscow: Fizmatlit, 2001)

\bibitem{AS-2}  Abramowitz M, Stegun I A (Eds) {\it Handbook of Mathematical
Functions with Formulas, Graphs, and Mathematical Tables} (New
York: Dover Publ., 1972)

\bibitem{whitham-74} Whitham G B {\it Linear and Nonlinear Waves} (New York: Wiley, 1974)

\bibitem{LL6} Landau L D, Lifshitz E M {\it Fluid Mechanics} (Oxford: Pergamon
Press, 1987); Translated from Russian: {\it Gidrodinamika} (Moscow:
Fizmatlit, 2001)

\bibitem{rozhd-yan} Rozdestvenskii B L, Janenko N N {\it Systems of Quasilinear Equations
and Their Applications to Gas Dynamics} (Providence, RI: American
Mathematical Society, 1983); Translated from Russian: {\it Sistemy
Kvazilineinykh Uravnenii i Ikh Prilozheniya k Gazovoi Dinamike} (Moscow: Nauka, 1978)

\bibitem{ggkm-67} Gardner S C et al.
{\it Phys. Rev. Lett.} {\bf 19} 1095 (1967)

\bibitem{gkm-89} Gurevich A V, Krylov A L, Mazur N G {\it Sov. Phys. JETP} {\bf 68} 966
(1989); {\it Zh. Eksp. Teor. Fiz.} {\bf 95} 1674 (1989)

\bibitem{gke-91} Gurevich A V, Krylov A L, El' G A {\it JETP Lett.} {\bf 54} 102 (1991);
{\it Pis'ma Zh. Eksp. Teor. Fiz.} {\bf 54} 104 (1991)

\bibitem{gkme-92} Gurevich A V et al. {\it Sov. Phys. Dokl.} {\bf 37} 198 (1992);
{\it Dokl. Ross. Akad. Nauk} {\bf 323} 876 (1992)

\bibitem{gke-92} Gurevich A V, Krylov A L, El' G A {\it Sov. Phys. JETP} {\bf 74} 957 (1992);
{\it Zh. Eksp. Teor. Fiz.} {\bf 101} 1797 (1992)

\bibitem{ek-93} El G A, Khodorovsky V V {\it Phys. Lett. A} {\bf 182} 49 (1993)

\bibitem{ks-90} Kudashev V P, Sharapov S E {\it Theor. Math. Phys.} {\bf 85} 1155 (1990);
{\it Teor. Matem. Fiz.} {\bf 85} 205 (1990)

\bibitem{ks-91} Kudashev V P, Sharapov S E {\it Theor. Math. Phys.} {\bf 87} 358 (1991);
{\it Teor. Matem. Fiz.} {\bf 87} 40 (1991)

\bibitem{kud-92} Kudashev V R {\it Phys. Lett. A} {\bf 166} 213 (1992)

\bibitem{kud-92b} Kudashev V R {\it Phys. Lett. A} {\bf 171} 335 (1992)

\bibitem{wright-93} Wright O C {\it Commun. Pure Appl. Math.} {\bf 46} 423 (1993)

\bibitem{tian-93} Tian F R {\it Commun. Pure Appl. Math.} {\bf 46} 1093 (1993)

\bibitem{tsarev-85} Tsarev S P {\it Sov. Math. Dokl.} {\bf 31} 488 (1985);
{\it Dokl. Akad. Nauk SSSR} {\bf 282} 534 (1985)

\bibitem{gs-86} Grimshaw R H J, Smyth N {\it J Fluid Mech.} {\bf 169}  429  (1986)

\bibitem{smyth-87} Smyth N {\it Pros. Roy. Soc. Lond. A} {\bf 409} 79 (1987)

\bibitem{potemin-88} Pot\"{e}min G V {\it Russ. Math. Surv.} {\bf 43} 252 (1988);
{\it Usp. Matem. Nauk} {\bf 43} 211 (1988)

\bibitem{kamch-19} Kamchatnov A M {\it Phys. Rev. E} {\bf 99} 012203 (2019)

\bibitem{gp-87} Gurevich A V, Pitaevskii L P Sov. {\it Phys. JETP} {\bf 66} 490 (1987);
{\it Zh. Eksp. Teor. Fiz.} {\bf 93} 871 (1987)

\bibitem{karpman-67} Karpman V I {\it Phys. Lett. A} {\bf 25} 708 (1967)

\bibitem{dvz-94} Deift P, Venakides S, Zhou Z {\it Commun. Pure Appl. Math.}
{\bf 47} 199 (1994)

\bibitem{ikp-19} Isoard M, Kamchatnov A M, Pavloff N {\it Phys. Rev. E} {\bf 99} 012210 (2019)

\bibitem{akn-87} Avilov V V, Krichever I M, Novikov S P {\it Sov. Phys. Dokl.} {\bf 32} 564
(1987); {\it Dokl. Akad. Nauk SSSR} {\bf 295} 345 (1987)

\bibitem{gp-91} Gurevich A V, Pitaevskii L P {\it Sov. Phys. JETP} {\bf 72} 821 (1991);
{\it Zh. Eksp. Teor. Fiz.} {\bf 99} 1470 (1991)

\bibitem{mg-95} Myint S, Grimshaw R {\it Wave Motion} {\bf 22} 215 (1995)

\bibitem{kamch-04} Kamchatnov A M {\it Physica D} {\bf 188} 247 (2004)

\bibitem{kamch-16} Kamchatnov A M {\it Physica D} {\bf 333} 99 (2016)

\bibitem{johnson-70} Johnson R S {\it J Fluid Mech.} {\bf 42} 49 (1970)

\bibitem{gross} Gross E P {\it Nuovo Cimento} {\bf 20} 454 (1961)

\bibitem{pit-1} Pitaevskii L P {\it Sov. Phys. JETP} {\bf 13} 451 (1961);
{\it Zh. Eksp. Teor. Fiz.} {\bf 40} 646 (1961)

\bibitem{dps-99} Daldovo F, Pitaevskii L P, Stringari S {\it Rev. Mod. Phys.} {\bf 71} 463 (1999)

\bibitem{pit-2} Pitaevskii L P {\it Phys. Usp.} {\bf 41} 569 (1998);
{\it Usp. Fiz. Nauk} {\bf 168} 641 (1998)

\bibitem{talanov-65} Talanov V I {\it JETP Lett.} {\bf 2} 138 (1965);
{\it Pis'ma Zh. Eksp. Teor. Fiz.} {\bf 2} 218 (1965)

\bibitem{kelley-65} Kelley P A {\it Phys. Rev. Lett.} {\bf 15} 1005 (1965)

\bibitem{bn-67} Benney D J, Newell A C {\it J. Math. \& Phys.} {\bf 46} 133 (1967)

\bibitem{bogol-47} Bogolubov N N {\it J. Phys. USSR} {\bf 11} 23 (1947);
{\it Izv. Akad. Nauk SSSR Ser. Fiz.} {\bf 11} 77 (1947); {\it Usp. Fiz. Nauk} {\bf 93} 552 (1967)

\bibitem{tsuzuki-71} Tsuzuki T {\it J. Low Temp. Phys.} {\bf 4} 441 (1971)

\bibitem{FL-86} Forest M G, Lee J E
in {\it Oscillation Theory, Computation, and Methods of Compensated Compactness,}
{vol.~2,} (Eds. C. Dafermos et al)  (Springer, N.Y., 1986)  p.35

\bibitem{pavlov-87} Pavlov M V {\it Theor. Math. Phys.} {\it 71} 584 (1987);
{\it Teor. Matem. Fiz.} {\bf 71} 351 (1987)

\bibitem{zs-73} Zakharov V E, Shabat A B {\bf Sov. Phys. JETP} {\bf 37} 823 (1973);
{\it Zh. Eksp. Teor. Fiz.} {\bf 64} 1627 (1973)

\bibitem{ffm-80} Flaschka H, Forest M G, McLaughlin D W
{\it Commun. Pure Appl. Math.} {\bf 3,} 739 (1980)

\bibitem{gk-87} Gurevich A V, Krylov A L {\it Sov. Phys. JETP} {\bf 65} 944 (1987);
{\it Zh. Eksp. Teor. Fiz.} {\bf 92} 1684 (1987)

\bibitem{eggk-95} El G A et al. {\it Physica D} {\bf 87} 186 (1995)

\bibitem{xckmt-17} Xu G et al. {\it Phys. Rev. Lett.} {\bf 118} 254101 (2017)

\bibitem{hakim-07} Hakim V {\it Phys. Rev. E} {\bf 55} 2835 (1997)

\bibitem{legk-09} Leszczyszyn A M et al. {\it Phys. Rev. A} {\bf 79} 063608 (2009)

\bibitem{ea-07} Engels P, Atherton C {\it Phys. Rev. Lett.} {\bf 99} 160405 (2007)

\bibitem{kgk-04} Kamchatnov A M, Gammal A, Kraenkel R A
{\it Phys. Rev. A} {\bf 69} 063605 (2004)

\bibitem{hoefer-06} Hoefer M A et al. {\it Phys. Rev. A} {\bf 74} 023623 (2006)

\bibitem{wjf-07} Wan W, Jia S, Fleischer J W  {\it Nature Phys.} {\bf 3} 46 (2007)

\bibitem{hae-08} Hoefer M A, Ablowitz M J, Engels P
{\it Phys. Rev. Lett.} {\bf 100} 084504 (2008)

\bibitem{kk-10} Kamchatnov A M, Korneev S V {\it J. Exp. Theor. Phys.} {\bf 110} 170 (2010);
{\it Zh. Eksp. Teor. Fiz.} {\bf 137} 191 (2010)

\bibitem{kamch-18a} Kamchatnov A M {\it J. Exp. Theor. Phys.} {\bf 127} 903 (2018); 
{\it Zh. Eksp. Teor. Fiz.} {\bf 154} 1016 (2018)

\bibitem{kamch-20} Kamchatnov A M {\it Theor. Math. Phys.} {\bf 202} 363 (2020);
{\it Teor.Matem. Fiz.} {\bf 202} 415 (2020)

\bibitem{jlml-99} Jin S, Levermore C D, McLaughlin D W
{\it Comm. Pure Appl. Math.} {\bf 52} 613 (1999)

\bibitem{kku-02} Kamchatnov A M, Kraenkel R A, Umarov B A
{\it Phys. Rev. E} {\bf 66} 036609 (2002)

\bibitem{egkkk-07} El G A et al. {\it Phys. Rev. A} {\bf 76} 053813 (2007)

\bibitem{egs-08} El G A, Grimshaw R H J, Smyth N F {\it Physica D} {\bf 237} 2423 (2008)

\bibitem{te-99} Tyurina A V, El' G A {\it J. Exp. Theor. Phys.} {\bf 88} 615 (1999); 
{\it Zh. Eksp. Teor. Fiz.} {\bf 115} 1116 (1999)

\bibitem{el-05} El G A {\it Chaos} {\bf 15} 037103 (2005)

\bibitem{lax-68} Lax P D {\it Comm. Pure Appl. Math.} {\bf 21} 467 (1968)

\bibitem{nov-74} Novikov S P {\it Funct. Anal. Its Appl.} {\bf 8} 236 (1974); 
{\it Funkts. Analiz Ego Prilozh.} {\bf 8} (3) 54 (1974)

\bibitem{dmn-76} Dubrovin B A, Matveev V B, Novikov S P {\it Russ. Math. Surv.} {\bf 31} (1)
59 (1976); {\it Usp. Matem. Nauk} {\bf 31} (1) 55 (1976)

\bibitem{pt-06} Pierce V U, Tian F-R {\it Comm. Math. Sci.} {\bf 4} 799 (2006)

\bibitem{marchant-08} Marchant T R {\it Wave Notion} {\bf 45} 540 (2008)

\bibitem{kamch-12} Kamchatnov A M et al. {\it Phys. Rev. E} {\bf 86} 0636605 (2012)

\bibitem{kamch-90} Kamchatnov A M {\it Sov. Phys. JETP} {\bf 70} 80 (1990); 
{\it Zh. Eksp. Teor. Fiz.} {\bf 97} 144 (1990)

\bibitem{gke-92b}  Gurevich A V, Krylov A L, El' G A {\it Sov. Phys. JETP} {\bf 75} 825 (1992);
{\it Zh. Eksp. Teor. Fiz.} {\bf 102} 1524 (1992)

\bibitem{ik-17} Ivanov S K, Kamchatnov A M {\it Phys. Rev. A} {\bf 96} 053844 (2017)

\bibitem{kamch-18} Kamchatnov A M {\it J. Phys. Commun.} {\bf 2} 025027 (2018)

\bibitem{kamch-92} Kamchatnov A M {\it Sov. Phys. JETP} {\bf 75} 868 (1992); 
{\it Zh. Eksp. Teor. Fiz.} {\bf 102} 1606 (1992)

\bibitem{ikcp-17} Ivanov S K et al. {\it Phys. Rev. E} {\bf 96} 062202 (2017)

\bibitem{lkp-12} Larr\'{e} P-\'{E}, Kamchatnov A M, Pavloff N {\it Phys. Rev. B} {\bf 86} 165304 (2012)

\bibitem{ll-83a} Lax P D, Levermore C D {\it Comm. Pure Appl. Math.} {\bf 36} 253 (1983)

\bibitem{ll-83b} Lax P D, Levermore C D {\it Comm. Pure Appl. Math.} {\bf 36} 573 (1983)

\bibitem{ll-83c} Lax P D, Levermore C D {\it Comm. Pure Appl. Math.} {\bf 36} 809 (1983)

\bibitem{ven-85a} Venakides S {\it Comm. Pure Appl. Math.} {\bf 38} 125 (1985)

\bibitem{ven-85b} Venakides S {\it Comm. Pure Appl. Math.} {\bf 38} 883 (1985)

\bibitem{mazur-96} Mazur N G {\it Theor. Math. Phys.} {\bf 106} 35 (1996);
{\it Teor. Matem. Fiz.} {\bf 106} 44 (1996)

\bibitem{sul-94} Suleimanov B I {\it J. Exp. Theor. Phys.} {\bf 78} 583 (1994); 
{\it Zh. Eksp. Teor. Fiz.} {\bf 105} 1089 (1994)

\bibitem{gs-10} Garifullin R N, Suleimanov B I {\it J. Exp. Theor. Phys.} {\bf 110} 133 (2010);
{\it Zh. Eksp. Teor. Fiz.} {\bf 137} 149 (2010)

\bibitem{cg-10} Claeys T, Grava T {\it Comm. Pure Appl. Math.} {\bf 63} 0203 (2010)

\bibitem{gm-84} Gurevich A V, Meshcherkin A R {\it Sov. Phys. JETP} {\bf 60} 732 (1984);
{\it Zh. Eksp. Teor. Fiz.} {\bf 87} 1277 (1984)

\bibitem{egs-06} El G A,  Grimshaw R H J, Smyth N F {\it Phys. Fluids} {\bf 18} 027104 (2006)

\bibitem{ep-11} Esler J G,  Pearce J D  {\it J. Fluid Mech.} {\bf 667} 555 (2011)

\bibitem{lh-13} Lowman N K, Hoefer M A {\it J. Fluid Mech.} {\bf 718} 524 (2013)

\bibitem{hoefer-14} Hoefer M A  {\it J. Nonlinear Sci.} {\bf 24} 525 (2014)

\bibitem{ckp-16}  Congy T, Kamchatnov A M, Pavloff N
{\it SciPost Phys.} {\bf 1} 006 (2016)

\bibitem{hek-17} Hoefer M A, El G A, Kamchatnov A M
{\it SIAM J. Appl. Math.} {\bf 77} 1352 (2017)

\bibitem{ams-18}  An X, Marchant T R,  Smyth N F
{\it Proc. Roy. Soc. London A} {\bf 474} 0278 (2018)

\bibitem{ik-19} Ivanov S K, Kamchatnov A M {\it Phys. Fluids} {\bf 31} 057102 (2019)

\bibitem{smyth-16}  Smyth N F {\it Physica D} {\bf 333}  301 (2016)

\bibitem{es-16} El G A,  Smyth N F {\it Proc. R. Soc. London A} {\bf 472} 20150633 (2016)

\bibitem{hss-18} Hoefer M A, Smyth N F, Sprenger P {\it Stud. Appl. Math.} {\bf 142} 219 (2019)

\bibitem{bs-20} Baqer S, Smyth N F {\it Physica D} {\bf 403} 132334 (2020)

\bibitem{hm-06} Helfrich K R,  Melville W K {\it Annu. Rev. Fluid Mech.}
{\bf 38} 395 (2006)

\bibitem{ps-02} Porter A, Smyth N F {\it J. Fluid Mech.} {\bf 454} 1 (2002)

\bibitem{maiden-16} Maiden M D et al.  {\it Phys. Rev. Lett.} {\bf 116} 174501 (2016)

\bibitem{maiden-20} Maiden M D et al.  {\it J. Fluid Mech.} {\bf 883} A10 (2020)

\bibitem{bog-90} Bogaevskii V N {\it USSR Comput. Math. Math. Phys.} {\bf 30} (5) 148
(1990); {\it Zh. Vychisl. Matem. Matem. Fiz.} {\bf 30} 1487 (1990)

\bibitem{abr-18} Ablowitz M J, Biondini G, Rumanov I {\it J. Phys. A: Math. Theor.}
{\bf 51} 215501 (2018)

\bibitem{eh-16} El G A,  Hoefer M A {\it Physica D} {\bf 333} 11 (2016)

\bibitem{dm-80} Dobrokhotov S Yu, Maslov V P, in {\it Itogi Nauki i Tekhniki Ser.
Sovremennye Problemy Matematiki} (Results of Science and Tech-
nology. Ser. Contemporary Problems of Mathematics) Vol. 15
(Moscow: VINITI SSSR, 1980) p. 3

\bibitem{dn-89} Dubrovin B A, Novikov S P {\it Russ. Math. Surv.} {\bf 44} (6) 35 (1989);
{\it Usp. Matem. Nauk} {\bf 44} (6) 29 (1989)

\end{thebibliography}
\end{document}